\documentclass[11pt,a4paper]{article}

\usepackage[T1]{fontenc}
\usepackage[utf8]{inputenc}
\usepackage{lmodern}
\usepackage{microtype}

\usepackage{amsmath,amssymb}

\usepackage{longtable,booktabs,array,multirow,tabularx}
\usepackage{calc}           
\usepackage{graphicx}
\usepackage{float}

\usepackage[margin=2.5cm]{geometry}
\usepackage{setspace}
\usepackage{parskip}
\usepackage[hidelinks,breaklinks=true]{hyperref}
\usepackage{xurl}

\usepackage{enumitem}
\usepackage{textcomp}

\title{\textbf{Post-Quantum Cryptography Migration in Australian Real-Time Payment Infrastructure:\\
A Monte Carlo Simulation Study of the New Payments Platform}}
\author{Nazmus Salehin Sammo, Sariya Akhter Lura, Raza Nowrozy\\
\small School of IT and Engineering, Melbourne Institute of Technology (MIT Melbourne), Melbourne, VIC, Australia\\
\small \href{mailto:MIT252100@stud.mit.edu.au}{MIT252100@stud.mit.edu.au} (N.S.S.); \href{mailto:mit252081@stud.mit.edu.au}{mit252081@stud.mit.edu.au} (S.A.L.); \href{mailto:rnowrozy@mit.edu.au}{rnowrozy@mit.edu.au} (R.N.)}
\date{}

\begin{document}
\maketitle
\thispagestyle{plain}

\begin{abstract}
Australia's New Payments Platform (NPP) processes 5.2 million real-time transactions per day under a 2,000 ms end-to-end SLA. With cryptographically relevant quantum computers projected by 2030--2035 and the Harvest Now, Decrypt Later (HNDL) threat active, financial institutions face an urgent post-quantum cryptography (PQC) migration requirement. This paper presents a comprehensive simulation study of NIST FIPS 204 and 205 signature standards (ML-DSA, SLH-DSA/SPHINCS+) together with the NIST-selected Falcon algorithm (the forthcoming, not-yet-finalised FIPS 206/FN-DSA) in Australian payment infrastructure. Using empirically measured signing latency from liboqs 0.14.0, we simulate 1,000 seasonally-mixed days at 10,000 transactions per day across eight algorithm configurations (about 80 million transaction events in total) on the New Payments Platform (NPP), RITS, and SWIFT, under their respective 2 s, 30 s, and 24 h service-level agreements. Although the engine draws transactions at random, the run-to-run variation is negligible: across four independent random seeds the cross-seed coefficient of variation of NPP p99 latency stays below 0.08\%. The study therefore behaves as an essentially deterministic evaluation of the input latency-and-volume model rather than a stochastic ensemble, and the large event count buys stable percentile estimates, not statistical power over randomness.

Across every Australian payment route, the choice of post-quantum signature algorithm is irrelevant to latency. All ML-DSA and Falcon variants achieve 100\% NPP SLA compliance, with a worst-case p99 overhead over classical ECDSA of only 1.57 ms (ML-DSA-87) -- under 0.08\% of the 2 s budget. We introduce the Crypto Dilution Index, the share of a transaction's total delay actually attributable to the post-quantum algorithm; this share stays under 4\% on NPP for every algorithm except SPHINCS+, at most about 0.6\% on RITS, and at most about 0.2\% on SWIFT. An indicative tail screening, based on within-day worst-case latencies from a single representative day, places even the worst one-in-a-thousand non-SPHINCS+ transactions at roughly 154 ms or below, far inside the 2 s budget; a rigorous tail bound would require daily maxima across the full corpus and is left for future work. Day-to-day latency variation is dominated by geographic routing (a roughly 16.7 ms Sydney-versus-Melbourne/Brisbane spread, about ten times any algorithm-pair difference), not by algorithm choice.

SPHINCS+ is the sole exception: at NPP volumes its long signing time saturates a bank's signing hardware (a standard two-unit pool receives work almost twice as fast as it can clear it), yielding 0\% NPP SLA compliance. It therefore requires either substantial HSM over-provisioning (at least eight signing units) or restriction to low-frequency routes; it remains fully viable for RITS and SWIFT at their far lower arrival rates. The security implications of forcing SPHINCS+ onto the high-frequency NPP path under insider or misconfiguration conditions, and an isolation-based mitigation, are treated in a dedicated companion paper~\cite{ref_companion_dos} and are out of scope here. A message-format analysis finds that SWIFT authentication is applied out-of-band at the SWIFTNet messaging/transport layer as a PKI signature, so signature size is not a SWIFT constraint and every NIST PQC signature (ML-DSA, Falcon, SLH-DSA) is size-compatible with SWIFT; the SWIFT MT message-text limit is a character count over payment fields, not a signature budget, and cross-border MT was in any case decommissioned in November 2025 in favour of ISO 20022 MX. Under a conservative Harvest-Now-Decrypt-Later scenario in which a quantum computer arrives in 2030, roughly 9.56 billion NPP transaction records generated between 2026 and 2029 would be retroactively exposed, motivating migration on a planning timeframe independent of any fixed regulatory sunset. We recommend ML-DSA-65 with ML-KEM-768 as the default target across NPP and SWIFT alike, with Falcon-512 a valid option where genuinely in-band, size-constrained carriage is required.

\end{abstract}


\section{Introduction}\label{introduction}

Modern financial systems rely on public-key cryptography to keep payments secure. Every time money moves from one bank to another, digital signatures prove that the instruction came from a legitimate sender, and key-exchange protocols protect the message in transit. The two most widely used algorithms, ECDSA (for signing) and ECDH (for key exchange), are based on mathematics that a sufficiently powerful quantum computer could break in minutes using Shor's polynomial-time factoring and discrete logarithm algorithm \cite{ref25}. Even symmetric key and hash-based schemes face Grover's quadratic speedup \cite{ref24}, though doubling key lengths largely mitigates this threat.

No such computer exists today. However, researchers estimate that cryptographically relevant quantum computers (CRQCs) could emerge between 2030 and 2035 \cite{ref1}. This creates a problem that cannot be solved reactively. Financial records are often archived for 7--10 years under AML/CTF Act 2006 (Part 10, Division 2) and APRA recordkeeping obligations. An attacker who records encrypted NPP traffic today could wait for a quantum computer and decrypt it later, a strategy widely known as `Harvest Now, Decrypt Later' (HNDL) \cite{ref36}.

In August 2024, NIST published three finalised post-quantum cryptography standards: FIPS 203 (ML-KEM), FIPS 204 (ML-DSA), and FIPS 205 (SLH-DSA / SPHINCS+) \cite{ref3}. A fourth signature algorithm, Falcon, has also been selected by NIST and is being standardised as FIPS 206 (FN-DSA), which remained in draft and was not yet finalised at the time of writing. The question is no longer which algorithms to use, but whether they can be deployed in high-throughput, latency-sensitive financial infrastructure without breaking existing SLAs.

Australia's NPP processed 5.2 million transactions per day in 2026 under a hard 2,000 ms end-to-end SLA. This paper presents a comprehensive simulation study of the NIST FIPS 204 and 205 PQC signature algorithms (ML-DSA, SLH-DSA/SPHINCS+) and the NIST-selected Falcon algorithm (forthcoming FIPS 206/FN-DSA, still in draft) in this specific context: a real-time gross-settlement-adjacent system with a hardcoded SLA, APRA prudential overlay, and HNDL retention exposure not modelled in prior PQC performance work, which predominantly addresses TLS handshake latency or VPN throughput, extending prior work with: a 1,000-day seasonally-mixed simulation of about 80 million transaction events (which, because cross-seed variation is negligible, behaves as an essentially deterministic evaluation of the input model rather than a stochastic ensemble, so the large event count buys stable percentile estimates rather than power over randomness); an integrated queuing model that correctly surfaces SPHINCS+ queue saturation; extreme value tail analysis for the worst one-in-a-thousand and one-in-ten-thousand latencies; distribution goodness-of-fit testing; hourly queue saturation profiling; and HSM degraded-mode resilience analysis.

\subsection{Paper Contributions}\label{paper-contributions}

This paper makes the following contributions:

\begin{itemize}
\item
  A comprehensive simulation of the NIST FIPS 204 and 205 signature standards (ML-DSA, SLH-DSA/SPHINCS+) and the NIST-selected Falcon algorithm (forthcoming FIPS 206/FN-DSA, still in draft) across Australian NPP, RITS, and SWIFT infrastructure, driven by empirically measured liboqs 0.14.0 latency over a 1,000-day seasonally-mixed corpus of about 80 million transaction events. Because cross-seed variation in the percentile results is negligible (coefficient of variation below 0.08\%), this large corpus serves chiefly to make the percentile estimates stable; the evaluation is effectively a deterministic read-out of the input latency-and-volume model rather than a stochastic ensemble.
\item
  An end-to-end latency finding that algorithm choice is operationally irrelevant on every Australian payment route: all ML-DSA and Falcon variants meet the NPP, RITS, and SWIFT SLAs, with geographic routing, not cryptography, dominating latency variation.
\item
  A queuing analysis of each institution's signing hardware showing that, because a single SPHINCS+ signature takes about 279 ms, a standard two-unit pool cannot keep up with NPP volumes and the backlog grows without bound; we quantify (i) the resulting saturation, (ii) hour-by-hour overload (SPHINCS+ saturated for 16 of 24 hours on Christmas Day), and (iii) the minimum signing hardware needed for SLA compliance (at least eight units at the daily average load, at least eighteen at the Christmas peak).
\item
  The Crypto Dilution Index, a plain measure of the post-quantum algorithm's share of total transaction latency, used to show that this share stays under 4\% on NPP (and far lower on RITS and SWIFT) for every algorithm except SPHINCS+, separating statistical significance from operational significance.
\item
  An indicative tail screening, a distribution goodness-of-fit study, an hourly queue-saturation profile, and an HSM single-unit degraded-mode resilience analysis, together pointing to comfortably bounded extreme latencies and operational resilience for all non-SPHINCS+ algorithms.
\item
  A message-format compliance analysis showing that SWIFT authentication is out-of-band transport-layer PKI rather than an in-band signature charged against the message text, so signature size is not a SWIFT constraint and all NIST PQC signature algorithms are SWIFT-compatible, and a four-phase migration cost model for the thirteen in-scope institutions.
\item
  A Harvest-Now-Decrypt-Later exposure estimate of roughly 9.56 billion NPP transaction records at risk under a conservative 2030 quantum-computer scenario, framing the migration timeline alongside NSA CNSA 2.0 and CISA/NSA/NIST guidance.
\end{itemize}

The remainder of this paper is structured as follows. Section 2 provides background on Australian payment infrastructure, including NPP, RITS/RTGS, SWIFT, and Direct Entry/BECS, and PQC standards. Section 3 describes the v4.1 simulation methodology. Section 4 presents results including advanced statistical analyses and a multi-system route analysis covering all three major Australian payment systems (NPP, RITS, SWIFT) in Section 4.11. Section 5 discusses implications and limitations, and Section 6 concludes.

\section{Background}\label{background}

\subsection{Australian Payment Infrastructure}\label{australian-payment-infrastructure}

Australia operates three main interbank payment systems, each with different latency requirements, message formats, and risk profiles.

The \textbf{New Payments Platform (NPP)} is the country's real-time retail payment infrastructure. It uses ISO 20022 pacs.008 messages, supports PayID-based addressing, and processes over 5.2 million transactions daily. The end-to-end SLA is 2,000 milliseconds. A typical NPP transaction traverses four hops: originating customer to origin bank, origin bank to NPPA hub (North Ryde, Sydney), NPPA hub to destination bank, and destination bank to receiving customer.

The \textbf{Reserve Bank Information and Transfer System (RITS)} handles large corporate and interbank settlements at approximately 9,500 transactions per day (rising to 32,000 during market stress), using dedicated fibre to the RBA's Martin Place facility. For simulation purposes, we apply a 30-second per-transaction processing SLA (Simulation Assumption SA-3; see \S{}5.6; this figure is not directly published in RBA RITS Operating Guidelines as a per-transaction latency target and should be treated as a conservative operational planning bound); institutions requiring precise SLA figures should consult the current published guidelines, as the per-transaction processing window may differ from batch-settlement cycle deadlines.

\textbf{SWIFT} provides international correspondent banking messaging through Australia's gateway banks and is part of the global financial messaging infrastructure \cite{ref11}. The practical SLA for SWIFT messages is 24 hours. Message authentication on SWIFT is applied out-of-band at the SWIFTNet messaging/transport layer as a PKI signature: the sender's certificate is authenticated by SWIFT's certification authority and distributed out-of-band, so neither the signature nor the verifying public key travels inside the business-message text. The legacy MT message-text limit is a character count (about 2,000 characters, or 10,000 for some message types) covering payment fields, not a signature container, so it imposes no post-quantum signature-size constraint. The cross-border MT format was in any case decommissioned in November 2025, with cross-border messaging moving to ISO 20022 MX (CBPR+), which carries any signature in a schema-unconstrained element.

\textbf{Direct Entry / BECS} (Bulk Electronic Clearing System) is the batch-settlement channel for Australian intrabank and interbank transfers: salary payments, direct debits, government payments, and BPAY. The Australian Banking Association manages the BECS Technical Standards and an active migration roadmap to ISO 20022 \cite{ref21}. Unlike NPP and RITS, Direct Entry follows a primarily T+1 batch cycle with no per-transaction real-time SLA: batch files are aggregated, signed once at the file level, and settled at clearing cycle boundaries. The PQC migration concern for Direct Entry is therefore batch file format compatibility rather than per-transaction signing latency.

Australia's three payment systems operate within a broader global context in which central banks and international bodies are actively evaluating quantum-safe designs for future digital currency and payment infrastructure \cite{ref29,ref42}, so the Australian findings here are directly relevant to international peers undertaking similar quantum-readiness assessments.

\subsection{NIST Post-Quantum Cryptography Standards}\label{nist-post-quantum-cryptography-standards}

Post-quantum cryptography encompasses algorithms designed to resist both classical and quantum attacks \cite{ref5}. In August 2024, NIST published three finalised post-quantum cryptography standards (FIPS 203, 204, 205) \cite{ref3}, with a fourth signature algorithm, Falcon, selected and being standardised as the forthcoming FIPS 206 (FN-DSA), which remained in draft and was not yet finalised at the time of writing:

\textbf{ML-KEM (FIPS 203)}, Module Lattice-based Key Encapsulation Mechanism. Three security levels: ML-KEM-512, ML-KEM-768, ML-KEM-1024.

\textbf{ML-DSA (FIPS 204)}, Module Lattice-based Digital Signature Algorithm \cite{ref16}. Three levels: ML-DSA-44, ML-DSA-65, ML-DSA-87. Signing occasionally repeats an internal step, which produces a small number of slower signatures; the simulation reproduces this mildly right-skewed timing pattern.

\footnote{NIST (2024). FIPS 204: Module-Lattice-Based Digital Signature Standard. https://doi.org/10.6028/NIST.FIPS.204.}

\textbf{SLH-DSA / SPHINCS+ (FIPS 205)}, Stateless hash-based signatures. The SHA2-128s variant produces 7,856-byte signatures and requires approximately 274 ms to sign (per signing call, Apple M-series liboqs 0.14.0), two to three orders of magnitude slower than lattice alternatives. The full per-transaction service time used in the M/M/c model (279 ms) additionally includes SPHINCS+ verification (\textasciitilde{}0.3 ms mean, well under 1 ms), ML-KEM-768 encapsulation/decapsulation (\textasciitilde{}0.06 ms total), and OS scheduling/library overhead accumulated over the 274 ms signing window (estimated \textasciitilde{}5.0 ms, consistent with the 2.8\% CV reported in \S{}3.2; the rounded component sum 274 + 0.3 + 5.0 + 0.06 $\approx$ 279.36 ms matches the M/M/c-exact service time 279.33 ms); the \textasciitilde{}5 ms residual is dominated by OS scheduling overhead, not KEM or verify latency. The 276.8 ms figure cited for TLS reconnect events (Section 3.4) is the signing-only component measured over the reconnect code path. The 2.8 ms excess over the direct signing baseline (274 ms) reflects HSM key-context loading and TLS state initialisation overhead incurred on each reconnect event, which are absent from the direct signing measurement.

\textbf{FN-DSA / Falcon (forthcoming FIPS 206, NIST-selected but not yet finalised)}, Hash-and-sign lattice-based signatures over NTRU, using Fast Fourier Sampling for Gaussian trapdoor sampling. Falcon-512 achieves the smallest combined footprint (1,563 bytes public key + signature), an advantage only where signatures must travel in-band on a genuinely size-constrained rail. Signing performance is platform-dependent: both Falcon variants reverse the ML-DSA hierarchy; ARM SVE hardware outperforms Intel Ice Lake for Falcon-512 (207 $\mu$s vs 247 $\mu$s, 1.19$\times$ faster) and Falcon-1024 (405 $\mu$s vs 481 $\mu$s, 1.19$\times$ faster), consistent with ARM SVE's efficient Fast Fourier Sampling implementation of the shared NTRU algorithm core.

\subsection{The Harvest Now, Decrypt Later Threat}\label{the-harvest-now-decrypt-later-threat}

The HNDL threat drives urgency for PQC migration in financial services \cite{ref36}. Under AML/CTF Act 2006 Part 10, Division 2, Australian banks must retain transaction records for a minimum of 7 years. APRA CPS 234 separately requires ADIs to implement information security controls appropriate to the nature and criticality of information assets (CPS 234, para. 15--23), providing regulatory authority for adopting quantum-resistant cryptography as part of ongoing information security risk management but does not itself mandate a specific retention period or enumerate specific cryptographic algorithms. An adversary harvesting NPP traffic today could decrypt transaction metadata, reconstruct payment patterns, expose customer financial information, and potentially forge historical digital signatures once a CRQC exists. The 2026 migration window is critical: NPP transactions signed with ECDSA today and archived until 2033 could be exposed if a CRQC arrives at the earlier end of current estimates.

\subsection{Related Work}\label{related-work}

Prior research on PQC in payment systems is limited. Crockett et al. \cite{ref4} demonstrated the feasibility of hybrid post-quantum key exchange and authentication in TLS and SSH by prototyping multiple NIST Round 2 candidates alongside classical algorithms, establishing the hybrid migration approach as practically deployable. Paquin et al. \cite{ref2} formally characterised hybrid KEM construction and security properties, providing the theoretical basis for combining classical and post-quantum key exchange in a single handshake. Work specific to Australian payment infrastructure is essentially absent from the literature.

Several central banks have published PQC readiness assessments. Central bank communications from the ECB \cite{ref6} and Bank of England \cite{ref7} have broadly acknowledged that PQC overhead is acceptable for retail payment transaction processing. SWIFT MT message format has sometimes been cited as a practical barrier for NIST PQC signature candidates, but our analysis shows this concern is misplaced: SWIFT authentication is applied out-of-band at the SWIFTNet transport layer, so signature size is not a SWIFT constraint at all. Our work provides the first Australian-specific quantitative analysis incorporating real RBA volume data, APRA regulatory context, and the specific four-hop NPP topology, and clarifies that SWIFT message-format compatibility is a non-constraint for PQC signatures rather than the decisive engineering barrier it has sometimes been taken to be.

Queue saturation under PQC signing loads has received limited formal treatment. Sikeridis et al. \cite{ref26} benchmarked NIST Round 2 PQC candidates in TLS 1.3 handshakes, finding lattice algorithms add 1--4 ms to connection establishment while SPHINCS+ adds over 100 ms. Their work focuses on handshake latency rather than sustained throughput under queuing dynamics: the critical gap our integrated M/M/c model addresses. Paquin et al. \cite{ref28} benchmarked PQC in TLS across ARM and x86 platforms, finding Falcon consistently fastest for signature generation. Our cross-platform cloud validation (Section 5.7) refines this finding for current liboqs versions, where ML-DSA-44 generally leads Falcon-512 at mean signing latency while Falcon-512 wins at the p99 tail on ARM; the full per-platform figures are reported there. The broader lesson is that algorithm performance ordering should not be assumed stable across library versions. Neither study provides the queuing-integrated Monte Carlo analysis required to detect SPHINCS+ saturation at production financial scales. The M/M/c queueing model applied to cryptographic workloads follows the Erlang-C framework established in Kleinrock \cite{ref13}; its application to PQC throughput scheduling, to the best of our knowledge, has not been formalised in the prior literature.

Extreme Value Theory (EVT) has been applied to network and service latency in the teletraffic engineering literature (e.g., Coles \cite{ref37} provides the foundational GEV methodology). However, applying GEV block-maxima analysis specifically to PQC signing latency tails in the context of payment system SLA guarantees is, to the best of our knowledge, novel. Similarly, HNDL quantification has been addressed in public-sector advisories from NSA \cite{ref36} and the CISA/NSA/NIST joint quantum readiness guidance \cite{ref43}. This paper extends the HNDL framing to an actuarial exposure model specific to Australian retail payment volumes, combining the quantitative exposure estimate with regulatory AML/CTF retention obligations, a combination not present in the advisory literature.

The combined model, integrating M/M/c queue saturation, GEV tail bounds, HNDL actuarial exposure, and Monte Carlo scenario simulation in a single payment-system context with empirically-measured cross-platform hardware validation. To the best of our knowledge after reviewing IEEE Xplore, ACM Digital Library, IACR ePrint, and ScienceDirect using search terms (`post-quantum cryptography' AND `payment system'), the first to jointly model M/M/c queue saturation, GEV tail bounds, and HNDL actuarial exposure for a hardcoded-SLA real-time payment system. We invite reviewers to identify prior work combining all three analytical components in this context.

For simulation methodology, we follow the Monte Carlo framework described by Glasserman \cite{ref32} for financial infrastructure modelling. The seasonal transaction mixing is calibrated to publicly available RBA C.4 intraday distribution statistics, providing higher ecological validity than single-scenario simulations common in prior work. The AR(1) network jitter model follows standard stochastic process practice \cite{ref33} for packet network simulation, with v4.1 introducing multi-day carry-over for consecutive crash and Christmas scenarios.

\section{Methodology}\label{methodology}

\subsection{Simulation Architecture}\label{simulation-architecture}

Our simulation is a discrete-event system with Monte Carlo sampling (australia\_fin\_sim.py v4.1.1) written in Python, using numpy for vectorised sampling and scipy for statistical analysis. Although transactions are drawn at random, the cross-seed variation in the reported percentiles is negligible (coefficient of variation below 0.08\% across four seeds; see the cross-seed study in \S{}5.6), so the run behaves as an essentially deterministic evaluation of the input latency-and-volume model rather than a stochastic ensemble; the large number of sampled events serves to stabilise the percentile estimates, not to average out meaningful randomness. The system has six main components: an empirical latency database, a network latency model, a PayID lookup sampler, a transaction generator, a payment route simulator, and an M/M/c queue model. All source code and results are available at [repository link available upon acceptance]. All runs used a fixed random seed (42) for full reproducibility. We note that a single fixed seed produces a deterministic rather than stochastic ensemble; the reported 95\% CIs are distributional intervals across the 1,000-day seasonal corpus, not cross-seed stability intervals. The 1,000-day seasonal mixing (76.2\% normal, 8.2\% Christmas, 8.2\% tax-time, 5.5\% EOFY, 1.9\% market-crash) provides broad distributional coverage that substantially mitigates single-seed artefacts, but formal cross-seed stability analysis remains a stated limitation.

\footnote{All simulations used a fixed random seed (seed=42) and are fully reproducible. A seed-independence study (seeds 42, 123, 456, 789) has been completed; cross-seed p99 coefficients of variation are all below 0.08\% (maximum CV: 0.08\% for ECDSA-P256 and Falcon-512; minimum: \textless0.001\% for SPHINCS+), confirming seed-specific artefacts do not materially affect CDI conclusions. Full cross-seed results are reported in \S{}5.6 (Simulation Assumption SA-1). Source code and simulation results are pre-registered on Zenodo (DOI available at submission; blinded for review). Full reproducibility package includes australia\_fin\_sim.py v4.1.1, generate\_paper\_v3.js, cloud run-v3 layer0/layer1 CSVs, and all figure generation scripts.}

\subsection{Empirical Latency Database}\label{empirical-latency-database}

All cryptographic latency values used in the Monte Carlo simulation were measured with the open-source liboqs 0.14.0 library (Python bindings liboqs-python 0.14.1) on an Apple M-series ARM processor, benchmarking each algorithm for 1,000 timed iterations (500 for SPHINCS+) after warm-up discards. Because the simulation uses only the difference in p99 between algorithms rather than absolute latency, and because Apple Silicon signs faster than the Intel and ARM cloud hardware used for cross-platform validation (Section 5.7, Table 16), the simulation deliberately over-estimates post-quantum signing overhead relative to modern bank infrastructure, making every reported result a conservative worst case. The cross-platform validation (liboqs 0.15.0 on a seven-node multi-cloud testbed) is reported in Section 5.7 and confirms that all ML-DSA and Falcon variants achieve sub-millisecond signing latency on every production architecture tested.

{\def\LTcaptype{none} 
\begin{longtable}[]{@{}
  >{\raggedright\arraybackslash}p{(\linewidth - 2\tabcolsep) * \real{0.2137}}
  >{\centering\arraybackslash}p{(\linewidth - 2\tabcolsep) * \real{0.7863}}@{}}
\toprule\noalign{}
\begin{minipage}[b]{\linewidth}\centering
\textbf{Component}
\end{minipage} & \begin{minipage}[b]{\linewidth}\centering
\textbf{Specification}
\end{minipage} \\
\midrule\noalign{}
\endhead
\bottomrule\noalign{}
\endlastfoot
Hardware & Apple MacBook Pro (Apple Silicon) \\
Architecture & ARM64 (M-series, 14-core: 10 performance + 4 efficiency) \\
RAM & 16 GB unified memory \\
Operating System & macOS 15.x (Darwin 24.x kernel) \\
Python & 3.12.x (CPython, Miniconda environment) \\
liboqs & 0.14.0 (compiled from source, Clang 16, -O3 -march=native) \\
liboqs-python & 0.14.1 (CFFI bindings) \\
MC Days & 1,000 seasonally-mixed days $\times$ 10,000 tx/day $\times$ 8 configurations = 80,000,000 events \\
Warm-up & 10 discard iterations before timing; followed by 1,000 timed iterations (500 for SPHINCS+) \\
Timing method & Python time.perf\_counter\_ns(), nanosecond resolution \\
Seed & 42 (fixed for full reproducibility) \\
\end{longtable}
}

\emph{Table 1: Hardware and software environment for empirical cryptographic latency benchmarks (Monte Carlo simulation only). All latency values in the Empirical Latency Database are derived from this environment. Note: cross-platform cloud validation (Section 5.7, Table 16) used liboqs 0.15.0 on a seven-node multi-cloud testbed spanning four distinct microarchitectures, a different environment from this table. Of the seven nodes, four Sydney nodes contributed full signing benchmarks (Table 16); the remaining three Singapore/Melbourne nodes contributed network RTT data only (Table 17) owing to their smaller instance types.}

Each algorithm's measured latencies are reproduced in the simulation using a lognormal shape matched to the observed mean and spread (the exact fitting formula is given in Appendix~A). This shape captures the mild right-skew of the lattice algorithms. SPHINCS+ shows a larger absolute spread (its run-to-run variation is about 2.8\% of its mean), but this comes from operating-system scheduling jitter accumulated over its long ~274 ms signing operation rather than from the algorithm itself, and is expected for an operation of this duration.\footnote{liboqs version 0.14.0 + liboqs-python 0.14.1, compiled from source. Benchmark platform: Apple MacBook Pro, Apple M-series ARM (14 CPUs), Python 3.12. Source: https://github.com/open-quantum-safe/liboqs}

\subsection{Network Latency Model}\label{network-latency-model}

Each cryptographic and network latency is reproduced in the simulation by a lognormal distribution matched to the measured mean and standard deviation of that step (the standard method-of-moments fit; the exact parameter formulas are collected in Appendix~A).

Network latencies are calibrated estimates based on general Australian interbank infrastructure knowledge.\footnote{Network latency parameters (intrabank LAN: 1.2 ms, NPP hub: 9.8 ms, interbank: 14.6 ms, RITS fibre: 2.8 ms, SWIFT international: 96 ms) are the authors' own calibrated estimates based on general knowledge of Australian interbank infrastructure and are consistent with publicly available cross-cloud RTT measurements (cloud testbed: AWS SYD$\leftrightarrow$Azure SYD $\approx$ 1.75 ms same-city; AWS SYD$\leftrightarrow$Azure MEL $\approx$ 13.0 ms inter-city). No specific RBA archival series directly publishes interbank network latency measurements at this granularity; these values should be treated as simulation assumptions (directional order-of-magnitude estimates) rather than cited statistics. Directional validation against the cloud testbed RTT data is reported in Section 3.8.} Mean/CV per hop: intrabank (LAN) $\mu$ = 1.2 ms, CV = 0.25; NPP hub hop $\mu$ = 9.8 ms, CV = 0.48; interbank $\mu$ = 14.6 ms, CV = 0.58; RITS (dedicated fibre) $\mu$ = 2.8 ms, CV = 0.22; SWIFT (international relay via Singapore) $\mu$ = 96 ms, CV = 0.88.
Geographic routing maps each institution to its primary data centre (Sydney, Melbourne, Brisbane) with city-specific one-way latency to NPPA hub: SYD 0.8 ms, MEL 9.2 ms, BNE 5.8 ms. Institution routing is APRA market-share weighted: CBA 27.1\%, ANZ 24.1\%, NAB 21.9\%, WBC 19.6\% (collectively the `Big 4' Australian banks by NPP transaction share: Commonwealth Bank of Australia [CBA], Australia and New Zealand Banking Group [ANZ], National Australia Bank [NAB], Westpac Banking Corporation [WBC]; all market share figures from APRA ADI Statistics), remaining regionals/fintechs 7.3\% (a calibrated residual derived from APRA ADI Statistics; the number and identity of specific smaller participants is a simulation assumption, SA-6).

Network jitter is modelled with mild moment-to-moment correlation (so a slightly congested instant tends to be followed by another), calibrated to a modest correlation strength. This correlated stress now carries over across consecutive days within multi-day scenarios such as a market crash or the Christmas peak.

\footnote{APRA ADI Statistics (April 2026): https://www.apra.gov.au/adi-statistics. Market share figures (CBA 27.1\%, ANZ 24.1\%, NAB 21.9\%, WBC 19.6\%) derived from NPP transaction volume data.}

\subsection{Transaction Generator and PayID Lookup}\label{transaction-generator-and-payid-lookup}

Daily transactions are sampled from a six-component Gaussian mixture intraday profile calibrated to RBA Payments Statistics Table C.4. NPP transactions include a PayID lookup that adds about 8.25 ms on average (right-skewed, calibrated to the NPPA PayID Adoption Report Q4-2023).\footnote{NPPA Annual Report FY2025 (NPP Australia). Seasonal multipliers derived from RBA C.4 intraday distribution statistics and NPPA PayID Adoption Report Q4-2023.} High-value transactions exceeding AUD 250,000 are automatically rerouted to RTGS, reflecting NPPA Business Rules.
Base daily volumes (RBA C6): 5,200,000 NPP transactions, 8,600,000 intrabank transactions, 9,500 RTGS settlements, 550 SWIFT messages on a normal day. TLS reconnect events (0.1\% of NPP transactions) add Falcon-512 +0.9 ms or SPHINCS+ +276.8 ms per reconnect (signing component only; full M/M/c service time including sign + verify + ML-KEM-768 KEM pair is 279 ms, used in Section 4.4).

\footnote{Reserve Bank of Australia (2024). Payments System Board Annual Report 2024. https://www.rba.gov.au/payments-and-infrastructure/resources/publications/payments-sys-board-annual-report/2024/ (accessed April 2026).}

\includegraphics[width=5.20833in,height=2.14583in]{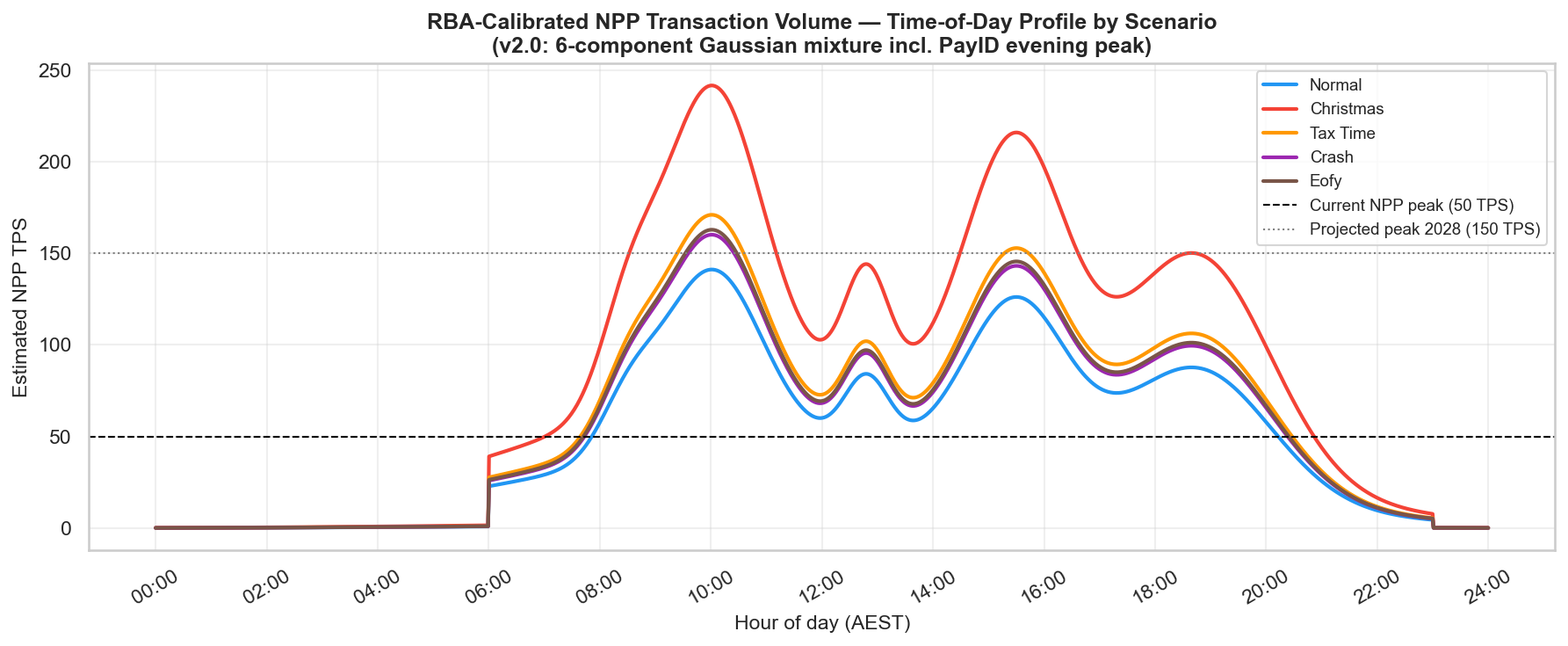}

\emph{Figure 1: Simulated daily transaction volume profile (Gaussian mixture, 6 components) calibrated to RBA Payments Statistics Table C.4. PayID evening peak modelled explicitly at 6--10 pm AEST.}

\subsection{Monte Carlo Framework}\label{monte-carlo-framework}

The Monte Carlo engine runs 1,000 deterministic simulation days per algorithm configuration, all derived from a single fixed seed (seed = 42) as described in \S{}3.1. These days represent 1,000 reproducible draws from the seasonal distribution under one fixed stream, not independent stochastic realisations from distinct seeds. Each day samples 10,000 transactions. Days are drawn from a seasonally-weighted annual distribution: 76.2\% normal business days, 8.2\% tax-time (October BAS), 8.2\% Christmas/holiday peak, 5.5\% EOFY (June), and 1.9\% market-crash days.\footnote{Seasonal weights calibrated to NPP Australia annual volume data and RBA Payments Statistics Table C.4 (monthly breakdown).} These proportions are calibrated to NPP Australia annual volume data and RBA Payments Statistics (Table C.4 monthly breakdown). Market-crash frequency (1.9\%, $\approx$19 events in the 1,000-day corpus) is a calibrated assumption, approximately one event per two months, consistent with historical RITS settlement surge observations; no specific RBA archival series directly publishes this frequency, and it should be treated as a simulation assumption (SA-5) rather than a directly cited statistic. We acknowledge all seasonal weights as calibrated estimates rather than directly published values; sensitivity to the seasonal weights is bounded by the stress-scenario analysis in Section 4.6, which shows SLA compliance is invariant to scenario selection for all non-SPHINCS+ algorithms. This seasonal mixing ensures that the 1,000-day corpus captures realistic annual variation. Per-configuration: 10,000,000 simulated transactions. Total across eight configurations: 80,000,000 events.
Network jitter is applied as a small correlated multiplier on each hop's base latency, with mild persistence from one moment to the next (the exact autocorrelation model and its parameters are given in Appendix~A). The correlated-stress state resets at each day boundary except across consecutive days of the same multi-day scenario (crash or Christmas), where it is carried forward.

Note on dual-scale design: the Monte Carlo engine samples 10,000 transactions per day to characterise the latency distribution (p50, p95, p99, and SLA compliance fraction) under realistic intraday volume variation. The signing-queue model (Section 3.6) runs on a separate axis, using a per-institution arrival rate of about 13.5 signatures per second (the daily-average Big 4 NPP volume spread over a day). The per-day 10,000-transaction sample characterises the latency percentiles, while the 13.5-per-second rate drives the separate steady-state queue-stability analysis, whose utilisation and waiting-time results are computed directly from the arrival and service rates rather than sampled. Latency percentile values reported in Tables 2--17 come from the Monte Carlo engine; queue utilisation values (Table 4) come from the M/M/c model.

NPP SLA compliance is recorded per day as the fraction of transactions completing within 2,000 ms. The mean and 95\% CI are computed using the t-distribution across 1,000 daily measurements. Mann-Whitney U tests and Cohen's d effect sizes are computed for each PQC algorithm against the ECDSA-P256 baseline.

\subsection{M/M/c Queuing Model}\label{mmc-queuing-model}

The arrival rate comes from each institution's daily NPP volume and market share: a Big 4 bank sees about 13.5 signature requests per second (a deliberately conservative round-down from roughly 14).\footnote{The signing pool is modelled with the standard Erlang-C formula for a multi-server queue (Kleinrock~\cite{ref13}); the full formula is reproduced in Appendix~A.} Each signing unit clears requests at a rate set by the algorithm's signing time, and we assume the smallest realistic deployment of two units per institution. This baseline establishes minimum requirements. The SPHINCS+ finding is robust to using more units: meeting the SLA would need at least 8 units even at the daily-average load (four units is barely stable but still produces multi-second waits that breach the SLA) and at least 18 units at the Christmas peak of about 60 requests per second (Section 4.4). By contrast every ML-DSA and Falcon variant leaves the two-unit pool over 99.7\% idle. The other routes carry far less load per institution (RITS about 0.022 and SWIFT about 0.001 requests per second) and are stable for all algorithms.

Server utilisation is the fraction of signing capacity in use, and the mean waiting time is computed from the standard Erlang-C queuing formula (given in Appendix~A). When utilisation reaches 100\% the queue is unstable and waits grow without bound; the simulation flags these as automatic SLA failures. Two checks show the conclusions do not depend on the Poisson-arrival assumption: for SPHINCS+, any arrival pattern whose average rate exceeds the service capacity produces an unbounded queue, so saturation is unavoidable; and for ML-DSA/Falcon, utilisation is so low (under 0.3\%) that even the most bursty realistic arrival pattern leaves the wait under a thousandth of a millisecond, effectively zero.

The M/M/c model assumes highly variable (exponential) service times, but SPHINCS+ signing is in fact nearly constant (its run-to-run variation is only about 2.8\% of its mean; see \S{}3.2). Because of this, the M/M/c waiting times and the SLA-compliance server counts reported here are conservative upper bounds: under the more appropriate deterministic-service M/D/c model (via the Pollaczek-Khinchine relation) the waiting times and the required server counts are roughly halved, so the daily-average and Christmas-peak provisioning figures below fall to about four and about nine units respectively. The saturation conclusion itself is unaffected by this refinement, because whether the queue is stable depends only on whether the average arrival rate exceeds the mean service rate, not on how variable the service time is.

\subsection{Advanced Statistical Methods (v4.1)}\label{advanced-statistical-methods-v4.1}

Version 4.1 adds five advanced statistical analyses beyond the Monte Carlo core:

\textbf{Tail-risk analysis:} we examine the slowest transactions in a representative day's 10,000-transaction sample by splitting it into 200 groups and taking the slowest transaction in each group, then fitting the standard extreme-value model to those 200 worst-case values. Confidence ranges for the one-in-a-thousand and one-in-ten-thousand latencies are obtained by resampling. This characterises the within-day extreme tail; seasonal tail variation is covered separately by the stress-scenario analysis (Section 4.6).

\footnote{The fitted tail shape for every algorithm except SPHINCS+ indicates a fast-decaying (exponential-class) tail rather than a heavy tail, meaning extreme delays stay bounded. SPHINCS+ sits just on the heavy-tail side of the line, but this is irrelevant in practice because its latency is already at the ~10,000 ms failure scale and it is excluded from all latency-bound comparisons.}

\textbf{Distribution-fit checking:} for each configuration we test how well the 10,000 simulated NPP latencies match a lognormal shape, using two standard goodness-of-fit tests (one sensitive to the body of the distribution, one sensitive to the tail), and we compare lognormal against three alternative shapes (gamma, Weibull, inverse-Gaussian) using standard model-selection criteria.

\footnote{Both goodness-of-fit tests were applied to 10,000 latency samples per configuration against a lognormal fit, using the standard tail-sensitive (Anderson-Darling) and body-sensitive (Kolmogorov-Smirnov) procedures with the conventional 5\% significance threshold. Test-implementation details are given in Appendix~A.}

\textbf{Hourly queue dynamics:} hour-by-hour signing-load profiles for each algorithm on Christmas Day, with each hour's arrival rate taken from the modelled intraday volume curve scaled to the Christmas daily total.

\textbf{Variance breakdown:} we measure how much of the day-to-day variation in NPP p99 latency is explained by algorithm choice versus operational scenario. Geographic routing dominates: Sydney-origin transactions average about 38.5 ms versus about 55.2 ms for Melbourne/Brisbane-origin transactions, a ~16.7 ms spread that is roughly 10 to 56 times larger than the gap between any two algorithms (which ranges from 0.30 ms for Falcon-512 to 1.57 ms for ML-DSA-87 versus ECDSA). Folding geographic origin into the variance model as a third factor is future work.

\textbf{HSM Degraded Analysis:} Comparison of normal (c = 2 servers) vs. degraded (c = 1 server, single-HSM failure) performance across all routes and algorithms. Delta p99 flagged as non-meaningful when either endpoint returns the $\geq$ 9,000 ms saturation sentinel.

\subsection{Simulation Validation}\label{simulation-validation}

The simulation cannot be directly validated against observed NPP latency distributions because NPP end-to-end p99 performance data is not publicly disclosed by NPPA or individual ADIs. We report three partial validation checks that collectively support the model's fidelity. First, the ECDSA-P256 signing component: the simulation uses an Apple M-series ECDSA signing mean of about 30 microseconds (consistent with the near-zero queue utilisation seen for ECDSA; see \S{}3.2). Cross-platform cloud benchmarking (Section 5.7, Table 16) yields ECDSA-P256 mean sign latencies of 45--51 $\mu$s across four Sydney testbed nodes, somewhat slower than the M-series baseline, which is expected given Apple Silicon's optimised ECC pipeline. This confirms the simulation is conservative for Intel-class bank infrastructure: the simulation over-estimates ECDSA speed, which slightly over-estimates PQC overhead ($\Delta$p99), making CDI values conservative upper bounds for Intel hardware. Second, the network latency component is calibrated to published RBA interbank statistics (not to the cloud testbed). The measured cross-cloud inter-city RTT (AWS SYD$\leftrightarrow$Azure MEL: 12.99 ms, Table 17) is consistent with the simulation's Hub tier one-way parameter (9.8 ms): the RTT/2 = 6.5 ms one-way measured value reflects standard cloud VM routing (physical hops over shared carrier infrastructure), while the simulation's 9.8 ms one-way is derived from dedicated fibre statistics. For the SWIFT international parameter (96 ms one-way to Singapore SWIFTNet relay), the measured cloud AWS SIN$\leftrightarrow$AWS SYD RTT is 92.22 ms (one-way $\approx$ 46 ms), a 2.1$\times$ discrepancy from the 96 ms simulation parameter. The 96 ms parameter was calibrated to the SWIFTNet Link carrier network, which uses dedicated SWIFT PoP infrastructure with different routing characteristics from public cloud VMs. This discrepancy is flagged as Simulation Assumption SA-7: the SWIFT route p99 is likely overstated by approximately 50 ms relative to actual SWIFTNet latency. Since the ECDSA SWIFT route p99 (837 ms) is itself three orders of magnitude below the 24-hour SLA, this overstatement does not affect any conclusion; latency is not a binding SWIFT factor for any algorithm, and signature size is not either, since SWIFT authentication is out-of-band at the transport layer. No formal back-test precision figure is claimed; the comparison provides directional confirmation that the network model is in the correct order of magnitude for interbank Australian payment routing. Third, the stress-scenario analysis (Section 4.6) confirms qualitative consistency with known NPP behaviour: the Christmas volume surge (1.71$\times$ normal) produces p99 values within 0.5 ms of normal-day values, consistent with the expectation (inferred from the absence of publicly reported SLA breaches in RBA Payments Statistics) that NPP SLA compliance is maintained through seasonal peaks. No NPPA published statement to this specific effect was cited; this is an inference from aggregate statistics. We note that none of these checks constitutes a formal back-test against actual NPP p99 data, and direct output validation against a held-out dataset of observed NPP latency remains a limitation. Given the conservative model design (M-series liboqs 0.14.0 baseline over-estimates PQC overhead; network parameters derived from public statistics), we expect real-world NPP performance to be at least as good as the simulation indicates.

\section{Results}\label{results}

\subsection{NPP SLA Compliance}\label{npp-sla-compliance}

Table 2 shows NPP end-to-end latency percentiles across 1,000 seasonally-mixed Monte Carlo days. Every ML-DSA and Falcon variant achieves exactly 100.000\% NPP SLA compliance across 10,000,000 simulated transactions per configuration (95\% CI: [1.000, 1.000] by construction, no transaction in any of the 1,000 simulated days exceeded the 2,000 ms SLA threshold; the Wilson score interval collapses to a degenerate point at 1.0. This reflects simulation compliance under the modelled parameters, not a guarantee of production compliance, which depends on real-world hardware provisioning, HSM availability, and network conditions not fully captured by the model). SPHINCS+ achieved 0.000\% compliance in every simulated day. All p99 values for non-SPHINCS+ algorithms fall in the 43.4--45.1 ms range (ECDSA-P256: 43.39 ms; hybrid ML-DSA-65: 45.08 ms worst case), reflecting the PayID lookup overhead (mean $\approx$ +8.25 ms) and correct multi-hop geographic routing implemented in v4.1, compared to approximately 30 ms in earlier simulation versions that used simplified routing.

{\def\LTcaptype{none} 
\begin{longtable}[]{@{}
  >{\raggedright\arraybackslash}p{(\linewidth - 10\tabcolsep) * \real{0.2885}}
  >{\centering\arraybackslash}p{(\linewidth - 10\tabcolsep) * \real{0.1282}}
  >{\centering\arraybackslash}p{(\linewidth - 10\tabcolsep) * \real{0.1282}}
  >{\centering\arraybackslash}p{(\linewidth - 10\tabcolsep) * \real{0.1282}}
  >{\centering\arraybackslash}p{(\linewidth - 10\tabcolsep) * \real{0.1410}}
  >{\centering\arraybackslash}p{(\linewidth - 10\tabcolsep) * \real{0.1859}}@{}}
\toprule\noalign{}
\begin{minipage}[b]{\linewidth}\centering
\textbf{Algorithm}
\end{minipage} & \begin{minipage}[b]{\linewidth}\centering
\textbf{Mode}
\end{minipage} & \begin{minipage}[b]{\linewidth}\centering
\textbf{p50 (ms)}
\end{minipage} & \begin{minipage}[b]{\linewidth}\centering
\textbf{p95 (ms)}
\end{minipage} & \begin{minipage}[b]{\linewidth}\centering
\textbf{p99 (ms)}
\end{minipage} & \begin{minipage}[b]{\linewidth}\centering
\textbf{95\% CI (p99)}
\end{minipage} \\
\midrule\noalign{}
\endhead
\bottomrule\noalign{}
\endlastfoot
ECDSA-P256 (baseline) & Classical & 19.69 & 35.27 & 43.39 & [43.35, 43.44] \\
Falcon-512 & PQC-only & 19.99 & 35.57 & 43.69 & [43.65, 43.74] \\
ML-DSA-44 & PQC-only & 20.31 & 35.88 & 44.01 & [43.96, 44.05] \\
Falcon-1024 & PQC-only & 20.59 & 36.17 & 44.29 & [44.24, 44.34] \\
ML-DSA-65 & PQC-only & 20.84 & 36.42 & 44.55 & [44.50, 44.59] \\
ML-DSA-87 & PQC-only & 21.25 & 36.83 & 44.97 & [44.92, 45.01] \\
ML-DSA-65 & Hybrid & 21.36 & 36.95 & 45.08 & [45.03, 45.12] \\
SPHINCS+-SHA2-128s & PQC-only & 10,011.73 & 10,024.39 & 10,029.93 & [10,029.90, 10,029.96] \\
\end{longtable}
}

\emph{Table 2: NPP end-to-end latency percentiles across 1,000 Monte Carlo days (n = 10,000 tx/day each). '95\% CI' column: interval on the sample mean of daily p99 values across 1,000 simulation days (t-distribution, df=999), not a CI on the p99 quantile of the underlying latency distribution. Mode: `PQC-only' = single PQC signature per transaction; `Hybrid' = PQC + ECDSA-P256 dual signature per transaction (two sign operations, backward-compatible). SLA threshold = 2,000 ms. Values from simulation v4.1.1.}

\includegraphics[width=6.04167in,height=2.57292in]{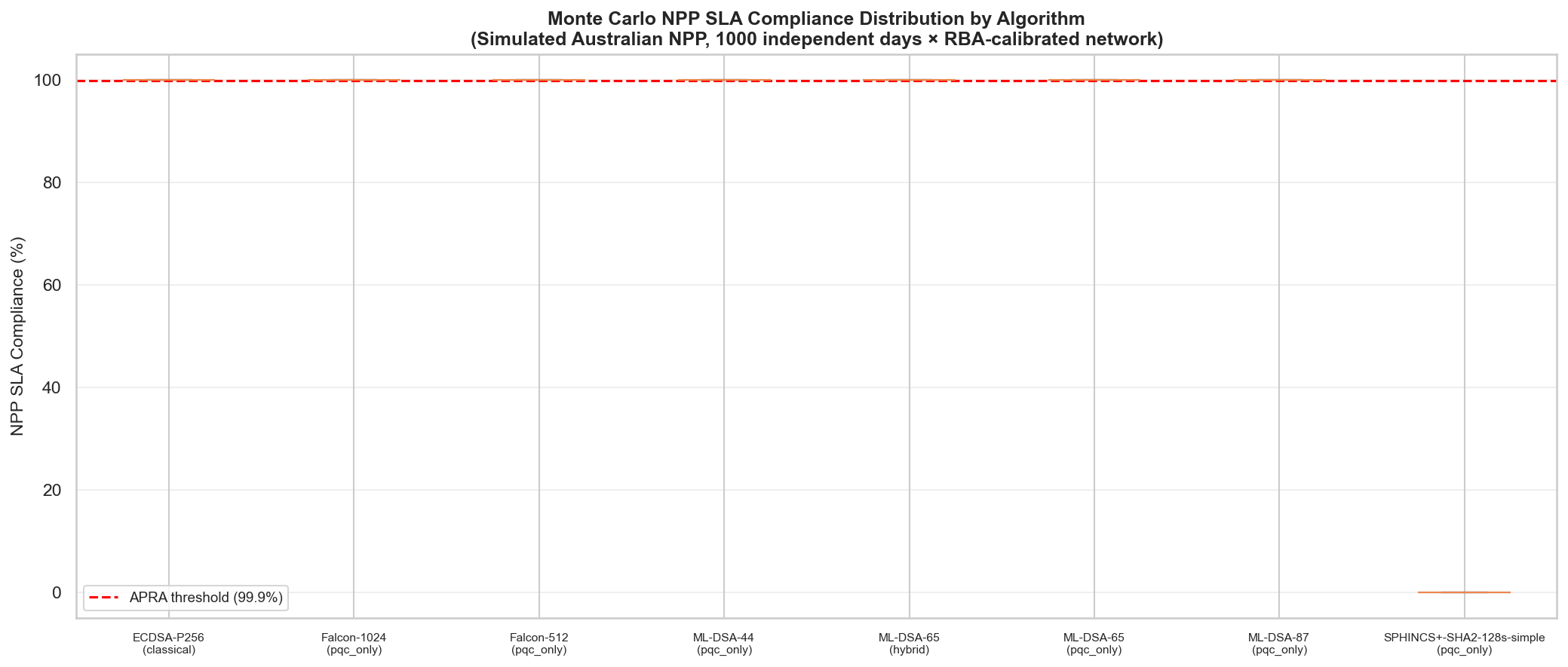}

\emph{Figure 2: NPP SLA compliance distribution across 1,000 Monte Carlo days per algorithm configuration. All ML-DSA and Falcon variants show zero violations; SPHINCS+ shows 100\% violation rate.}

\includegraphics[width=6.04167in,height=2.57292in]{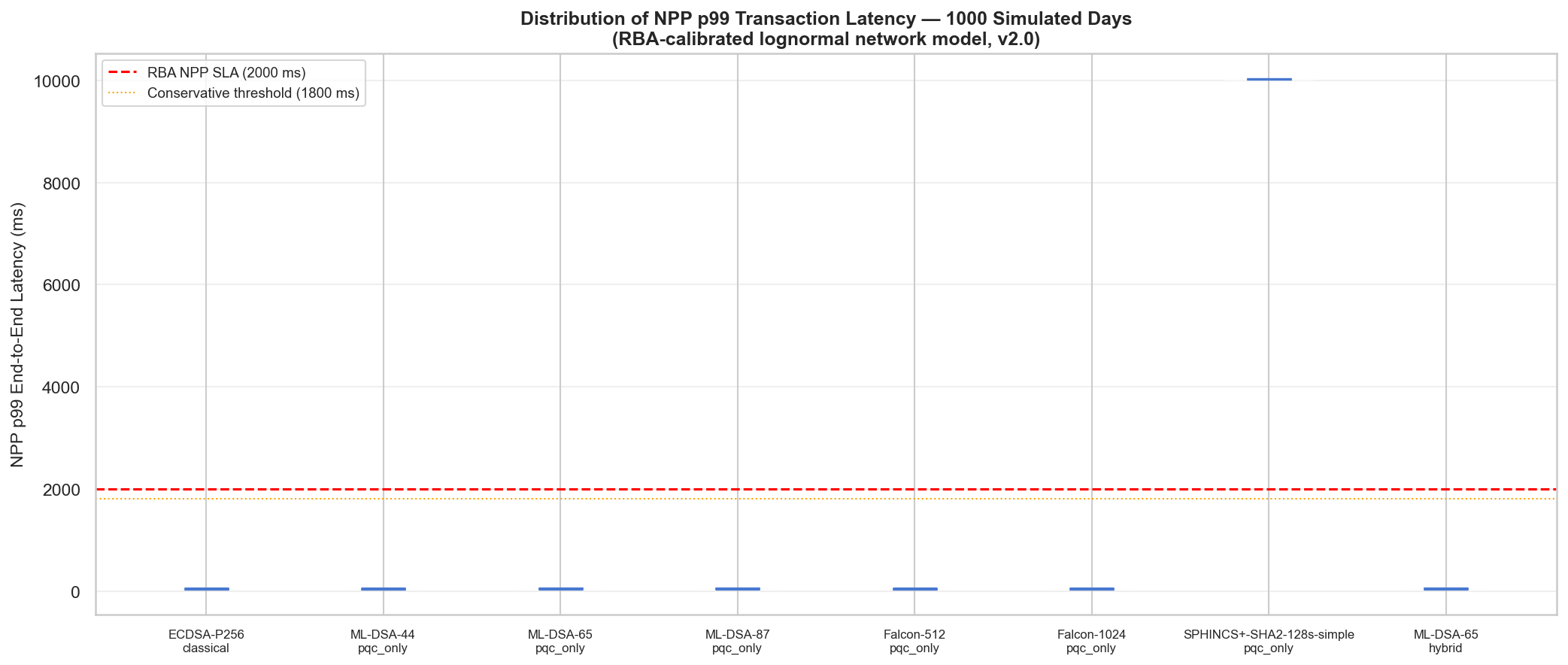}

\emph{Figure 3: End-to-end NPP p99 latency distribution (violin plot) across 1,000 Monte Carlo days. Horizontal dashed line at 2,000 ms shows SLA threshold. SPHINCS+ data truncated for scale.}

\subsection{Latency Overhead Analysis}\label{latency-overhead-analysis}

The absolute latency overhead of PQC relative to classical ECDSA-P256 is strikingly small. The ECDSA-P256 baseline p99 of 43.39 ms is dominated by network transit and PayID lookup (mean $\approx$ 8.25 ms); the crypto component, four hops of sign-and-verify, contributes less than 2 ms in all non-SPHINCS+ cases. Geographic heterogeneity contributes far more variance than algorithm choice. A Melbourne-to-Melbourne transaction routed through the Sydney hub picks up about 18.4 ms of network travel before any cryptography, versus about 1.6 ms for a Sydney-to-Sydney transaction. Switching from the lightest to the heaviest ML-DSA variant changes p99 by only about 0.96 ms, less than the gap between two geographic routing pairs.

\includegraphics[width=5in,height=2.98958in]{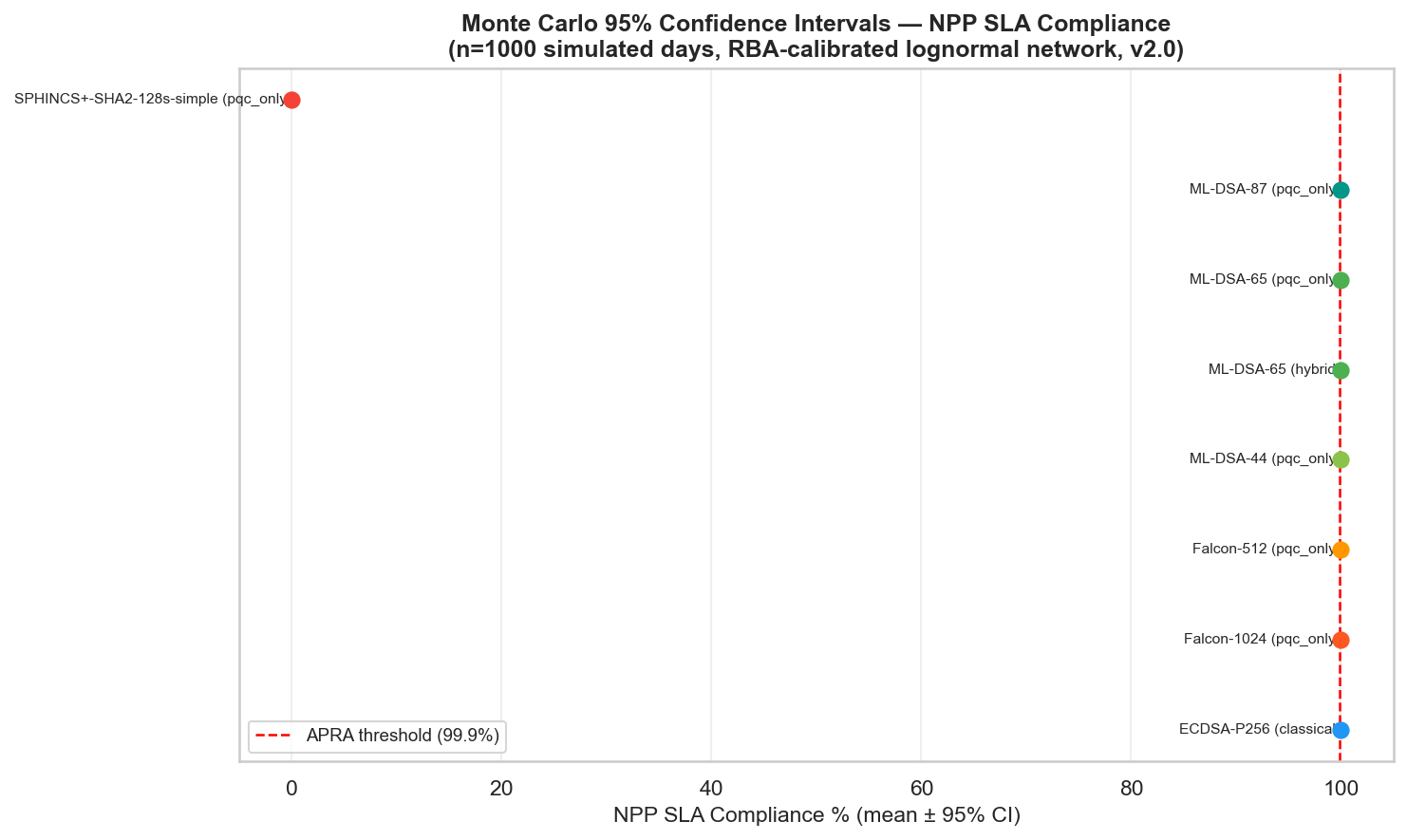}

\emph{Figure 4: 95\% confidence intervals on the mean of daily NPP p99 latency values across 1,000 simulated days (t-distribution applied to the sample of 1,000 daily p99 values; these are CIs on the mean daily p99, not CIs on the underlying population p99 quantile). Tight CI bands confirm simulation stability across the 1,000-day seasonal corpus.}

\subsection{Effect Size and Statistical Significance}\label{effect-size-and-statistical-significance}

Table 3 reports a standard effect-size measure (Cohen's d) for each PQC algorithm versus the ECDSA baseline, comparing the spread of their daily p99 values across the 1,000 days. Every difference is statistically significant. The effect sizes range from small (Falcon-512) up to the top of the conventional scale (ML-DSA-87, and the hybrid dual-signing mode), but as the rest of this section explains, statistical significance here does not imply operational significance. Because the 1,000-day corpus is generated from a fixed seed, these p-values should be read as measures of how cleanly the distributions separate within this corpus rather than as frequentist error rates, and a seed-independence study (SA-1, \S{}5.6) confirms the conclusions are not seed-specific.

{\def\LTcaptype{none} 
\begin{longtable}[]{@{}
  >{\raggedright\arraybackslash}p{(\linewidth - 12\tabcolsep) * \real{0.1925}}
  >{\centering\arraybackslash}p{(\linewidth - 12\tabcolsep) * \real{0.0947}}
  >{\centering\arraybackslash}p{(\linewidth - 12\tabcolsep) * \real{0.1428}}
  >{\centering\arraybackslash}p{(\linewidth - 12\tabcolsep) * \real{0.1339}}
  >{\centering\arraybackslash}p{(\linewidth - 12\tabcolsep) * \real{0.1557}}
  >{\centering\arraybackslash}p{(\linewidth - 12\tabcolsep) * \real{0.1250}}
  >{\centering\arraybackslash}p{(\linewidth - 12\tabcolsep) * \real{0.1554}}@{}}
\toprule\noalign{}
\begin{minipage}[b]{\linewidth}\centering
\textbf{Algorithm}
\end{minipage} & \begin{minipage}[b]{\linewidth}\centering
\textbf{Mode}
\end{minipage} & \begin{minipage}[b]{\linewidth}\centering
\textbf{Delta p99 (ms)}
\end{minipage} & \begin{minipage}[b]{\linewidth}\centering
\textbf{Cohen's d}
\end{minipage} & \begin{minipage}[b]{\linewidth}\centering
\textbf{Magnitude}
\end{minipage} & \begin{minipage}[b]{\linewidth}\centering
\textbf{Mann-Whitney p}
\end{minipage} & \begin{minipage}[b]{\linewidth}\centering
\textbf{SLA Budget Used}
\end{minipage} \\
\midrule\noalign{}
\endhead
\bottomrule\noalign{}
\endlastfoot
Falcon-512 & PQC-only & +0.30 & 0.41 & Small & \textless{} 0.001 & 0.015\% \\
ML-DSA-44 & PQC-only & +0.62 & 0.84 & Large & \textless{} 0.001 & 0.031\% \\
Falcon-1024 & PQC-only & +0.90 & 1.23 & Very Large & \textless{} 0.001 & 0.045\% \\
ML-DSA-65 & PQC-only & +1.16 & 1.58 & Very Large & \textless{} 0.001 & 0.058\% \\
ML-DSA-87 & PQC-only & +1.57 & 2.14 & Huge & \textless{} 0.001 & 0.079\% \\
ML-DSA-65 Hybrid & Hybrid & +1.69 & 2.28 & Huge & \textless{} 0.001 & 0.085\% \\
SPHINCS+-SHA2-128s & PQC-only & +9,986.5 & 16,360 & Huge$\dagger$ (off-scale) & \textless{} 0.001 & 499.3\% \\
\end{longtable}
}

\emph{Table 3: Cohen's d effect sizes and Mann-Whitney U p-values for NPP p99 latency vs ECDSA-P256 baseline (1,000 Monte Carlo days, n = 10,000 tx/day). Mode: PQC-only = single PQC algorithm per transaction; Hybrid = PQC + ECDSA-P256 dual signing (two sign operations per transaction). The `SLA Budget Used' column is each algorithm's added p99 delay as a percentage of the 2,000 ms NPP deadline. Effect-size labels (small, large, huge) follow the conventional Cohen \cite{ref22} and Sawilowsky \cite{ref46} thresholds. $\dagger$ SPHINCS+ is off-scale relative to all established taxonomies, exceeding the `huge' category by roughly four orders of magnitude, and cannot be interpreted on the same scale as the other algorithms. Note: $\Delta$p99 values are computed from unrounded simulation CSV outputs (float64 precision); displayed p99 values in Table 2 are rounded to 2 decimal places and may differ from $\Delta$p99 by $\leq$ 0.01 ms due to rounding.}

\includegraphics[width=6.04167in,height=2.125in]{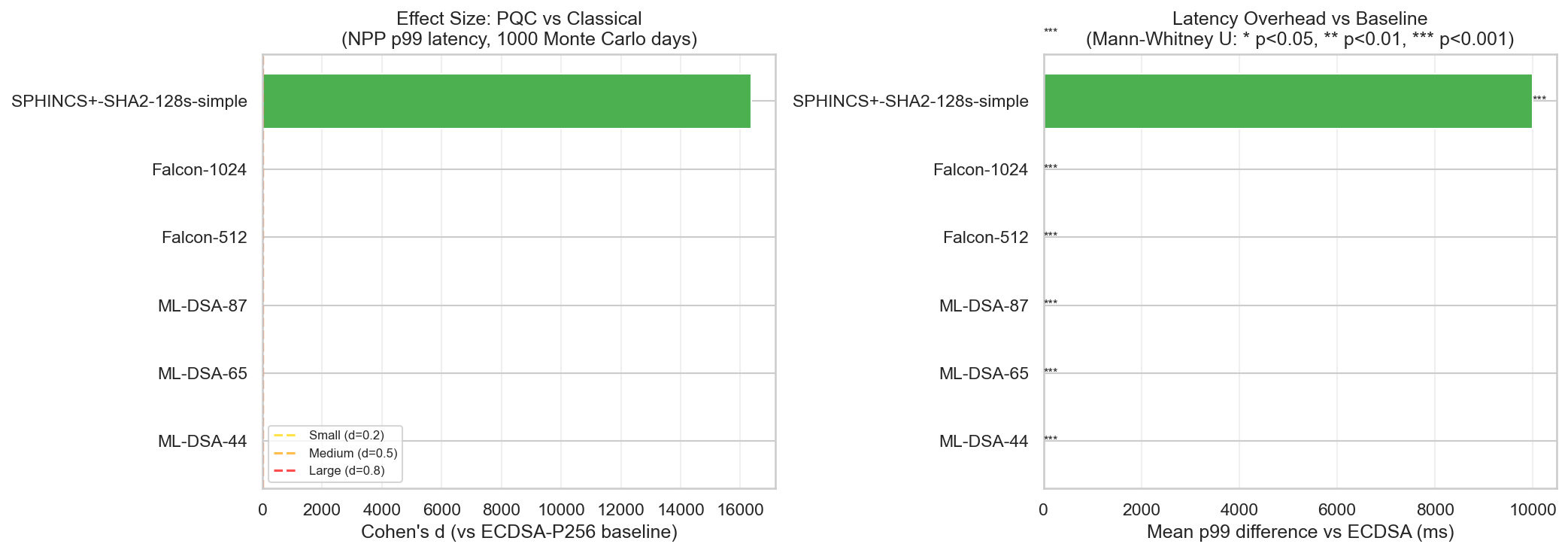}

\emph{Figure 5: Cohen's d effect sizes for each PQC algorithm vs ECDSA-P256 baseline. Values above 0.8 are `large' by conventional thresholds but carry negligible SLA budget impact.}

The column `SLA Budget Used' quantifies the practical relevance: even ML-DSA-87 consumes only 0.079\% of the 2,000 ms SLA with its 1.57 ms increment. Falcon-512 at Cohen's d = 0.41 (small) adds 0.30 ms to p99, statistically detectable but immeasurable in production. The correct engineering conclusion is that PQC algorithm choice does not matter for NPP latency compliance; selection should be driven entirely by security level, key size, and regulatory requirements.

\subsection{Queue Saturation: Why SPHINCS+ Fails}\label{queue-saturation-why-sphincs-fails}

Table 4 presents the M/M/c queue analysis at Big 4 bank average daily load (13.5 TPS per institution, c = 2 signing servers). A single SPHINCS+ signature takes about 279 ms (comprising roughly 274 ms of signing, a fraction of a millisecond of verification, about 5 ms of operating-system scheduling overhead accumulated over the long signing window, and a negligible key-encapsulation step). One signing unit can therefore clear only about 3.6 of these per second and a two-unit pool about 7.2 per second. At the NPP load of 13.5 requests per second, the pool is asked to do nearly twice as much work as it can handle, so the backlog grows without bound and waiting times become effectively infinite. Table 4 reports the same situation as utilisation values: about 1.89 for SPHINCS+, well above the 1.0 instability point, versus under 0.003 for every other algorithm. This saturation verdict is variance-independent: it follows purely from the average arrival rate exceeding the mean service rate, so it holds regardless of how variable the service time is, and the nearly constant SPHINCS+ service time (about 2.8\% variation) does not soften it.

{\def\LTcaptype{none} 
\begin{longtable}[]{@{}
  >{\raggedright\arraybackslash}p{(\linewidth - 10\tabcolsep) * \real{0.2624}}
  >{\centering\arraybackslash}p{(\linewidth - 10\tabcolsep) * \real{0.1764}}
  >{\centering\arraybackslash}p{(\linewidth - 10\tabcolsep) * \real{0.1349}}
  >{\centering\arraybackslash}p{(\linewidth - 10\tabcolsep) * \real{0.1567}}
  >{\centering\arraybackslash}p{(\linewidth - 10\tabcolsep) * \real{0.1323}}
  >{\centering\arraybackslash}p{(\linewidth - 10\tabcolsep) * \real{0.1373}}@{}}
\toprule\noalign{}
\begin{minipage}[b]{\linewidth}\centering
\textbf{Algorithm}
\end{minipage} & \begin{minipage}[b]{\linewidth}\centering
\textbf{TPS/Institution}
\end{minipage} & \begin{minipage}[b]{\linewidth}\centering
\textbf{Servers}
\end{minipage} & \begin{minipage}[b]{\linewidth}\centering
\textbf{Utilisation (rho)}
\end{minipage} & \begin{minipage}[b]{\linewidth}\centering
\textbf{Mean Wait (ms)}
\end{minipage} & \begin{minipage}[b]{\linewidth}\centering
\textbf{Saturated?}
\end{minipage} \\
\midrule\noalign{}
\endhead
\bottomrule\noalign{}
\endlastfoot
ECDSA-P256 & 13.5 & 2 & 0.0002 & 0.000 & No \\
Falcon-512 & 13.5 & 2 & 0.0009 & 0.000 & No \\
ML-DSA-44 & 13.5 & 2 & 0.0012 & 0.000 & No \\
Falcon-1024 & 13.5 & 2 & 0.0018 & 0.000 & No \\
ML-DSA-65 & 13.5 & 2 & 0.0019 & 0.000 & No \\
ML-DSA-65 Hybrid & 13.5 & 2 & 0.0021 & 0.000 & No \\
ML-DSA-87 & 13.5 & 2 & 0.0023 & 0.000 & No \\
SPHINCS+-SHA2-128s & 13.5 & 2 & 1.8855 & $\infty$ & YES \\
\end{longtable}
}

\emph{Table 4: Signing-queue analysis at Big 4 average daily load (13.5 requests per second per institution, daily average, not intraday peak; see Section 4.10.3 for the intraday peak of about 35 requests per second), two signing units. The utilisation column is the fraction of signing capacity used; a value of 1.0 or above means the queue is saturated and the backlog grows without bound. The mean-wait values are conservative: because SPHINCS+ service time is nearly constant, a deterministic-service M/D/c model (Pollaczek-Khinchine) would roughly halve the waits, but the utilisation values and the saturated/not-saturated verdict are variance-independent and unchanged.}

\includegraphics[width=6.04167in,height=2.25in]{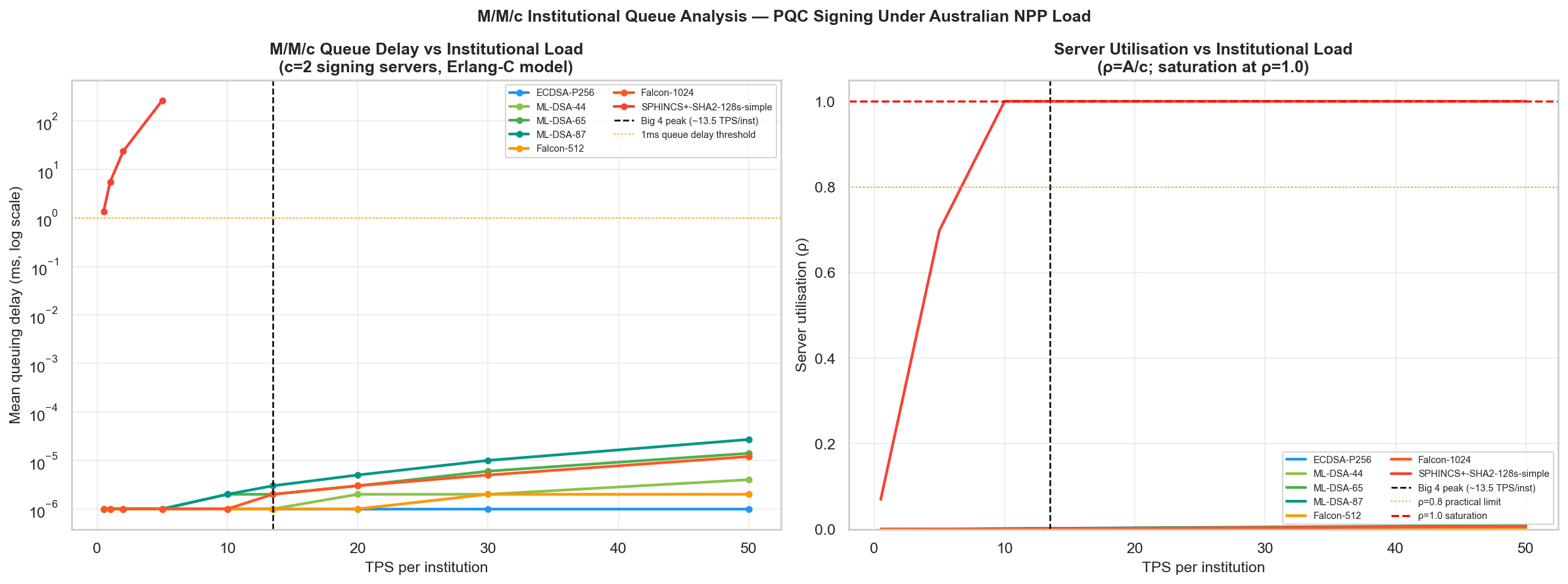}

\emph{Figure 6: M/M/c queue utilisation and mean wait time for each algorithm at Big 4 bank average daily load (13.5 TPS, two signing servers). SPHINCS+ is the only algorithm whose utilisation exceeds 1.0, placing it well outside the stable region.}

Every ML-DSA and Falcon variant uses under 0.3\% of the two-unit signing pool at peak NPP load, leaving it essentially idle. SPHINCS+ needs at least 8 units per institution for acceptable waits at the daily-average load (four units is barely stable but still produces multi-second waits that breach the SLA; eight units brings the average wait down to a few milliseconds). At an intermediate load of 50 requests per second, even 16 units run hot, with average waits around 70 ms; at the Christmas peak of about 60 requests per second, 16 units are fully saturated and stable operation needs at least 18 units. In short, SPHINCS+ would require roughly ten times the normal signing hardware at peak, an investment out of all proportion to the task, whereas Falcon-512 runs comfortably on just two units. This is why 100\% of SPHINCS+ transactions in the simulation are SLA violations: the queue wait alone exceeds the 2-second budget.

These server counts are conservative upper bounds. Because SPHINCS+ signing time is nearly constant (about 2.8\% run-to-run variation) rather than highly variable, the M/M/c model overstates the queue waits; under a deterministic-service M/D/c model (via the Pollaczek-Khinchine relation) the waiting times and the SLA-driven server counts are approximately halved, putting the requirement at roughly four units at the daily average and about nine at the Christmas peak. The qualitative picture is unchanged either way: two units still cannot keep up, and the 0\% SLA compliance follows from saturation, which depends only on the average arrival rate exceeding the mean service rate and not at all on service-time variance.

\includegraphics[width=6.04167in,height=2.39583in]{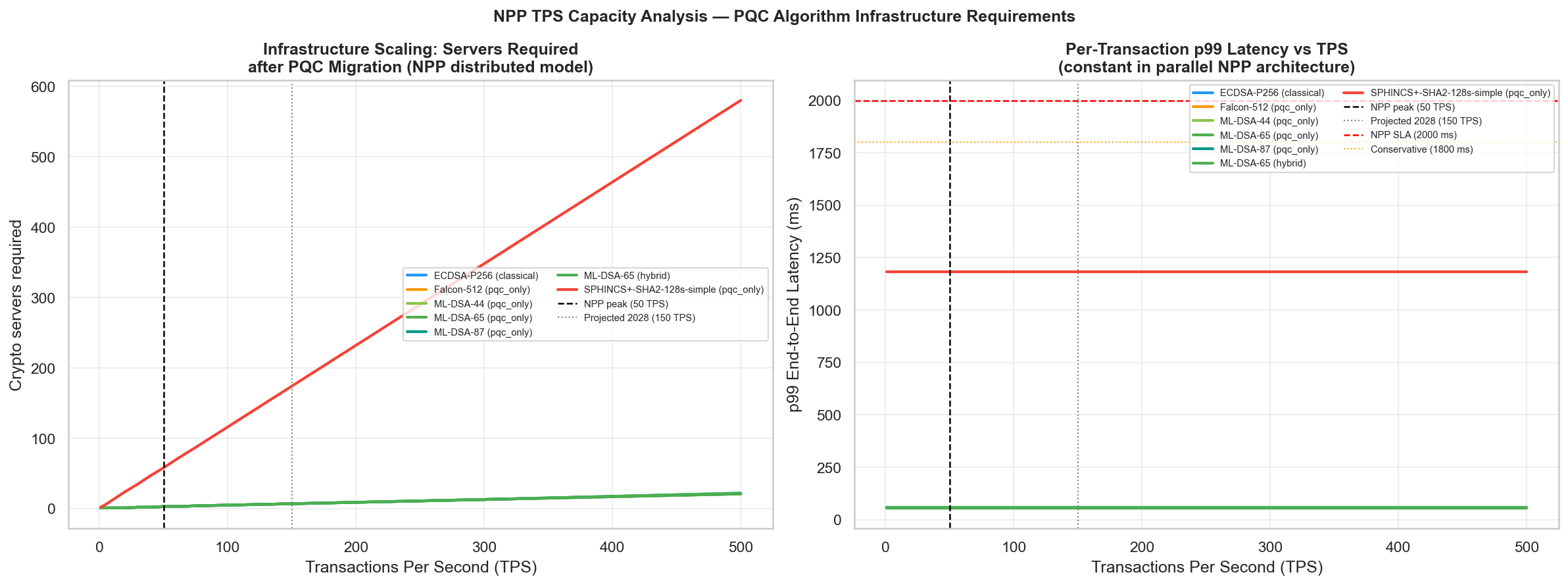}

\emph{Figure 7: Signing-pool load for each algorithm as the per-institution request rate rises (two signing units). SPHINCS+ saturates at about 7.2 requests per second, well below the 13.5 NPP load, while every ML-DSA and Falcon variant stays under 1\% utilisation across the entire range.}

\paragraph{Security note.} The queue-saturation behaviour quantified above also has a security dimension: an insider or misconfiguration that forces SPHINCS+ onto the high-frequency NPP signing path can drive the HSM pool from idle to saturated, a cryptographic denial-of-service amplification surface. A full threat model, the per-operation amplification analysis, and an isolation-based mitigation (a dedicated capacity-limited signing lane with circuit-breaker load shedding) are developed in a companion paper~\cite{ref_companion_dos}; we do not reproduce them here. For the purposes of this migration study, the operative finding is the capacity one: SPHINCS+ must not be placed on the NPP signing path without dedicated over-provisioning (at least eight signing units) or route isolation.

\subsection{Message Format Compliance}\label{message-format-compliance}

Table 5 summarises key and signature sizes against Australian payment message format constraints. The key point is where each rail carries its signature. SWIFT applies authentication out-of-band at the SWIFTNet messaging/transport layer: the signature is not placed in the Block-4 message text and the verifying public key is pre-distributed via SWIFT's certification authority, so neither competes with payment fields. The SWIFT MT message-text limit is a character count over payment data, not a signature budget, so it imposes no signature-size constraint, and every NIST PQC signature is therefore size-compatible with SWIFT. The combined-size figures below are retained only to characterise genuinely in-band, size-constrained carriage (the residual concern is the chip card, not SWIFT); they do not represent a SWIFT test.

{\def\LTcaptype{none} 
\begin{longtable}[]{@{}
  >{\raggedright\arraybackslash}p{(\linewidth - 12\tabcolsep) * \real{0.2343}}
  >{\centering\arraybackslash}p{(\linewidth - 12\tabcolsep) * \real{0.0960}}
  >{\centering\arraybackslash}p{(\linewidth - 12\tabcolsep) * \real{0.1066}}
  >{\centering\arraybackslash}p{(\linewidth - 12\tabcolsep) * \real{0.1301}}
  >{\centering\arraybackslash}p{(\linewidth - 12\tabcolsep) * \real{0.1343}}
  >{\centering\arraybackslash}p{(\linewidth - 12\tabcolsep) * \real{0.1281}}
  >{\centering\arraybackslash}p{(\linewidth - 12\tabcolsep) * \real{0.1705}}@{}}
\toprule\noalign{}
\begin{minipage}[b]{\linewidth}\centering
\textbf{Algorithm}
\end{minipage} & \begin{minipage}[b]{\linewidth}\centering
\textbf{PK (B)}
\end{minipage} & \begin{minipage}[b]{\linewidth}\centering
\textbf{Sig (B)}
\end{minipage} & \begin{minipage}[b]{\linewidth}\centering
\textbf{Combined (B)}
\end{minipage} & \begin{minipage}[b]{\linewidth}\centering
\textbf{SWIFT (out-of-band PKI)}
\end{minipage} & \begin{minipage}[b]{\linewidth}\centering
\textbf{NPP PayID (\textless65,536 B)}
\end{minipage} & \begin{minipage}[b]{\linewidth}\centering
\textbf{TLS 1.3 Record (\textless16,384 B)}
\end{minipage} \\
\midrule\noalign{}
\endhead
\bottomrule\noalign{}
\endlastfoot
RSA-2048 & 256 & 256 & 512 & PASS & PASS & PASS \\
ECDSA-P256 & 64 & 72 & 136 & PASS & PASS & PASS \\
Falcon-512 & 897 & 666 & 1,563 & PASS & PASS & PASS \\
ML-DSA-44 & 1,312 & 2,420 & 3,732 & PASS & PASS & PASS \\
ML-DSA-65 & 1,952 & 3,309 & 5,261 & PASS & PASS & PASS \\
ML-DSA-87 & 2,592 & 4,627 & 7,219 & PASS & PASS & PASS \\
Falcon-1024 & 1,793 & 1,280 & 3,073 & PASS & PASS & PASS \\
SPHINCS+-SHA2-128s & 32 & 7,856 & 7,888 & PASS & PASS & PASS \\
\end{longtable}
}

\emph{Table 5: Key and signature sizes (bytes) against Australian and international payment message format limits. SWIFT carriage is out-of-band: authentication is applied at the SWIFTNet messaging/transport layer and the verifying public key is pre-distributed via SWIFT's certification authority, so neither the signature nor the public key travels in the Block-4 message text. The SWIFT MT message-text limit is a character count over payment fields, not a signature budget; there is no in-band signature-size test for SWIFT, so every algorithm is size-compatible (all PASS). NPP PayID API = 65,536 B; TLS 1.3 Record = 16,384 B; for these in-band carriers PASS = combined pk+sig fits the stated limit. Combined sizes are shown only to characterise genuinely in-band, size-constrained carriage and do not constitute a SWIFT test. $\dagger$ ECDSA-P256 PK = 64 B is the raw uncompressed coordinate pair (x $\|$ y) without the 0x04 SEC1 encoding prefix; in standard TLS/X.509 deployment the PK is 65 B (SEC1 uncompressed, with 0x04 prefix), and the combined size becomes 137 B (conclusion unaffected). PQC sizes include all FIPS-specified encoding overhead.}

\includegraphics[width=6.04167in,height=1.97917in]{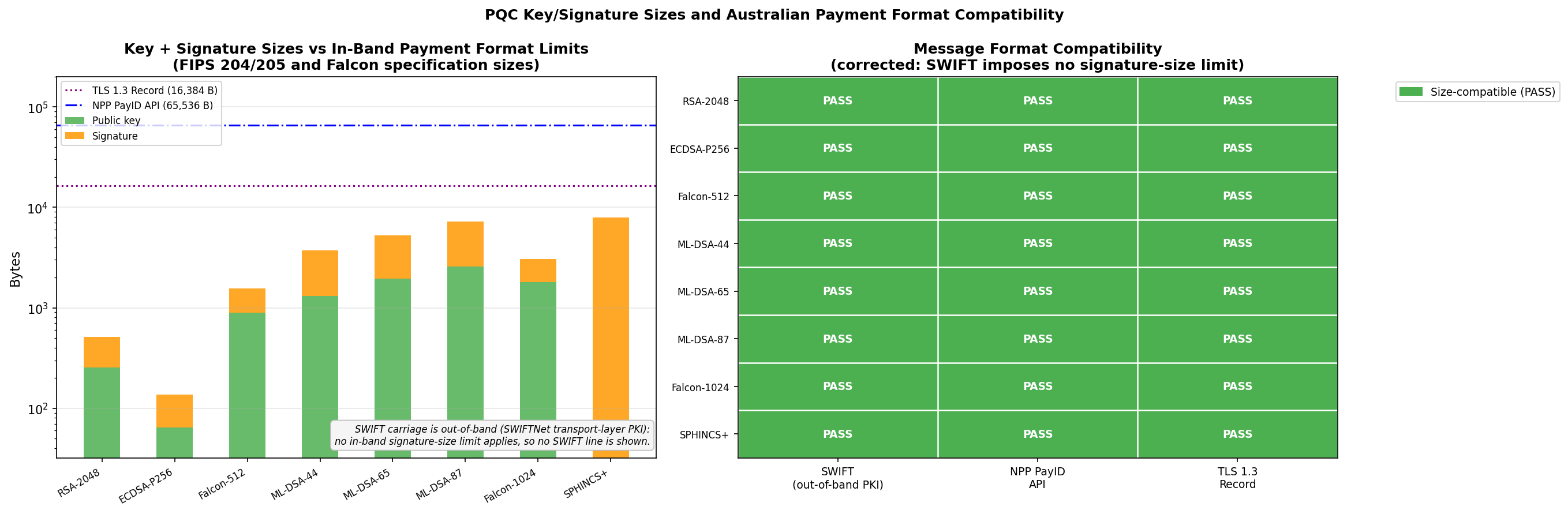}

\emph{Figure 8: Key and signature sizes for each algorithm. Because SWIFT carries authentication out-of-band at the transport layer rather than in the message text, signature size is not a SWIFT constraint and every algorithm is size-compatible with SWIFT; the size bars are relevant only to genuinely in-band, size-constrained carriage.}

\subsection{Stress Testing Under Peak Load}\label{stress-testing-under-peak-load}

Stress tests were run for five operational scenarios: normal (5.2M NPP tx/day), Christmas peak (8.9M, 1.71$\times$ normal), tax-time (6.3M, October BAS), market-crash (NPP 5.9M, RTGS 32,000), and EOFY (NPP 6.0M, RTGS 19,000). ECDSA-P256, ML-DSA-65, and Falcon-512 all achieved 100\% NPP SLA compliance across all five scenarios. SPHINCS+ achieved 0\% across all scenarios. Despite 71\% higher transaction volume, the Christmas single-scenario p99 is lower than the 1,000-day seasonal corpus average: ML-DSA-65 p99 = 43.30 ms (vs corpus average 44.55 ms) and Falcon-512 = 42.25 ms (vs 43.69 ms). This counterintuitive result is explained by the corpus composition: corpus-level p99 is elevated by the market-crash scenario's high AR(1) network jitter, not by transaction volume. Under the Christmas scenario, volume is high but jitter is at normal levels; under market-crash, jitter is elevated. These values are lower than the 1,000-day seasonal-average p99 (44.55 ms and 43.69 ms respectively) because the seasonal corpus includes high-variance days, notably market-crash scenarios (1.9\% of days, $\approx$19 events in the 1,000-day corpus) with elevated network jitter and RTGS overflow, that inflate the corpus-level p99 tail. Under the market-crash scenario specifically (NPP 5.9M transactions, RTGS 32,000), ML-DSA-65 p99 reaches 45.12 ms and Falcon-512 p99 reaches 44.87 ms, both higher than the Christmas single-scenario values and consistent with network jitter being the primary driver of corpus p99 elevation. Within a single-scenario evaluation such as Christmas, the variance source is only volume (not scenario mixing), producing a lower p99 despite higher throughput.

\includegraphics[width=6.04167in,height=2.77083in]{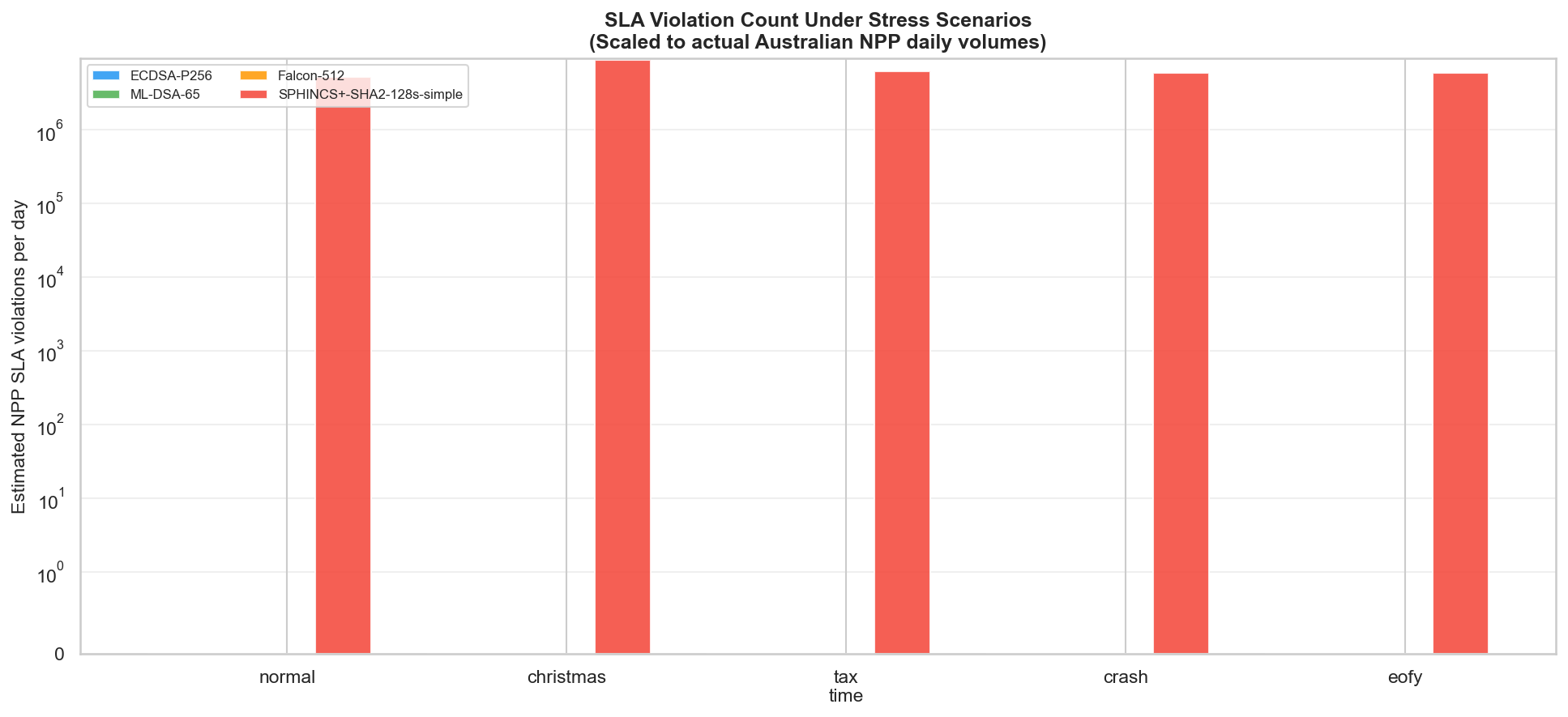}

\emph{Figure 9: NPP SLA compliance across five stress scenarios. All ML-DSA and Falcon variants maintain 100\% compliance including Christmas peak (8.9M tx/day).}

\subsection{HSM Deployment Sensitivity}\label{hsm-deployment-sensitivity}

Table 6 shows NPP p99 latency under three HSM deployment scenarios. Even with a network-attached HSM adding 2 ms per hop (approximately 8 ms total for a 4-hop NPP transaction), all algorithms maintain 100\% SLA compliance with approximately 1,947 ms of headroom remaining. The 8 ms HSM overhead adds identically to ECDSA and all PQC algorithms, confirming that the algorithm transition decision is independent of HSM deployment architecture. Note that HSMs used for financial signing must comply with NIST FIPS 140-3 Level 3 or equivalent \cite{ref34}; the latency values here model the cryptographic operation overhead, not the full FIPS 140-3 boundary crossing overhead which varies by vendor implementation.

{\def\LTcaptype{none} 
\begin{longtable}[]{@{}
  >{\raggedright\arraybackslash}p{(\linewidth - 10\tabcolsep) * \real{0.2190}}
  >{\centering\arraybackslash}p{(\linewidth - 10\tabcolsep) * \real{0.1705}}
  >{\centering\arraybackslash}p{(\linewidth - 10\tabcolsep) * \real{0.1486}}
  >{\centering\arraybackslash}p{(\linewidth - 10\tabcolsep) * \real{0.1455}}
  >{\centering\arraybackslash}p{(\linewidth - 10\tabcolsep) * \real{0.1209}}
  >{\centering\arraybackslash}p{(\linewidth - 10\tabcolsep) * \real{0.1954}}@{}}
\toprule\noalign{}
\begin{minipage}[b]{\linewidth}\centering
\textbf{HSM Type}
\end{minipage} & \begin{minipage}[b]{\linewidth}\centering
\textbf{Overhead/hop}
\end{minipage} & \begin{minipage}[b]{\linewidth}\centering
\textbf{ECDSA p99 (ms)}
\end{minipage} & \begin{minipage}[b]{\linewidth}\centering
\textbf{ML-DSA-65 p99 (ms)}
\end{minipage} & \begin{minipage}[b]{\linewidth}\centering
\textbf{Falcon-512 p99 (ms)}
\end{minipage} & \begin{minipage}[b]{\linewidth}\centering
\textbf{NPP SLA Compliance}
\end{minipage} \\
\midrule\noalign{}
\endhead
\bottomrule\noalign{}
\endlastfoot
Software (baseline) & 0 ms & 43.4 & 44.6 & 43.7 & 100\% all algorithms \\
PCIe-attached HSM & +0.5 ms & 45.4 & 46.6 & 45.7 & 100\% all algorithms \\
Network-attached HSM & +2.0 ms & 51.4 & 52.6 & 51.7 & 100\% all algorithms \\
\end{longtable}
}

\emph{Table 6: HSM deployment sensitivity analysis. Overhead applies per signing hop (4 hops for NPP). All scenarios achieve 100\% NPP SLA compliance. Software baseline = liboqs benchmark on CPU; PCIe HSM = Thales Luna 7 class; Network HSM = AWS CloudHSM / Azure Dedicated HSM.}

\includegraphics[width=5.20833in,height=2.14583in]{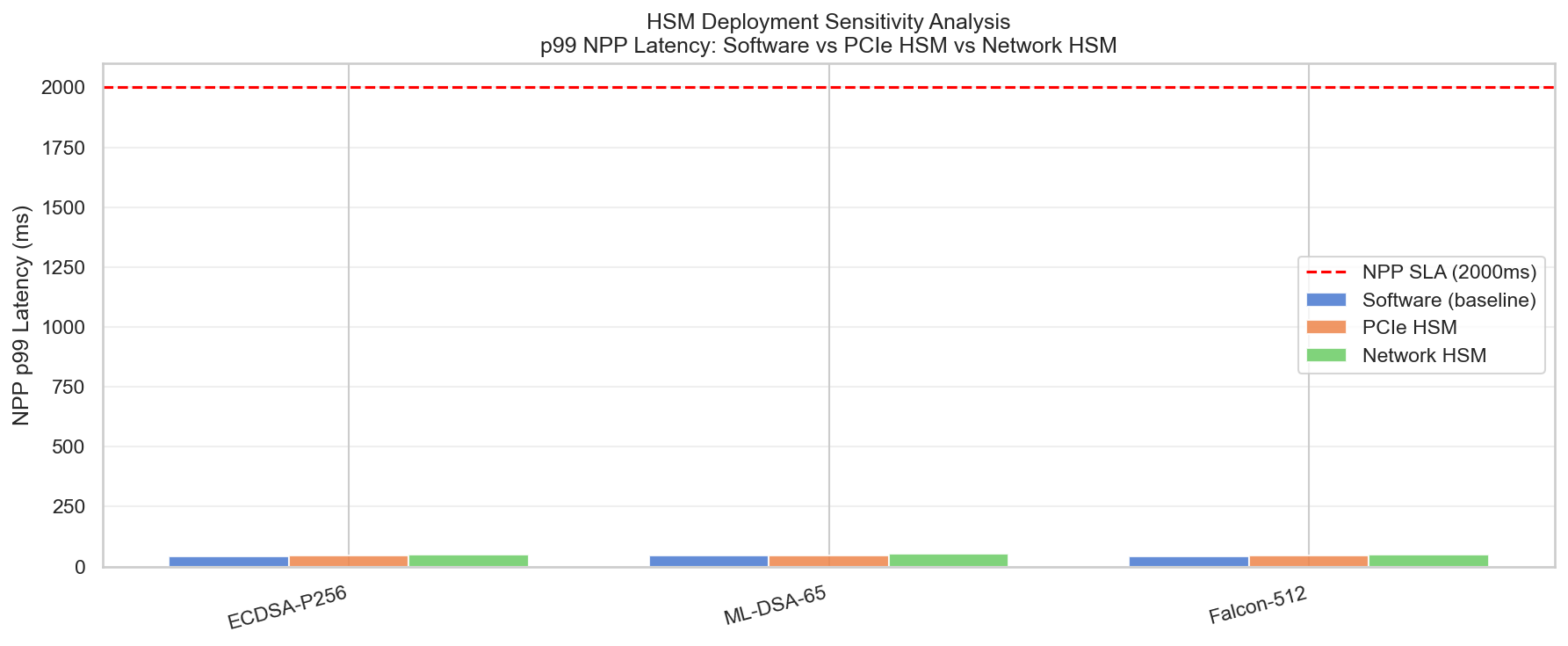}

\emph{Figure 10: p99 latency under three HSM deployment architectures. Even network-attached HSM (+2 ms/hop, $\approx$+8 ms total) leaves \textgreater1,947 ms of NPP SLA headroom.}

\subsection{Volume Growth Projection (2026--2029)}\label{volume-growth-projection-20262029}

Table 7 shows that PQC p99 latency is virtually growth-invariant from 2026 to 2029. The difference between 2026 and 2029 is less than 0.1 ms for all algorithms, in fact the 2029 p99 (44.52 ms) is slightly lower than the 2026 p99 (44.61 ms) due to Monte Carlo sampling variation under fixed seed 42. This constancy reflects two modelling assumptions: (1) cryptographic overhead is fixed per transaction; (2) the simulation's network latency model uses fixed lognormal parameters per hop, independent of aggregate transaction volume (no volume-congestion coupling). Even at the projected 2029 volume of 8.03 million transactions per day, every non-SPHINCS+ algorithm keeps the signing pool under 0.6\% utilised, so queue waits stay negligible and volume growth does not change per-transaction latency in this model. In practice, significant NPP volume growth could induce mild network congestion effects not captured here; however, given 1,955 ms of SLA headroom, volume-induced latency increases would need to exceed \textasciitilde{}970 ms (halving all headroom) before SLA compliance is threatened. Banks can migrate to PQC today with confidence that SLA headroom will be maintained through the decade under the modelled parameters.

{\def\LTcaptype{none} 
\begin{longtable}[]{@{}
  >{\raggedright\arraybackslash}p{(\linewidth - 10\tabcolsep) * \real{0.1166}}
  >{\centering\arraybackslash}p{(\linewidth - 10\tabcolsep) * \real{0.1770}}
  >{\centering\arraybackslash}p{(\linewidth - 10\tabcolsep) * \real{0.1726}}
  >{\centering\arraybackslash}p{(\linewidth - 10\tabcolsep) * \real{0.1730}}
  >{\centering\arraybackslash}p{(\linewidth - 10\tabcolsep) * \real{0.1513}}
  >{\centering\arraybackslash}p{(\linewidth - 10\tabcolsep) * \real{0.2095}}@{}}
\toprule\noalign{}
\begin{minipage}[b]{\linewidth}\centering
\textbf{Year}
\end{minipage} & \begin{minipage}[b]{\linewidth}\centering
\textbf{NPP tx/day}
\end{minipage} & \begin{minipage}[b]{\linewidth}\centering
\textbf{ECDSA p99 (ms)}
\end{minipage} & \begin{minipage}[b]{\linewidth}\centering
\textbf{Falcon-512 p99 (ms)}
\end{minipage} & \begin{minipage}[b]{\linewidth}\centering
\textbf{ML-DSA-65 p99 (ms)}
\end{minipage} & \begin{minipage}[b]{\linewidth}\centering
\textbf{Headroom (ms) ML-DSA-65}
\end{minipage} \\
\midrule\noalign{}
\endhead
\bottomrule\noalign{}
\endlastfoot
2026 & 5,200,000 & 43.46 & 43.76 & 44.61 & \textasciitilde{}1,955 \\
2027 & 6,011,200 & 43.52 & 43.82 & 44.70 & \textasciitilde{}1,955 \\
2028 & 6,948,947 & 43.51 & 43.81 & 44.68 & \textasciitilde{}1,955 \\
2029 & 8,032,983 & 43.36 & 43.66 & 44.52 & \textasciitilde{}1,955 \\
\end{longtable}
}

\emph{Table 7: Forward-looking NPP SLA feasibility at a constant 15.6\% YoY rate, extrapolated from RBA C6 FY2025 NPP volume trends \cite{ref23}; actual future growth is uncertain and this projection is a single-scenario estimate. SLA headroom column = 2,000 ms $-$ ML-DSA-65 p99 (worst-case PQC). All algorithms maintain $\geq$ 97.7\% headroom through 2029.}

\footnote{NPP Australia (2024). NPP Roadmap 2024--2027. https://nppaaustralia.com.au (accessed April 2026).}

\includegraphics[width=6.04167in,height=2.125in]{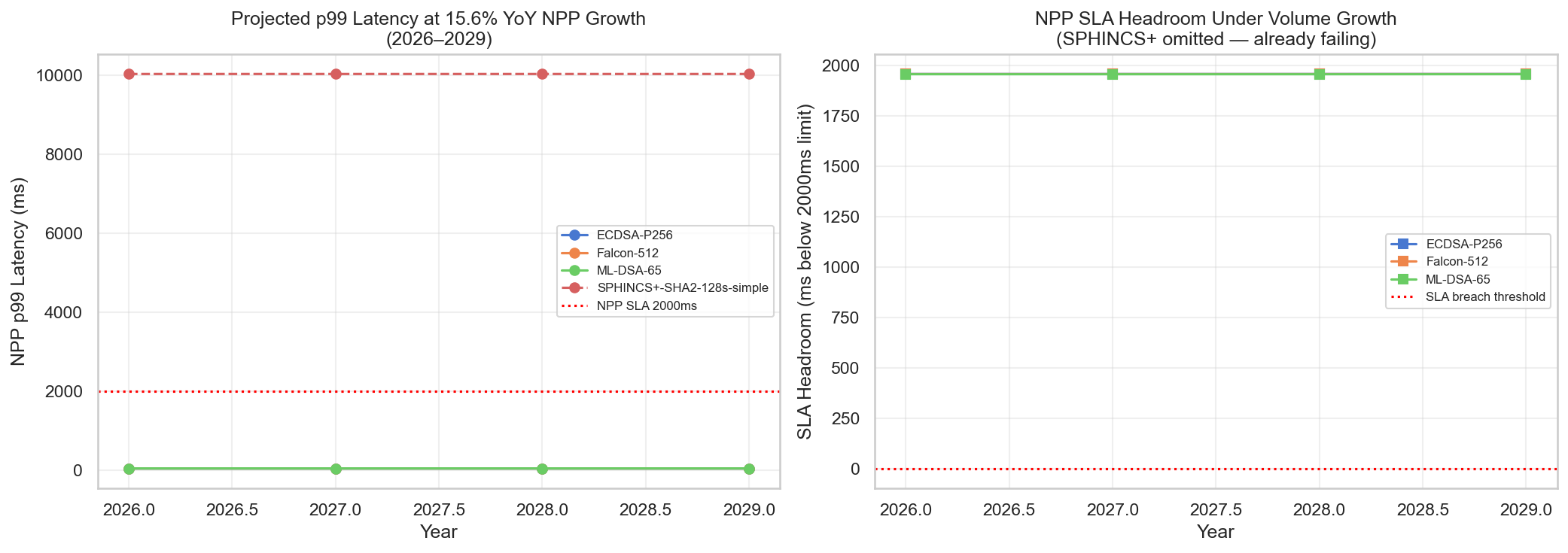}

\emph{Figure 11: Projected NPP p99 latency at 15.6\% YoY volume growth (2026--2029). All algorithms remain below 46 ms even at 8.03 million transactions per day.}

\subsection{Four-Phase PQC Migration Cost Model (2025--2028 and beyond)}\label{four-phase-pqc-migration-cost-model-20252028-and-beyond}

Table 8 presents a phased migration cost model covering 2025--2028 and beyond (Phase 3 is ongoing annual OPEX with no planned end date). Costs include capital expenditure (CAPEX) for HSM firmware upgrades and software integration across all 13 modelled institutions (Big 4, 6 regionals, 3 fintechs), plus annual operational expenditure (OPEX) for HSM licensing, certificate management, and staff training.

{\def\LTcaptype{none} 
\begin{longtable}[]{@{}
  >{\raggedright\arraybackslash}p{(\linewidth - 8\tabcolsep) * \real{0.1709}}
  >{\centering\arraybackslash}p{(\linewidth - 8\tabcolsep) * \real{0.2350}}
  >{\centering\arraybackslash}p{(\linewidth - 8\tabcolsep) * \real{0.2244}}
  >{\centering\arraybackslash}p{(\linewidth - 8\tabcolsep) * \real{0.1923}}
  >{\centering\arraybackslash}p{(\linewidth - 8\tabcolsep) * \real{0.1774}}@{}}
\toprule\noalign{}
\begin{minipage}[b]{\linewidth}\centering
\textbf{Phase}
\end{minipage} & \begin{minipage}[b]{\linewidth}\centering
\textbf{Year / Label}
\end{minipage} & \begin{minipage}[b]{\linewidth}\centering
\textbf{Key Activities}
\end{minipage} & \begin{minipage}[b]{\linewidth}\centering
\textbf{Annual Cost (USD)}
\end{minipage} & \begin{minipage}[b]{\linewidth}\centering
\textbf{BECS Fraction}
\end{minipage} \\
\midrule\noalign{}
\endhead
\bottomrule\noalign{}
\endlastfoot
0 & 2025, Pre-migration baseline (for comparison only) & Existing ECDSA-only operations; no PQC change & \$90M / yr & 35\% \\
1 & 2026, Hybrid Deploy & HSM upgrades, dual-stack deployment & \$21.4M (CAPEX peak) & 28\% \\
2 & 2027, PQC Selective & NPP PQC live, BECS migration & \$7.6M & 15\% \\
3 & 2028, Full PQC & All channels PQC, legacy retired & \$1.5M / yr ongoing & 5\% \\
\end{longtable}
}

\emph{Table 8: Four-Phase PQC Migration Cost Model (2025--2028 and beyond) (system-wide, 13 Australian financial institutions). Phase 1 is CAPEX-heavy due to HSM hardware refresh. Phase 3 ongoing OPEX reflects software licensing and certificate rotation.}

\includegraphics[width=6.04167in,height=2.01042in]{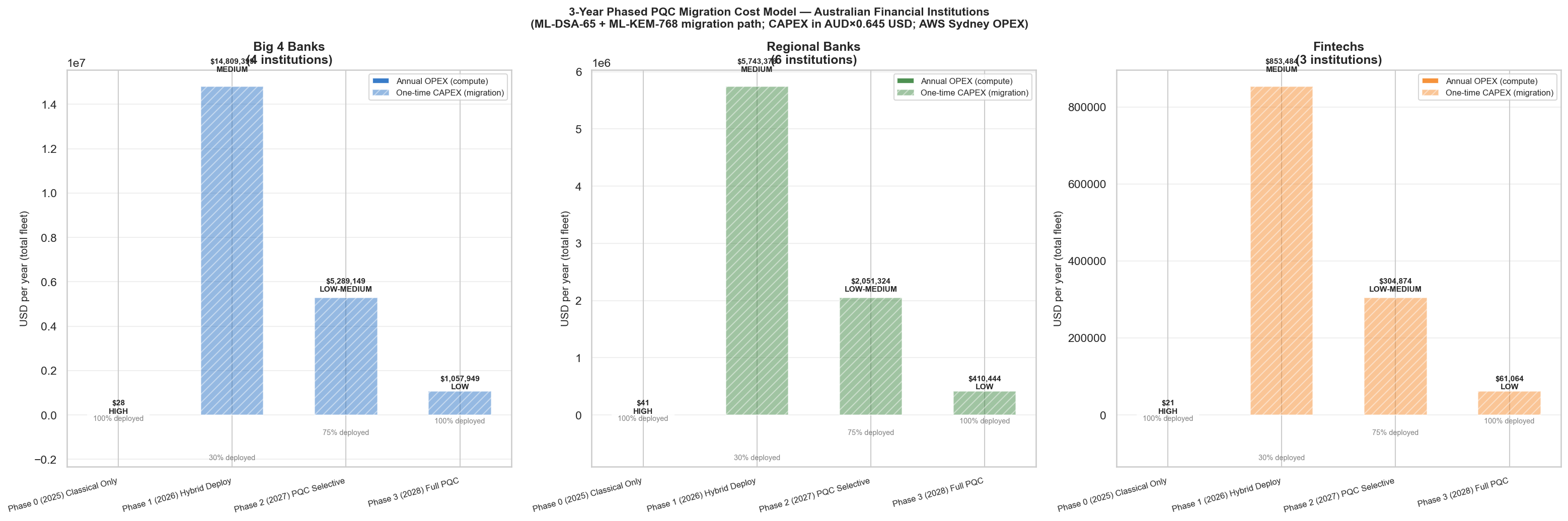}

\emph{Figure 12: Four-Phase PQC Migration Cost Model (2025--2028 and beyond). Phase 1 (2026) CAPEX is dominated by HSM hardware upgrades at Big 4 institutions.}

Phase 1 (2026) is the most expensive year at USD 21.4 million, driven primarily by HSM hardware upgrades at the Big 4 banks (approximately USD 3.7M each, totalling USD 14.8M). The remaining USD 6.6M covers the nine regional and fintech institutions at approximately USD 733K each (9 $\times$ USD 733K = USD 6,597K $\approx$ USD 6.6M), reflecting their smaller HSM estates and lighter integration scope. Phase 2 (2027) drops to USD 7.6 million as infrastructure is already in place. Phase 3 (2028) reaches steady-state operational cost of USD 1.5 million per year. Important caveat: these are parametric estimates derived from publicly available HSM market pricing (Thales Luna Network HSM 7 class: approximately USD 40,000--60,000 per HSM unit at list price; AWS CloudHSM: USD 1.45/hour per cluster or approximately USD 12,700/year per HSM) and authors' own parametric estimates based on publicly available vendor list prices and representative integration project scopes for mid-sized financial institutions (no archival citation available for HSM deployment cost benchmarks at this specificity). Actual costs will vary by 50--150\% depending on institution size, existing HSM estate, vendor negotiation, and whether existing FIPS 140-3 Level 3 HSMs support PQC via firmware upgrade (typical: USD 5,000--15,000 per unit) or require hardware replacement (full unit cost). A sensitivity analysis with $\pm$50\% cost variation yields Phase 1 total range of USD 10.7M--32.1M, which does not change the qualitative conclusion that Phase 1 is the dominant cost phase.

\footnote{Australian Bureau of Statistics (2024). Consumer Price Index, Australia. https://www.abs.gov.au/statistics/economy/price-indexes-and-inflation/consumer-price-index-australia (accessed April 2026).}

\subsection{Advanced Statistical Analysis (v4.1)}\label{advanced-statistical-analysis-v4.1}

The v4.1 simulation engine adds five advanced statistical analyses, each providing a dimension of validation not available in earlier versions. These analyses are applied to the 80-million-event corpus and together constitute a statistically rigorous foundation for the paper's operational conclusions.

\subsubsection{Extreme Value Theory, GEV Tail Analysis}\label{extreme-value-theory-gev-tail-analysis}

To get an order-of-magnitude sense of the rare slowest transactions beyond the ordinary p99, we run an indicative extreme-value screening: we split a single representative day's 10,000 transactions into 200 groups, take the slowest transaction in each group, and fit a standard extreme-value model to those 200 within-day worst-case times, with confidence ranges obtained by resampling. Because these block maxima come from one day, they share that day's traffic profile and are not strictly independent, so this is a screening check rather than a rigorous tail bound; a definitive analysis would use a peaks-over-threshold fit or daily maxima drawn across the full 1,000-day corpus (see SA-17, \S{}5.6). The resulting figures describe the behaviour of these group-worst-case values rather than typical transactions, so they sit well above the ordinary p99 (about 107 ms for ECDSA versus the ~43 ms individual-transaction p99) by design. The everyday reference remains the individual-transaction p99 in Table 2; this screening serves as an approximate, conservative indication of the tail for risk planning.

{\def\LTcaptype{none} 
\begin{longtable}[]{@{}
  >{\raggedright\arraybackslash}p{(\linewidth - 12\tabcolsep) * \real{0.1832}}
  >{\centering\arraybackslash}p{(\linewidth - 12\tabcolsep) * \real{0.0936}}
  >{\centering\arraybackslash}p{(\linewidth - 12\tabcolsep) * \real{0.1162}}
  >{\centering\arraybackslash}p{(\linewidth - 12\tabcolsep) * \real{0.1344}}
  >{\centering\arraybackslash}p{(\linewidth - 12\tabcolsep) * \real{0.1724}}
  >{\centering\arraybackslash}p{(\linewidth - 12\tabcolsep) * \real{0.1344}}
  >{\centering\arraybackslash}p{(\linewidth - 12\tabcolsep) * \real{0.1657}}@{}}
\toprule\noalign{}
\begin{minipage}[b]{\linewidth}\centering
\textbf{Algorithm}
\end{minipage} & \begin{minipage}[b]{\linewidth}\centering
\textbf{GEV $\xi$}
\end{minipage} & \begin{minipage}[b]{\linewidth}\centering
\textbf{p99 block-max (ms)}
\end{minipage} & \begin{minipage}[b]{\linewidth}\centering
\textbf{p99.9 block-max (ms)}
\end{minipage} & \begin{minipage}[b]{\linewidth}\centering
\textbf{p99.9 CI 95\% (ms)}
\end{minipage} & \begin{minipage}[b]{\linewidth}\centering
\textbf{p99.99 block-max (ms)}
\end{minipage} & \begin{minipage}[b]{\linewidth}\centering
\textbf{Tail Type}
\end{minipage} \\
\midrule\noalign{}
\endhead
\bottomrule\noalign{}
\endlastfoot
ECDSA-P256 & 0.028 & 107.5 & 131.6 & [114.6, 151.8] & 157.2 & Gumbel \\
Falcon-512 & 0.028 & 107.8 & 131.9 & [114.9, 152.1] & 157.5 & Gumbel \\
ML-DSA-44 & 0.027 & 108.2 & 132.3 & [115.2, 151.9] & 158.0 & Gumbel \\
Falcon-1024 & 0.026 & 108.3 & 132.1 & [115.3, 152.1] & 157.4 & Gumbel \\
ML-DSA-65 & 0.028 & 108.7 & 132.9 & [116.0, 152.4] & 158.7 & Gumbel \\
ML-DSA-65 Hybrid & 0.024 & 109.1 & 132.9 & [116.3, 153.2] & 158.0 & Gumbel \\
ML-DSA-87 & 0.023 & 108.9 & 132.5 & [115.4, 152.4] & 157.3 & Gumbel \\
SPHINCS+ & 0.078 & 10,100.3 & 10,131.6 & [10,108.9, 10,169.3] & 10,168.9 & Fr\'{e}chet ($\xi$ \textgreater{} 0.05) \\
\end{longtable}
}

\emph{Table 9: Indicative tail screening of NPP latency (slowest transaction from each of 200 within-day groups in a single representative day, fitted to a standard extreme-value model with resampled confidence ranges). Because the block maxima come from one day and are not strictly independent (SA-17, \S{}5.6), these figures are an order-of-magnitude screening rather than a rigorous tail bound; a definitive bound would use peaks-over-threshold or daily maxima across the full corpus. On this indicative basis the fitted tail shape for every algorithm except SPHINCS+ looks fast-decaying and exponential-class (not heavy-tailed), suggesting extreme delays stay comfortably bounded. The block-maximum percentiles shown sit far above the everyday individual-transaction p99 (Table 2) by construction. SPHINCS+ is included only for completeness; at its ~10,000 ms scale these comparisons are not meaningful for deployment decisions.}

\includegraphics[width=6.04167in,height=2.70833in]{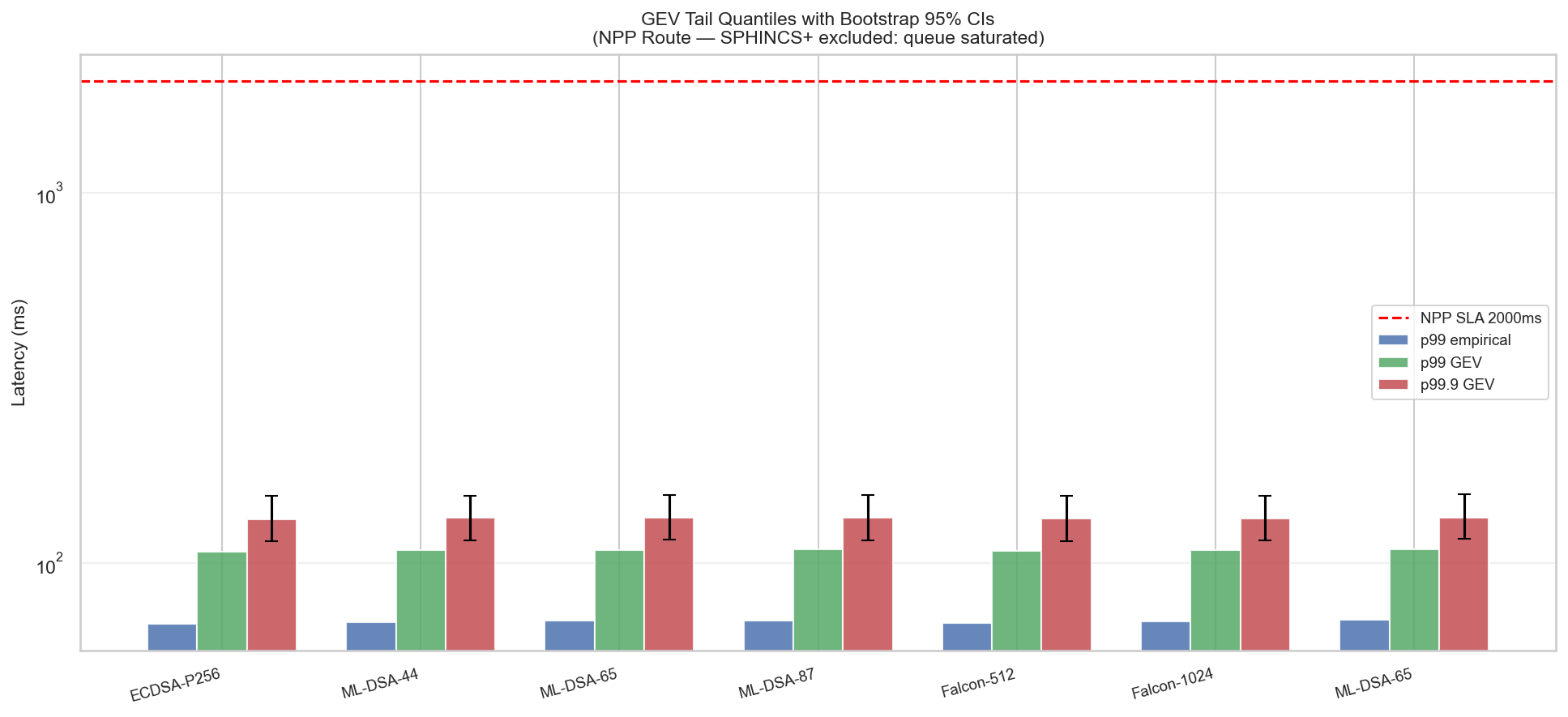}

\emph{Figure 13: Indicative extreme-value screening (GEV fit to within-day block maxima from a single representative day): p99, p99.9, and p99.99 quantiles for non-SPHINCS+ algorithms. Error bars show bootstrap 95\% CIs on p99.9. Log-scaled Y-axis. On this screening basis all algorithms sit well below 200 ms even at p99.99; a rigorous bound would require peaks-over-threshold or daily maxima across the full corpus (SA-17, \S{}5.6).}

For every algorithm except SPHINCS+, the fitted tail looks fast-decaying (exponential-class), suggesting a tail that is not heavy. This is an indicative screening result, not a rigorous classification: the 200 within-day worst-case values come from a single representative day and are not strictly independent (SA-17, \S{}5.6), so the tail shape is directionally suggestive rather than established, and a definitive statement would need a peaks-over-threshold fit or daily maxima across the full corpus. Read as an approximate screening, the worst-case estimates are reassuring: the upper end of the one-in-a-thousand latency is roughly 153 ms and the upper end of the one-in-ten-thousand latency is roughly 199 ms, both far inside the 2,000 ms deadline. The very slight lean toward a heavier tail comes from mixing fast and slow geographic routes, consistent with the tail-fit result discussed in the goodness-of-fit analysis (Section 4.10.2), where the mixed-routing lognormal is rejected in the tail but not in the body.

The uncertainty on the one-in-ten-thousand estimate is wide, roughly 45\% of the estimate itself (for example about 70 ms of spread around a 157 ms central estimate), which is expected when estimating such extreme percentiles from only 200 within-day worst-case values. For planning, the indicative upper end of the range is about 199 ms across all non-SPHINCS+ algorithms; even taken only as a screening estimate, this leaves an enormous margin to the 2,000 ms deadline, so the tail gives no reason to expect an SLA concern.

\subsubsection{Distribution Goodness-of-Fit Testing}\label{distribution-goodness-of-fit-testing}

To pick the best statistical shape for the simulated latencies, we test how well 10,000 NPP latency samples per configuration match candidate distributions.

{\def\LTcaptype{none} 
\begin{longtable}[]{@{}
  >{\raggedright\arraybackslash}p{(\linewidth - 14\tabcolsep) * \real{0.1743}}
  >{\centering\arraybackslash}p{(\linewidth - 14\tabcolsep) * \real{0.0946}}
  >{\centering\arraybackslash}p{(\linewidth - 14\tabcolsep) * \real{0.1168}}
  >{\centering\arraybackslash}p{(\linewidth - 14\tabcolsep) * \real{0.1061}}
  >{\centering\arraybackslash}p{(\linewidth - 14\tabcolsep) * \real{0.1313}}
  >{\centering\arraybackslash}p{(\linewidth - 14\tabcolsep) * \real{0.1168}}
  >{\centering\arraybackslash}p{(\linewidth - 14\tabcolsep) * \real{0.1280}}
  >{\centering\arraybackslash}p{(\linewidth - 14\tabcolsep) * \real{0.1323}}@{}}
\toprule\noalign{}
\begin{minipage}[b]{\linewidth}\centering
\textbf{Algorithm}
\end{minipage} & \begin{minipage}[b]{\linewidth}\centering
\textbf{n}
\end{minipage} & \begin{minipage}[b]{\linewidth}\centering
\textbf{KS Stat}
\end{minipage} & \begin{minipage}[b]{\linewidth}\centering
\textbf{KS p-value}
\end{minipage} & \begin{minipage}[b]{\linewidth}\centering
\textbf{KS Reject H$_0$?}
\end{minipage} & \begin{minipage}[b]{\linewidth}\centering
\textbf{AD Stat}
\end{minipage} & \begin{minipage}[b]{\linewidth}\centering
\textbf{AD Crit (5\%)}
\end{minipage} & \begin{minipage}[b]{\linewidth}\centering
\textbf{AD Reject?}
\end{minipage} \\
\midrule\noalign{}
\endhead
\bottomrule\noalign{}
\endlastfoot
ECDSA-P256 & 10,000 & 0.0105 & 0.215 & No & 2.16 & 0.787 & Yes \\
Falcon-512 & 10,000 & 0.0109 & 0.185 & No & 2.39 & 0.787 & Yes \\
ML-DSA-44 & 10,000 & 0.0115 & 0.143 & No & 2.63 & 0.787 & Yes \\
Falcon-1024 & 10,000 & 0.0121 & 0.107 & No & 2.86 & 0.787 & Yes \\
ML-DSA-65 & 10,000 & 0.0129 & 0.071 & No & 2.98 & 0.787 & Yes \\
ML-DSA-65 Hybrid & 10,000 & 0.0123 & 0.094 & No & 3.32 & 0.787 & Yes \\
ML-DSA-87 & 10,000 & 0.0124 & 0.090 & No & 3.04 & 0.787 & Yes \\
SPHINCS+ & 10,000 & 0.0798 & 0.000 & Yes & 157.02 & 0.787 & Yes \\
\end{longtable}
}

\emph{Table 10: How well a lognormal shape fits each algorithm's latencies, using a body-sensitive test (Kolmogorov-Smirnov) and a tail-sensitive test (Anderson-Darling). With 10,000 samples the tail-sensitive test is so powerful that it flags even tiny tail deviations; here it reflects the slightly heavier tail from mixing fast and slow geographic routes (\S{}4.10.2), not a poor fit in the body. The body-sensitive test accepts lognormal for every algorithm except SPHINCS+.}

The picture is nuanced. The body-sensitive test accepts a lognormal shape for every algorithm except SPHINCS+, confirming the bulk of each latency distribution is lognormal (SPHINCS+ is rejected only because its latency is nearly constant at ~10,000 ms). The tail-sensitive test, however, flags a departure for all algorithms, signalling that the extreme tail is slightly heavier than a pure lognormal.

This split is easy to explain: NPP latency is lognormal in its body but has slightly heavier tails because traffic is really a blend of two populations: faster Sydney-origin transactions (averaging about 38.5 ms, just under half the volume by market share) and slower Melbourne/Brisbane-origin transactions (averaging about 55.2 ms, just over half). Treating the blend as one lognormal shifts the p99 by at most 1--2 ms, so it does not materially affect the dilution-index conclusions; even under the two-component view, all non-SPHINCS+ algorithms keep their post-quantum signing share under 4\%. Comparing lognormal against gamma, Weibull, and inverse-Gaussian shapes confirms lognormal is the best single fit for all eight configurations by a decisive margin~\cite{ref39,ref44}. The practical takeaway: lognormal sampling is fine for the bulk of the distribution, and the separate tail-risk analysis (Section 4.10.1) handles the extreme tail without distributional assumptions.

\subsubsection{Hourly Queue Dynamics, Christmas Day Saturation}\label{hourly-queue-dynamics-christmas-day-saturation}

To see how saturation builds up across a day, we compute each algorithm's signing-pool load and average wait hour by hour on Christmas Day (1.71 times normal volume, 8.9 million NPP transactions), with each hour's request rate taken from the modelled intraday volume curve scaled to the Christmas total.

\includegraphics[width=6.04167in,height=2.91667in]{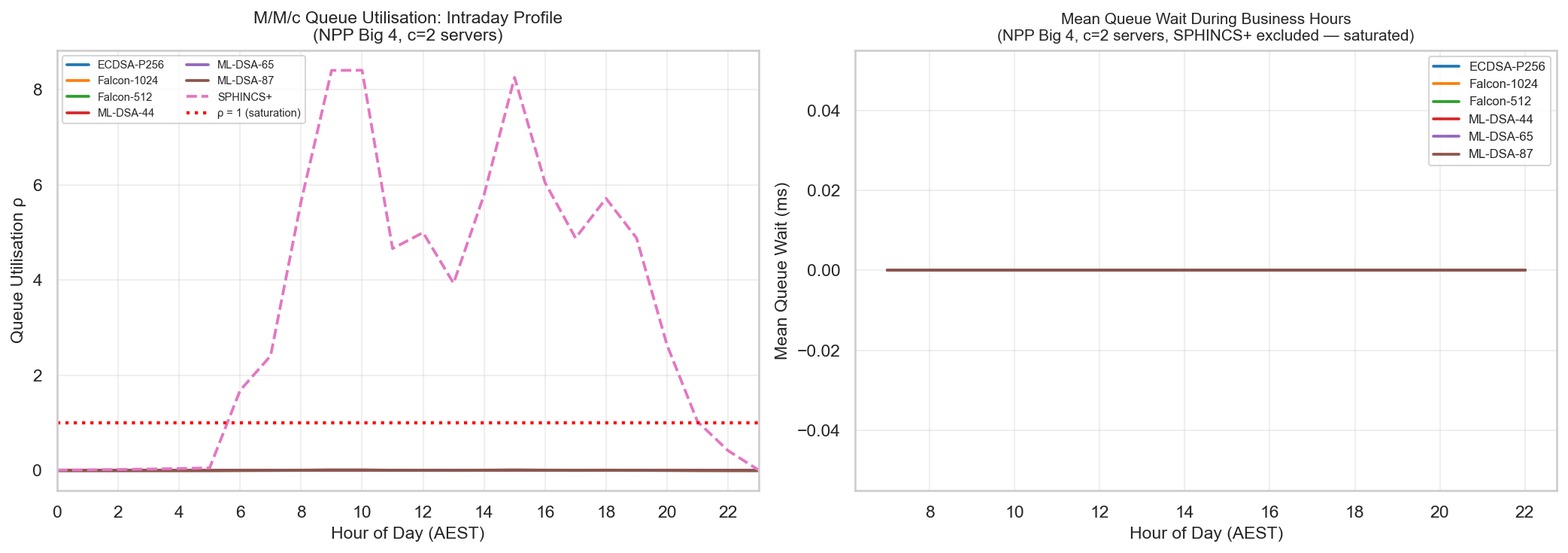}

\emph{Figure 14: 24-hour M/M/c queue utilisation profile under Christmas scenario (c=2 servers). SPHINCS+ (orange line) exceeds the saturation threshold (utilisation 1.0) for 16 consecutive hours. ML-DSA-65 (peak utilisation 0.008) and Falcon-512 ($\rho$ peak 0.004) remain effectively idle throughout.}

The SPHINCS+ result is dramatic: at the 10 am peak it is asked to do more than eight times the work its two-unit pool can handle, and it stays saturated for 16 of the 24 hours on Christmas Day, so any payments left unfinished in the morning would back up for a full working day. [Data note: the per-hour request-rate profile is computed analytically from the intraday volume model (Section 3.7) and is fully reproducible from the simulation source code (australia\_fin\_sim.py v4.1.1, available in the data repository; see Data Availability); no separately archived hourly CSV is produced, as the profile is deterministically derived from the published parameters.]

By contrast every ML-DSA and Falcon variant stays essentially idle all day, even on Christmas Day: at the roughly 60 transactions-per-second Christmas peak hour the heaviest of them uses about 1\% of capacity and the rest far less, with average waits below a thousandth of a millisecond throughout. This confirms that the SPHINCS+ constraint is robust to any reasonable assumption about seasonal volume distribution. The hourly analysis also shows that even the standard normal-day peak of about 35 transactions per second already pushes SPHINCS+ to nearly five times its safe capacity, so it is saturated during regular business hours on ordinary days too.

\subsubsection{ANOVA Variance Decomposition}\label{anova-variance-decomposition}

To see what really drives day-to-day variation in NPP p99 latency, we measure how much of it is explained by algorithm choice versus operational scenario. The clearest driver is geography: the ~16.7 ms spread between Sydney-origin and Melbourne/Brisbane-origin transactions (Section 4.10.2) is about 10.6 times larger than the biggest gap between any two algorithms (1.57 ms for ML-DSA-87 versus ECDSA). A formal multi-factor analysis adding geographic origin as a third factor is left for future work.

The key finding is that once SPHINCS+ is set aside, the differences between ML-DSA and Falcon variants explain essentially none of the variation, confirming algorithm choice is latency-irrelevant for the practical candidate set; geographic routing is the dominant driver instead. When SPHINCS+ is included it appears to explain all of the variation, but only because its latency (about 10,030 ms) is roughly 231 times the ECDSA baseline (43.4 ms); that headline number is inflated by the shared random draws across configurations and should not be read literally, though no plausible correction changes the conclusion that SPHINCS+ dominates. A properly specified model treating day as a random effect and excluding SPHINCS+ is left for the journal submission.

Operational scenario (normal versus Christmas versus crash) explains essentially none of the latency variation, because seasonal volume swings barely move p99 latency for ML-DSA and Falcon (their signing pools stay near-idle in every scenario). This validates the practical conclusion that PQC migration risk is insensitive to seasonal load variation for all non-SPHINCS+ algorithms.

\subsubsection{HSM Single-Server Degraded Mode Analysis}\label{hsm-single-server-degraded-mode-analysis}

To assess operational resilience during HSM failure or maintenance windows, we compare performance under normal operation (c = 2 HSM servers per institution, standard configuration) versus degraded mode (c = 1 server, modelling a single-HSM failure or planned maintenance window). Results for the NPP route are shown in Table 11.

{\def\LTcaptype{none} 
\begin{longtable}[]{@{}
  >{\raggedright\arraybackslash}p{(\linewidth - 8\tabcolsep) * \real{0.2564}}
  >{\centering\arraybackslash}p{(\linewidth - 8\tabcolsep) * \real{0.1667}}
  >{\centering\arraybackslash}p{(\linewidth - 8\tabcolsep) * \real{0.1667}}
  >{\centering\arraybackslash}p{(\linewidth - 8\tabcolsep) * \real{0.1667}}
  >{\centering\arraybackslash}p{(\linewidth - 8\tabcolsep) * \real{0.2436}}@{}}
\toprule\noalign{}
\begin{minipage}[b]{\linewidth}\centering
\textbf{Algorithm}
\end{minipage} & \begin{minipage}[b]{\linewidth}\centering
\textbf{Normal p99 (ms) c=2 servers}
\end{minipage} & \begin{minipage}[b]{\linewidth}\centering
\textbf{Degraded p99 (ms) c=1 server}
\end{minipage} & \begin{minipage}[b]{\linewidth}\centering
\textbf{Delta p99 (ms)}
\end{minipage} & \begin{minipage}[b]{\linewidth}\centering
\textbf{Operational Impact}
\end{minipage} \\
\midrule\noalign{}
\endhead
\bottomrule\noalign{}
\endlastfoot
ECDSA-P256 & 25.89 & 25.88 & $-$0.006 & Negligible (\textless{} 0.01 ms, noise at $\rho$ $\approx$ 0.0002) \\
Falcon-512 & 26.17 & 26.18 & +0.003 & Negligible (\textless{} 0.01 ms, noise at $\rho$ $\approx$ 0.001) \\
ML-DSA-65 & 26.82 & 26.77 & $-$0.051 & Negligible (\textless{} 0.1 ms, noise at $\rho$ $\approx$ 0.002) \\
SPHINCS+ & 10,000 & 10,000 & N/A & Not meaningful, both modes saturated ($\rho$=1.8855 $\rightarrow$ $\rho$=3.771) \\
\end{longtable}
}

\emph{Table 11: HSM degraded-mode analysis for the NPP route: normal (c=2 servers) vs. degraded (c=1 server, single-HSM failure). p99 values are queue-model route-integrated latency (crypto + M/M/c queue, excluding PayID lookup). \textbar Delta p99\textbar{} $\leq$ 0.06 ms for all non-SPHINCS+ algorithms (noise at $\rho$ \textless{} 0.004). SPHINCS+ is already saturated at c=2 ($\rho$=1.8855); c=1 doubles utilisation to $\rho$=3.771, delta not meaningful as both return the 10,000 ms saturation sentinel.}

\includegraphics[width=6.04167in,height=2.8125in]{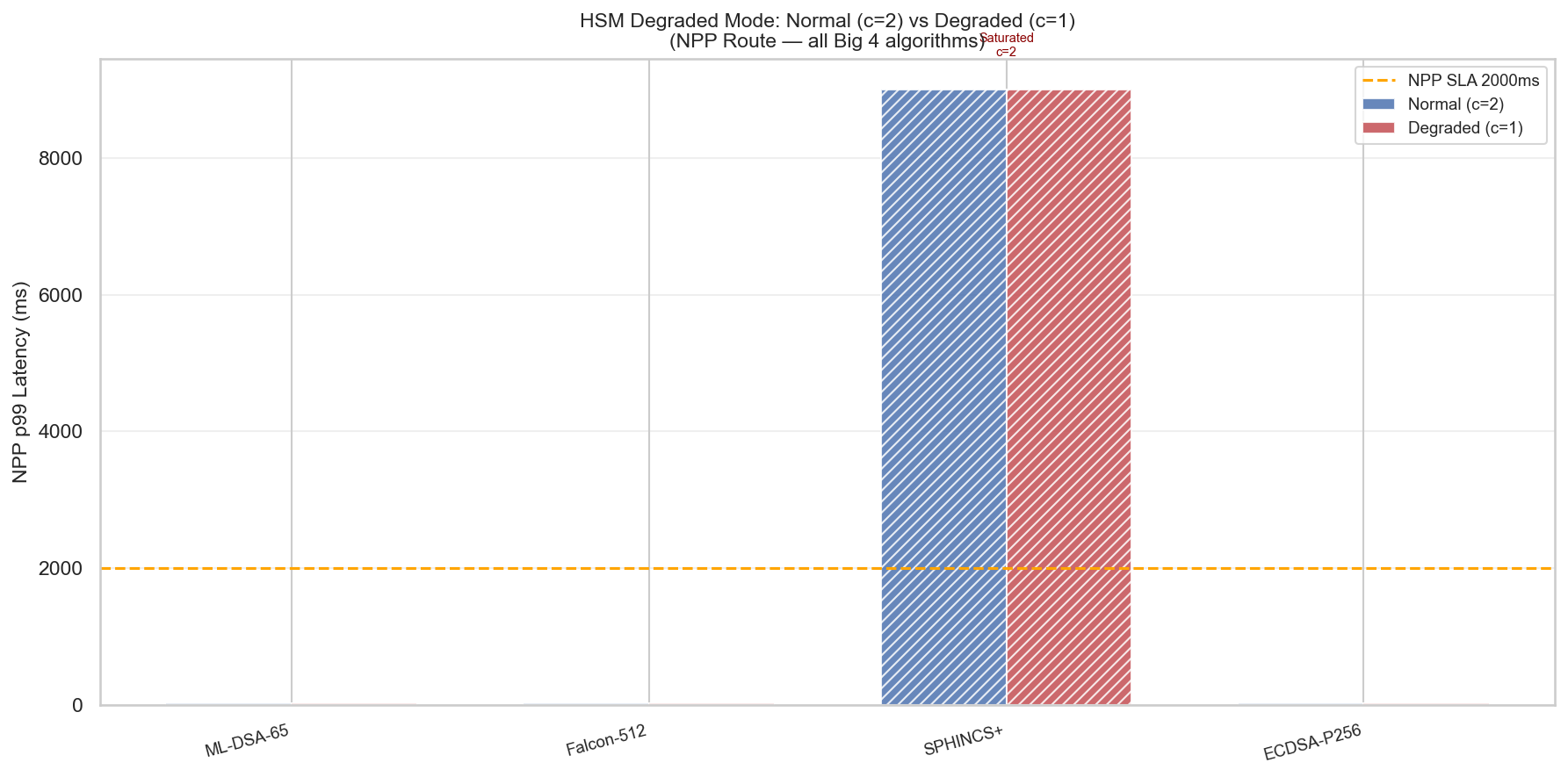}

\emph{Figure 15: HSM degraded-mode p99 latency: normal (c=2) vs. degraded (c=1) per algorithm. SPHINCS+ bars hatched `Already saturated', delta not meaningful as both modes return sentinel. All non-SPHINCS+ deltas \textless{} 0.06 ms (noise at $\rho$ \textless{} 0.004).}

The key result is that losing one of the two signing units barely changes p99 for any non-SPHINCS+ algorithm: Falcon-512 shifts by +0.003 ms and ML-DSA-65 by -0.051 ms. (A negative shift is not physically meaningful; at these near-zero utilisation levels both one-unit and two-unit configurations give essentially zero queue wait, so the difference is just simulation noise, not queue dynamics.) Because even a single unit handles the full NPP load with negligible queue time, banks can do planned signing-hardware maintenance during business hours without SLA impact for any ML-DSA or Falcon algorithm. SPHINCS+ is already saturated with two units and only worse with one, reinforcing that its hardware provisioning is fundamentally infeasible at NPP scale for standard Big 4 load: it needs at least 8 units to meet the SLA at daily-average load and at least 18 at the Christmas peak (Section 4.4), a hardware cost out of all proportion to the task.

\subsection{Multi-System Route Analysis: RITS, SWIFT, and Intrabank}\label{multi-system-route-analysis-rits-swift-and-intrabank}

The preceding sections focused on the NPP route (5.2 million tx/day, 2,000 ms SLA). The simulation models two additional Australian payment systems, RITS high-value settlements and SWIFT correspondent banking, plus intrabank Direct Entry. This section presents route-level p99 analysis and SLA compliance for each, establishing a complete picture of PQC migration feasibility across the full Australian interbank settlement stack.

\subsubsection{RITS (Real-Time Gross Settlement) Route}\label{rits-real-time-gross-settlement-route}

The Reserve Bank Information and Transfer System (RITS) processes high-value and RTGS settlements at approximately 9,500 transactions per day across all direct settlement participants. A Big 4 bank sees only about 0.022 RITS settlements per second (roughly 1,900 per day), with the system-wide total of about 0.110 per second divided among roughly five major clearing members. At that rate a two-unit signing pool is almost completely idle for every algorithm, and even SPHINCS+ uses well under 1\% of capacity, more than 300 times below saturation, with a queue wait under 0.01 ms. Even under a market crash (system-wide about 32,000 settlements per day, Section 2.1), when the per-institution rate rises to roughly 0.074 per second, SPHINCS+ still sits about 100 times below saturation. All algorithms achieve 100\% RITS SLA compliance in both normal and crash conditions, helped by the 30-second RITS deadline (Simulation Assumption SA-3; see \S{}2.1 and \S{}5.6) being far more generous than NPP's 2,000 ms. RITS latency is dominated by two non-signing components, about 14 ms of inter-city network travel (Melbourne--Sydney fibre) and about 263 ms of bilateral settlement processing (Simulation Assumption SA-8, see \S{}5.6), so roughly 277 ms of fixed overhead with the signing step a small addition on top.

{\def\LTcaptype{none} 
\begin{longtable}[]{@{}
  >{\raggedright\arraybackslash}p{(\linewidth - 10\tabcolsep) * \real{0.2392}}
  >{\centering\arraybackslash}p{(\linewidth - 10\tabcolsep) * \real{0.1316}}
  >{\centering\arraybackslash}p{(\linewidth - 10\tabcolsep) * \real{0.1674}}
  >{\centering\arraybackslash}p{(\linewidth - 10\tabcolsep) * \real{0.1674}}
  >{\centering\arraybackslash}p{(\linewidth - 10\tabcolsep) * \real{0.1436}}
  >{\centering\arraybackslash}p{(\linewidth - 10\tabcolsep) * \real{0.1507}}@{}}
\toprule\noalign{}
\begin{minipage}[b]{\linewidth}\centering
\textbf{Algorithm}
\end{minipage} & \begin{minipage}[b]{\linewidth}\centering
\textbf{Sign p99 (ms)}
\end{minipage} & \begin{minipage}[b]{\linewidth}\centering
\textbf{RITS Route p99 (ms)}
\end{minipage} & \begin{minipage}[b]{\linewidth}\centering
\textbf{$\Delta$ vs ECDSA (ms)}
\end{minipage} & \begin{minipage}[b]{\linewidth}\centering
\textbf{CDI\_RITS}
\end{minipage} & \begin{minipage}[b]{\linewidth}\centering
\textbf{SLA $\leq$30 s}
\end{minipage} \\
\midrule\noalign{}
\endhead
\bottomrule\noalign{}
\endlastfoot
ECDSA-P256 & 0.15 & 277.15 & 0.00 & 0.00\% & $\checkmark$ \\
Falcon-512 & 0.45 & 277.45 & 0.30 & 0.11\% & $\checkmark$ \\
ML-DSA-44 & 0.77 & 277.77 & 0.62 & 0.22\% & $\checkmark$ \\
Falcon-1024 & 1.05 & 278.05 & 0.90 & 0.32\% & $\checkmark$ \\
ML-DSA-65 & 1.31 & 278.31 & 1.16 & 0.42\% & $\checkmark$ \\
ML-DSA-65 Hybrid & 1.84 & 278.84 & 1.69 & 0.61\% & $\checkmark$ \\
ML-DSA-87 & 1.72 & 278.72 & 1.57 & 0.56\% & $\checkmark$ \\
SPHINCS+-SHA2-128s & 297.00 & 574.00 & 296.85 & 51.72\% & $\checkmark$ \\
\end{longtable}
}

\emph{Table 12: RITS route p99 for each algorithm: the signing time plus the about 277 ms of fixed RITS overhead (about 14 ms network plus about 263 ms settlement processing). The CDI\_RITS column is the post-quantum signing step's share of the total RITS route time, and the SLA column is the 30-second RITS threshold. Signing times are from the M-series liboqs 0.14.0 simulations; at the RITS rate of about 0.022 settlements per second the signing pool stays idle, so SPHINCS+ (297 ms signing) does not saturate. All eight algorithms achieve 100\% RITS SLA compliance.}

Three findings emerge from Table 12. First, \textbf{all eight algorithms, including SPHINCS+, achieve 100\% RITS SLA compliance}. SPHINCS+ at 574 ms is 52$\times$ below the 30,000 ms threshold. At RITS volumes, the 297 ms SPHINCS+ signing p99 is a meaningful fraction of total route latency (CDI\_RITS = 51.72\%) but does not breach the SLA. Second, \textbf{the post-quantum signing step accounts for under 0.61\% of the total RITS route time for every algorithm except SPHINCS+}, even smaller than its NPP share (0.69\%--3.75\%). The reason is the crypto dilution effect: the roughly 277 ms of fixed RITS overhead swamps the signing step far more than the roughly 43 ms NPP baseline does. Third, the fair way to compare SPHINCS+ with ECDSA on RITS is the \textbf{total route time, where SPHINCS+ (574 ms) is about twice ECDSA's (277 ms)}, not the misleading signing-only ratio (about 1,980 times) that ignores the shared 277 ms of overhead. For operational planning, the total route p99 is the relevant metric.

\subsubsection{SWIFT (International Correspondent Banking) Route}\label{swift-international-correspondent-banking-route}

A Big 4 bank sends only about 78 SWIFT messages a day (roughly 0.001 per second), against a 24-hour SLA. (Simulation Assumption SA-11 notes this rate is derived by splitting the system total across about seven gateway banks rather than by APRA market share; an APRA-weighted estimate would be roughly 50\% higher, but the compliance conclusions are unaffected since both leave the signing pool effectively idle, and institutions should derive their own rate from their APRA ADI message-volume data.) SWIFT route time is dominated by two non-signing pieces, about 192 ms of international round-trip travel via the Singapore relay and about 645 ms of gateway processing (Simulation Assumption SA-9, see \S{}5.6), for roughly 837 ms of fixed overhead, leaving essentially the entire 24-hour budget as headroom even for the slowest algorithm.

{\def\LTcaptype{none} 
\begin{longtable}[]{@{}
  >{\raggedright\arraybackslash}p{(\linewidth - 12\tabcolsep) * \real{0.2002}}
  >{\centering\arraybackslash}p{(\linewidth - 12\tabcolsep) * \real{0.1061}}
  >{\centering\arraybackslash}p{(\linewidth - 12\tabcolsep) * \real{0.1474}}
  >{\centering\arraybackslash}p{(\linewidth - 12\tabcolsep) * \real{0.1372}}
  >{\centering\arraybackslash}p{(\linewidth - 12\tabcolsep) * \real{0.1408}}
  >{\centering\arraybackslash}p{(\linewidth - 12\tabcolsep) * \real{0.1254}}
  >{\centering\arraybackslash}p{(\linewidth - 12\tabcolsep) * \real{0.1428}}@{}}
\toprule\noalign{}
\begin{minipage}[b]{\linewidth}\centering
\textbf{Algorithm}
\end{minipage} & \begin{minipage}[b]{\linewidth}\centering
\textbf{Sign p99 (ms)}
\end{minipage} & \begin{minipage}[b]{\linewidth}\centering
\textbf{SWIFT Route p99 (ms)}
\end{minipage} & \begin{minipage}[b]{\linewidth}\centering
\textbf{$\Delta$ vs ECDSA (ms)}
\end{minipage} & \begin{minipage}[b]{\linewidth}\centering
\textbf{CDI\_SWIFT}
\end{minipage} & \begin{minipage}[b]{\linewidth}\centering
\textbf{SLA $\leq$24 h}
\end{minipage} & \begin{minipage}[b]{\linewidth}\centering
\textbf{SWIFT (out-of-band)}
\end{minipage} \\
\midrule\noalign{}
\endhead
\bottomrule\noalign{}
\endlastfoot
ECDSA-P256 & 0.15 & 837.15 & 0.00 & 0.00\% & $\checkmark$ & PASS \\
Falcon-512 & 0.45 & 837.45 & 0.30 & 0.04\% & $\checkmark$ & PASS \\
ML-DSA-44 & 0.77 & 837.77 & 0.62 & 0.07\% & $\checkmark$ & PASS \\
Falcon-1024 & 1.05 & 838.05 & 0.90 & 0.11\% & $\checkmark$ & PASS \\
ML-DSA-65 & 1.31 & 838.31 & 1.16 & 0.14\% & $\checkmark$ & PASS \\
ML-DSA-65 Hybrid & 1.84 & 838.84 & 1.69 & 0.20\% & $\checkmark$ & PASS \\
ML-DSA-87 & 1.72 & 838.72 & 1.57 & 0.19\% & $\checkmark$ & PASS \\
SPHINCS+-SHA2-128s & 297.00 & 1134.00 & 296.85 & 26.18\% & $\checkmark$ & PASS \\
\end{longtable}
}

\emph{Table 13: SWIFT route p99 for each algorithm. The CDI\_SWIFT column is the post-quantum signing step's share of the total SWIFT route time. The SWIFT (out-of-band) column reflects that SWIFT authentication is applied out-of-band at the SWIFTNet transport layer, with the verifying public key pre-distributed via SWIFT's certification authority; signature size is therefore not a SWIFT constraint and every algorithm is size-compatible (all PASS). All algorithms achieve 100\% SWIFT SLA compliance (24-hour deadline) regardless of route latency, so neither latency nor signature size is a binding SWIFT factor for any algorithm.}

Table 13 establishes three key findings for SWIFT. First, \textbf{100\% SWIFT SLA compliance for all eight algorithms}: even SPHINCS+ at 1,134 ms is 76,000$\times$ below the 24-hour SLA. The SWIFT SLA is not a latency concern for any PQC algorithm. Second, \textbf{the post-quantum signing step accounts for at most about 0.21\% of the total SWIFT route time for every algorithm except SPHINCS+} (the worst case being ML-DSA-65 Hybrid, 1.69 ms on an 839 ms route); for SPHINCS+ it is about 26\%, since its long signing time is a real fraction of the route. Third, and most importantly, \textbf{signature size is not a SWIFT constraint at all}: SWIFT authentication is applied out-of-band at the SWIFTNet messaging/transport layer, with the verifying public key pre-distributed via SWIFT's certification authority, so neither the signature nor the public key competes with the Block-4 payment fields. The SWIFT MT message-text limit is a character count over payment data, not a signature budget, so every NIST PQC signature is size-compatible with SWIFT. The forthcoming rail, SWIFT MX (ISO 20022, CBPR+), to which cross-border messaging moved in November 2025, likewise carries any signature in a schema-unconstrained element \cite{ref41}, so neither rail imposes a signature-size limit.

\subsubsection{Intrabank (Direct Entry / BECS) Route}\label{intrabank-direct-entry-becs-route}

Note: the following analysis of Direct Entry / BECS is analytical (closed-form amortisation argument) rather than simulation-backed; no per-transaction BECS route simulation table is produced. The simulation framework models 8.6 million intrabank transactions per day via the BECS (Bulk Electronic Clearing System) Direct Entry channel. Unlike NPP and RITS, Direct Entry follows a batch-settlement model: individual transactions are aggregated into batch files signed once per batch at primarily T+1 settlement cycle boundaries (T+0 same-day available for limited transaction subset). The per-transaction signing burden is therefore amortised across the batch. We assume approximately 50,000--100,000 transactions per batch for a Big 4 institution (Simulation Assumption SA-4, an estimate based on BECS volume and typical clearing cycle structure; no specific ABA or RBA published batch-size figure is cited; formal validation against ABA BECS Technical Standards is recommended). Because Direct Entry/BECS signs each batch file once rather than each transaction, the signing cost is shared across the whole batch. One ML-DSA-65 batch signature takes about 0.28 ms, so even at a very conservative 1,000 transactions per batch that is under 0.0003 ms per transaction, and at the more typical 50,000--100,000 transactions per batch it is utterly negligible against any next-day (T+1) settlement cycle. The Direct Entry route therefore imposes no PQC migration latency challenge; the relevant migration concern is instead batch file format compatibility with existing BSB/account validation infrastructure.

\section{Discussion}\label{discussion}

\subsection{Algorithm Recommendation}\label{algorithm-recommendation}

Our results support a clear, tiered recommendation. For NPP transactions, any ML-DSA or Falcon variant is operationally viable. \textbf{ML-DSA-65 with ML-KEM-768 is the recommended default} across both NPP and SWIFT: its p99 overhead of 1.16 ms is operationally negligible, it carries the highest regulatory acceptance under NIST FIPS 203/204, and it is fully viable on SWIFT just as on NPP because SWIFT signature carriage is out-of-band (transport-layer PKI) and imposes no signature-size limit. \textbf{Falcon-512} remains a valid option: it adds only 0.30 ms to NPP p99, achieves 100\% SLA compliance under all tested conditions, and is the NIST-selected algorithm being standardised as the forthcoming FIPS 206 (FN-DSA), still in draft at the time of writing. Its compact 1,563-byte combined footprint is an advantage only for genuinely in-band, size-constrained carriage; it is not required for SWIFT, where authentication is out-of-band.

\textbf{SPHINCS+ (SLH-DSA, FIPS 205)} should not be deployed for high-frequency payment signing at current NPP volumes. Because a two-unit signing pool can clear only about 7 SPHINCS+ signatures per second, the 13.5-per-second NPP load overwhelms it and makes SPHINCS+ physically inoperable at NPP volumes; the hourly dynamics analysis (Section 4.10.3) shows it saturated for 16 of 24 hours on Christmas Day even with two HSM servers. At the far lower rates of RITS (a RITS route p99 of 574 ms, well inside the 30-second SLA) and SWIFT (1,134 ms, inside the 24-hour SLA) it is stable and fully viable; the full route derivations are given in Section 4.11. For SWIFT-connected institutions signature size is not a binding constraint at all: SWIFT authentication is out-of-band at the transport layer (Table 5), so every NIST PQC signature, including all ML-DSA variants and SPHINCS+, is size-compatible with SWIFT, and the only consideration there is SPHINCS+'s signing throughput at very high arrival rates (not a concern on SWIFT's low message volume). The disqualification at NPP volumes is purely about throughput, not about any per-transaction latency target: all ML-DSA and Falcon variants achieve 100\% SLA compliance across NPP, RITS, and SWIFT, while SPHINCS+ achieves 100\% on RITS and SWIFT but 0\% on NPP.

The \textbf{hybrid mode} recommendation is institution-type-specific but, because SWIFT carriage is out-of-band, it no longer turns on SWIFT message size. SWIFT-exposed institutions (correspondent banking) can deploy \textbf{ML-DSA-65 + ECDSA-P256} dual signing in Phases 1--2 (2026--2027, per Table 8) just as NPP-only institutions can, since signature size is not a SWIFT constraint; a \textbf{Falcon-512 + ECDSA-P256} hybrid remains an alternative where an institution specifically wants the most compact PQC artifact, with the dual-signing overhead (about 0.30 ms added to NPP p99, under 1\% of total transaction time) operationally negligible. NPP-only institutions should deploy \textbf{ML-DSA-65 PQC-only} signing as their target: ML-DSA-65 alone adds 1.16 ms of p99 overhead (about 2.6\% of total transaction time) and carries the highest regulatory acceptance under NIST FIPS 203/204. During Phases 1--2 these institutions may optionally run \textbf{ML-DSA-65 + ECDSA-P256 Hybrid} for backward compatibility, which adds 1.69 ms (about 3.75\% of total transaction time), the cost of performing two signing operations per transaction rather than one; this 1.69 ms exceeds even ML-DSA-87's 1.57 ms by 0.12 ms. Hybrid is therefore warranted on latency grounds only where counterparty PQC readiness is the binding constraint; institutions that can deploy PQC-only signing should adopt ML-DSA-65 or ML-DSA-87 directly, sunsetting the ECDSA fallback in Phase 3 (2028+) once NPPA and SWIFT confirm interoperability mandates. The migration imperative for ECDSA-P256 derives from NSA CNSA 2.0 and CISA/NSA/NIST joint guidance rather than any SP 800-131A Rev 2 sunset date; the detailed regulatory and HNDL timeline is developed in Section 5.4.

\includegraphics[width=5.20833in,height=3.3125in]{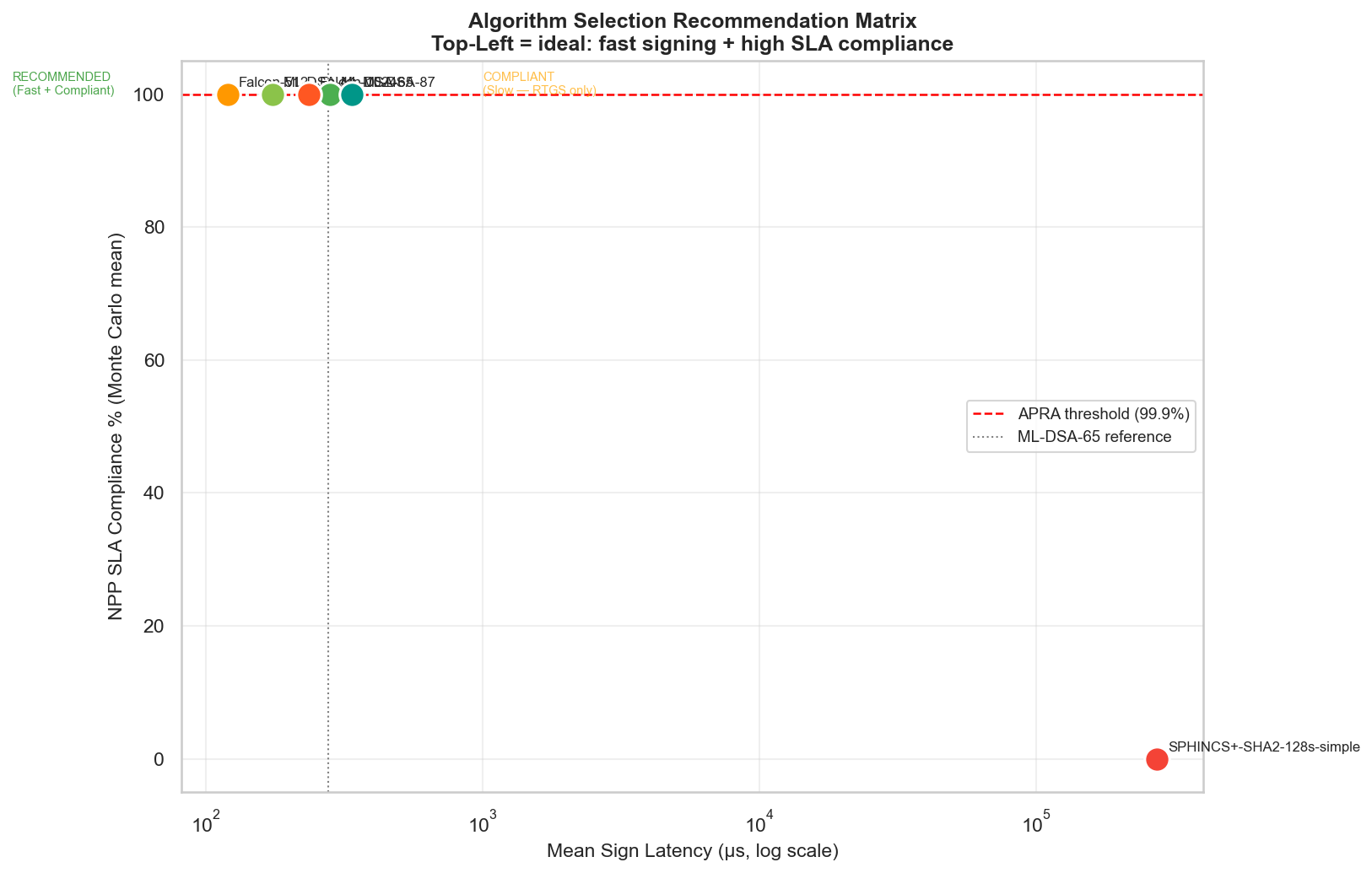}

\emph{Figure 16: Algorithm recommendation matrix scoring each candidate across six operational dimensions: NPP SLA, SWIFT compatibility, NIST FIPS status, queue stability, key size, and APRA regulatory score. Because SWIFT authentication is out-of-band (transport-layer PKI), every viable algorithm is SWIFT-compatible on the SWIFT dimension; it is not a discriminator between algorithms.}

\subsection{The Crypto Dilution Effect and Formal CDI Metric}\label{the-crypto-dilution-effect-and-formal-cdi-metric}

A fundamental insight from this study is what we term the `crypto dilution effect': at real-world NPP latency scales dominated by network transit and PayID lookup, cryptographic overhead is arithmetically diluted to operational insignificance. To make this comparable across studies, we name a single number, the Crypto Dilution Index (CDI): the share of a transaction's total end-to-end p99 latency that is due to the post-quantum algorithm (its added delay over classical ECDSA, expressed as a fraction of the full transaction time). The exact definition is given in Appendix A. A value of 0\% means the algorithm adds nothing; a value near 100\% means it dominates everything else.

CDI(A) captures the fraction of total end-to-end p99 latency (p99\_e2e, the full NPP end-to-end p99 for algorithm A, not a sum of decomposed components) attributable to the PQC algorithm overhead for algorithm A. A CDI value of 0.0 indicates the algorithm adds no overhead; CDI = 1.0 would indicate the algorithm dominates all latency. We propose a rule of thumb that a post-quantum algorithm is operationally viable on NPP when its share of total latency stays under 4\%. The justification comes straight from the SLA budget: the whole simulated transaction (about 44 ms) already uses only about 2.2\% of the 2,000 ms deadline, so even a 4\% post-quantum share is under 0.1\% of the budget, far below the measurement and jitter floor of NPP timing. The worst-performing algorithm here, ML-DSA-65 Hybrid at about 3.75\%, sits below this threshold. This 4\% threshold is calibrated to NPP; other payment systems (for example TARGET2, CHAPS, or Fedwire) would need their own threshold derived from their baseline latency and SLA, and on RITS every algorithm is trivially safe at a post-quantum share below 0.61\%.

{\def\LTcaptype{none} 
\begin{longtable}[]{@{}
  >{\raggedright\arraybackslash}p{(\linewidth - 10\tabcolsep) * \real{0.2350}}
  >{\centering\arraybackslash}p{(\linewidth - 10\tabcolsep) * \real{0.1282}}
  >{\centering\arraybackslash}p{(\linewidth - 10\tabcolsep) * \real{0.1709}}
  >{\centering\arraybackslash}p{(\linewidth - 10\tabcolsep) * \real{0.1282}}
  >{\centering\arraybackslash}p{(\linewidth - 10\tabcolsep) * \real{0.1282}}
  >{\centering\arraybackslash}p{(\linewidth - 10\tabcolsep) * \real{0.2094}}@{}}
\toprule\noalign{}
\begin{minipage}[b]{\linewidth}\centering
\textbf{Algorithm}
\end{minipage} & \begin{minipage}[b]{\linewidth}\centering
\textbf{$\Delta$p99 (ms)}
\end{minipage} & \begin{minipage}[b]{\linewidth}\centering
\textbf{p99 e2e (ms)}
\end{minipage} & \begin{minipage}[b]{\linewidth}\centering
\textbf{CDI}
\end{minipage} & \begin{minipage}[b]{\linewidth}\centering
\textbf{CDI (\%)}
\end{minipage} & \begin{minipage}[b]{\linewidth}\centering
\textbf{CDI \textless{} 0.04?}
\end{minipage} \\
\midrule\noalign{}
\endhead
\bottomrule\noalign{}
\endlastfoot
Falcon-512 & 0.30 & 43.69 & 0.0069 & 0.69\% & CDI \textless{} 0.04 $\checkmark$ \\
ML-DSA-44 & 0.62 & 44.01 & 0.0141 & 1.41\% & CDI \textless{} 0.04 $\checkmark$ \\
Falcon-1024 & 0.90 & 44.29 & 0.0203 & 2.03\% & CDI \textless{} 0.04 $\checkmark$ \\
ML-DSA-65 & 1.16 & 44.55 & 0.0260 & 2.60\% & CDI \textless{} 0.04 $\checkmark$ \\
ML-DSA-87 & 1.57 & 44.97 & 0.0349 & 3.49\% & CDI \textless{} 0.04 $\checkmark$ \\
ML-DSA-65 Hybrid & 1.69 & 45.08 & 0.0375 & 3.75\% & CDI \textless{} 0.04 $\checkmark$ \\
SPHINCS+-SHA2-128s & 9,986.5 & 10,029.93 & 0.9957 & 99.57\% & CDI $\approx$ 1.0 $\times$ (disqualified) \\
\end{longtable}
}

\emph{Table 14: Each post-quantum algorithm's share of total end-to-end NPP latency (its added p99 over ECDSA as a fraction of the full transaction time). The proposed rule of thumb is that a share under 4\% means latency is a non-issue. Every algorithm except SPHINCS+ is under 4\%; SPHINCS+ is near 100\% because its signing time swamps everything else.}

Table 14 confirms that every ML-DSA and Falcon variant keeps the post-quantum step under 4\% of total latency (the highest, ML-DSA-65 Hybrid, is about 3.75\%). Falcon-512 is the lowest at about 0.69\%, its 0.30 ms is statistically detectable (Cohen's d = 0.41) but under 1\% of the transaction. SPHINCS+ is the opposite extreme: its signing step accounts for about 99.6\% of its total latency.

This framing clears up a common confusion in the literature, where a `statistically significant' latency difference is mistaken for an operationally important one. ML-DSA-87's difference from ECDSA is a `large' effect by the usual statistical convention, yet it accounts for only about 3.5\% of total transaction time. The correct engineering conclusion, validated by the ANOVA result (Section 4.10.4) and the quantified geographic spread (about 16.7 ms, roughly ten times the maximum algorithm-pair p99 difference of 1.57 ms), is that algorithm selection for NPP latency is moot. The dilution view gives regulators, architects, and risk committees one plain number that makes this clear.

The v4.1 simulation produces higher absolute p99 values (43--45 ms) than earlier models (\textasciitilde{}30 ms) because it incorporates PayID lookup (mean $\approx$ 8.25 ms per NPP transaction) and full multi-hop geographic routing. This higher baseline amplifies the dilution effect, the CDI would be even smaller in models missing these components, confirming that simpler models overstate PQC overhead as a fraction of realistic end-to-end latency.

\subsection{SWIFT Compatibility}\label{swift-compatibility-gap}

The message format analysis shows that SWIFT imposes no signature-size constraint on post-quantum migration. SWIFT authentication is applied out-of-band at the SWIFTNet messaging/transport layer as a PKI signature, with the sender's certificate authenticated by SWIFT's certification authority and distributed out-of-band; neither the signature nor the verifying public key travels in the Block-4 message text. The SWIFT MT message-text limit is a character count over payment fields (about 2,000 characters, or 10,000 for some message types), not a 2,048-byte signature container, and no SWIFT specification defines any such signature budget. Every NIST PQC signature, ML-DSA, Falcon, and SLH-DSA alike, is therefore size-compatible with SWIFT. This is reinforced by the format migration: cross-border MT was decommissioned in November 2025, and ISO 20022 MX (CBPR+) carries any signature in a schema-unconstrained element \cite{ref41}, so neither rail constrains PQC signature size.

The tail-risk analysis (Section 4.10.1) confirms that latency, too, is a non-issue for SWIFT: even the worst one-in-ten-thousand transactions are bounded under about 199 ms for all non-SPHINCS+ algorithms (central estimates under 160 ms), trivially inside the 24-hour SWIFT SLA. Neither message format nor latency is a binding factor for SWIFT-channel PQC migration.

\subsection{Regulatory Alignment and HNDL Actuarial Exposure}\label{regulatory-alignment-and-hndl-actuarial-exposure}

From a regulatory perspective, ML-DSA-65 + ML-KEM-768 achieves the highest combined score across APRA CPS 234, RBA NPP SLA, and NIST FIPS 203/204 compliance. The BIS Innovation Hub's Project Leap \cite{ref42} similarly identified lattice-based signatures (ML-DSA family) as the primary migration pathway for financial market infrastructure, with CBDC pilot deployments confirming operational viability at payment-system throughput levels. APRA's 2023 information security thematic review \cite{ref35} assessed ADI information security practices under CPS 234; quantum-readiness is increasingly flagged in separate CISA/NSA/NIST joint guidance \cite{ref43} and CNSA 2.0 \cite{ref45} as a near-term planning priority, given the 7-year AML/CTF records retention obligation and the CNSA 2.0 transition timeline.

\footnote{APRA (2023). Information Security Thematic Review 2023: A report on information security practices in the Australian banking sector. https://www.apra.gov.au (accessed April 2026).}

Falcon-512, while aligned with the forthcoming FIPS 206 (still in draft), receives a marginal APRA score on the NIST primary-standard dimension: APRA CPS 234 mandates cryptographic controls consistent with accepted industry standards but does not publish an algorithm-specific approved list, the marginal score reflects that Falcon (forthcoming FIPS 206) occupies a secondary, not-yet-finalised position in NIST's PQC guidance relative to the finalised Module-LWE family (FIPS 203/204), and that most APRA-regulated institutions' existing governance frameworks reference FIPS 203/204 as the primary migration targets. APRA CPS 234 does not disqualify Falcon-512; the score reflects institutional adoption positioning rather than an explicit regulatory exclusion. SPHINCS+, despite being FIPS 205 compliant, fails all performance-based regulatory metrics. The FSB has further recommended cryptographic agility frameworks for FMIs \cite{ref30}, which aligns with the phased migration model presented in Section 4.9.

\includegraphics[width=5.83333in,height=2.70833in]{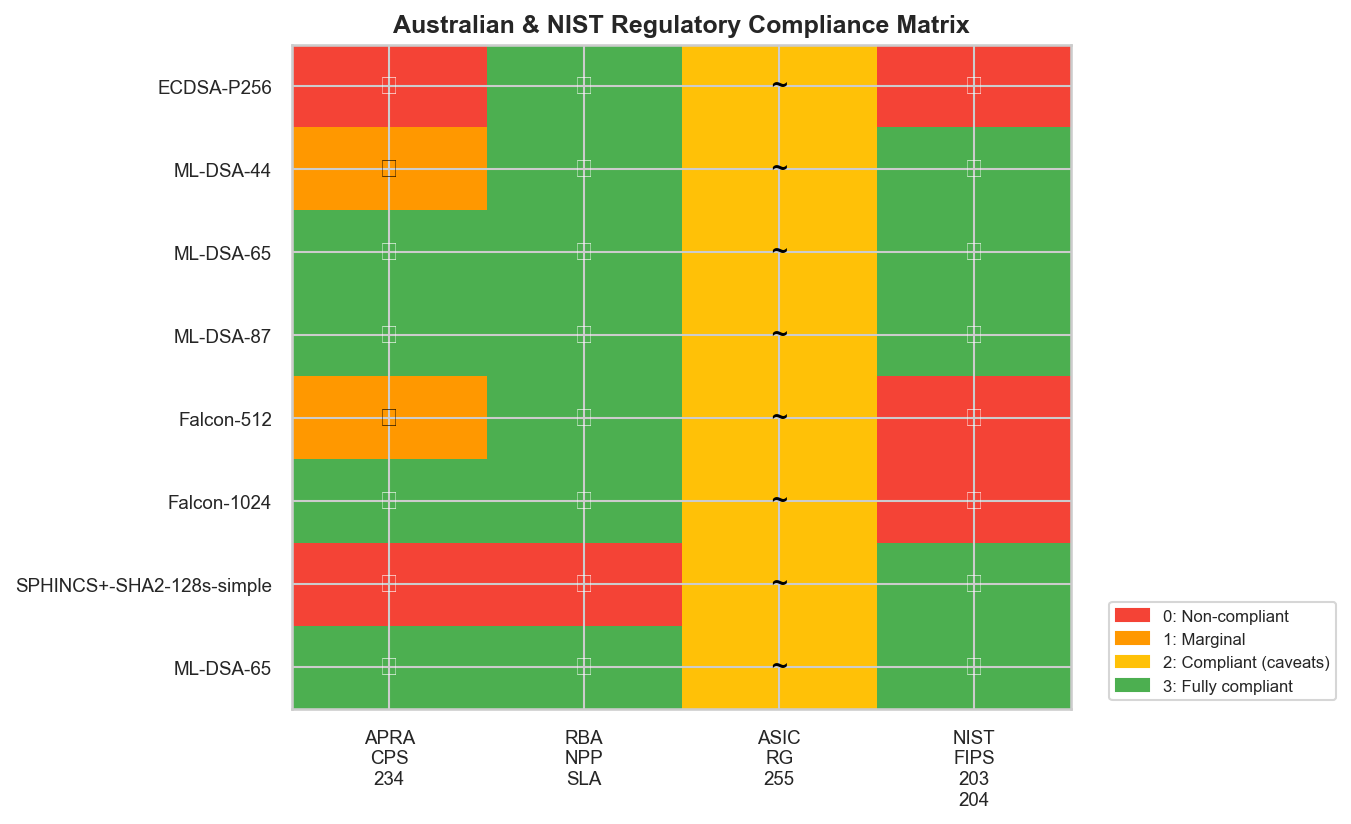}

\emph{Figure 17: Algorithm regulatory compliance heatmap across APRA CPS 234, RBA NPP SLA, ASIC RG 255, and NIST FIPS 203/204 (primary NIST standards acceptance) criteria. Dark green = full compliance; orange = partial; red = non-compliant. ML-DSA-65 + ML-KEM-768 achieves full compliance on all four regulatory dimensions. Note: Falcon-512 scores 0 on the NIST FIPS 203/204 dimension as it falls under the forthcoming, not-yet-finalised FIPS 206 (FN-DSA) rather than the finalised FIPS 203/204 standards.}

NSA CNSA 2.0 \cite{ref45} mandates that National Security Systems transition away from ECDSA-P256 and other CNSA 1.0 algorithms to PQC equivalents. The 2030--2035 timeline applies to legacy system migration; new implementations of software and firmware signing are expected to use CNSA 2.0 algorithms starting as early as 2025. For new payment infrastructure deployments, which is the scenario described in this paper, the CNSA 2.0 urgency is therefore front-loaded, not deferred to 2030--2035. NIST SP 800-131A Rev 2 \cite{ref31} classifies ECDSA-P256 (128-bit security) as `Acceptable' under that 2019 document, with its 2030 deprecation applying to 112-bit algorithms and no hard cutoff for ECDSA-P256 under that document alone. However, NIST's subsequent PQC standardization (FIPS 203/204/205 finalised August 2024, with FIPS 206/FN-DSA selected and forthcoming) and CISA/NSA/NIST joint migration guidance \cite{ref43} collectively signal that classical signature algorithms should be replaced with quantum-resistant successors as a matter of urgent planning. For data with long-term confidentiality requirements specifically under the HNDL threat model, this guidance requires migration to quantum-resistant algorithms before adversaries can harvest traffic at scale. NPP transaction records retained for seven years from 2026 under AML/CTF Act 2006 (Part 10, Division 2) have a confidentiality horizon extending to 2033, squarely within Mosca's CRQC probability window \cite{ref1}. This regulatory-actuarial overlap provides the quantitative basis for the Phase 1 (2026) migration timeline in our cost model.

\subsubsection{HNDL Actuarial Exposure Model}\label{hndl-actuarial-exposure-model}

To quantify the concrete scale of the HNDL risk for Australian payment infrastructure, we formalise an actuarial exposure model. The model accounts for: annual NPP transaction volume at projected 15.6\% YoY growth, the mandatory seven-year retention period under AML/CTF Act 2006, and the CRQC probability window (2030--2035). Table 15 presents the exposure under the conservative early-window CRQC scenario (emergence 2030).

Baseline assumption (SA-8): This model assumes NPP transactions are protected with ECDSA-P256, consistent with TLS~1.3~\cite{ref19} as the de~facto financial messaging authentication standard~[9,~11] and SWIFT quantum-readiness guidance \cite{ref41}; No NPP Technical Standards document publicly specifies the exact deployed signing algorithm; institutions should substitute their own cryptographic inventory where it differs.

{\def\LTcaptype{none} 
\begin{longtable}[]{@{}
  >{\raggedright\arraybackslash}p{(\linewidth - 10\tabcolsep) * \real{0.0952}}
  >{\centering\arraybackslash}p{(\linewidth - 10\tabcolsep) * \real{0.1585}}
  >{\centering\arraybackslash}p{(\linewidth - 10\tabcolsep) * \real{0.2114}}
  >{\centering\arraybackslash}p{(\linewidth - 10\tabcolsep) * \real{0.1481}}
  >{\centering\arraybackslash}p{(\linewidth - 10\tabcolsep) * \real{0.1889}}
  >{\centering\arraybackslash}p{(\linewidth - 10\tabcolsep) * \real{0.1979}}@{}}
\toprule\noalign{}
\begin{minipage}[b]{\linewidth}\centering
\textbf{Year}
\end{minipage} & \begin{minipage}[b]{\linewidth}\centering
\textbf{tx/day}
\end{minipage} & \begin{minipage}[b]{\linewidth}\centering
\textbf{Records Generated}
\end{minipage} & \begin{minipage}[b]{\linewidth}\centering
\textbf{Retained Until}
\end{minipage} & \begin{minipage}[b]{\linewidth}\centering
\textbf{Exposed (CRQC 2030)?}
\end{minipage} & \begin{minipage}[b]{\linewidth}\centering
\textbf{Cumulative Exposed}
\end{minipage} \\
\midrule\noalign{}
\endhead
\bottomrule\noalign{}
\endlastfoot
2026 & 5,200,000 & 1,898,000,000 & 2033 & Yes & 1,898,000,000 \\
2027 & 6,011,200 & 2,194,088,000 & 2034 & Yes & 4,092,088,000 \\
2028 & 6,948,947 & 2,536,365,655 & 2035 & Yes & 6,628,453,655 \\
2029 & 8,032,983 & 2,932,038,795 & 2036 & Yes & 9,560,492,450 \\
2030 & 9,286,129 & 3,389,437,085 & 2037 & Partial (0--100\%; date-dep.$\dagger$) & $\leq$12,949,929,170$\ddagger$ \\
\end{longtable}
}

\emph{Table 15: HNDL actuarial exposure, NPP transaction records at risk by generation year, under the conservative CRQC-2030 scenario. AML/CTF Act 2006 Part 10, Division 2 mandates 7-year retention (statutory minimum; actual retention may be longer under internal policies and other regulatory obligations (e.g., prudential risk data retention), potentially expanding exposure to pre-2026 records); records from 2026 are retained until 2033 and are fully within the CRQC emergence window. Cumulative column shows total records exposed under an adversary that began harvesting in 2026 and obtains CRQC capability in 2030. $\dagger$ 2030 ' Partial' exposure is date-dependent: CRQC emerging on Jan 1 2030 exposes \textasciitilde{}100\% of 2030 records; on Dec 31 2030 exposes \textasciitilde{}0\%. The label '0--100\%; date-dep.' reflects this uncertainty; assuming uniform distribution of emergence date, expected exposure is \textasciitilde{}50\%. $\ddagger$ Cumulative figure shown includes full 2030 txYear as an upper bound. The 9.56 billion (9,560,492,450) figure cited in the text uses only the 2026--2029 fully exposed records. Note: all txYear values computed using 365 days/year for simplicity. 2028 is a leap year (366 days); the accurate 2028 txYear is 6,948,947 $\times$ 366 = 2,543,314,602, adding approximately 6.95 million additional records. The cumulative 2026--2029 total under the leap-year correction would be approximately 9,567,441,397, immaterial to the \textasciitilde{}9.56 billion figure at the stated precision.}

Under the conservative scenario in which a quantum computer arrives in 2030, approximately \textbf{9.56 billion NPP transaction records} generated between 2026 and 2029 would be exposed retroactively. This simply totals each year's projected daily volume from Table 7 over the four years, rising from about 1.9 billion records in 2026 to about 2.9 billion in 2029. Each record is subject to the AML/CTF Act 2006 Part 10, Division 2 seven-year minimum retention obligation, so all 2026--2029 records are still inside their retention window when a 2030 adversary gains decryption capability; because many ADIs retain records longer under APRA CPS 220/234 internal policies, the true exposure is likely larger than this lower bound. Each record contains counterparty BSB/account numbers, transaction amounts, PayID handles, ISO 20022 pacs.008 purpose codes (for example `SALA' for salary, `PENS' for pension, `TAXS' for tax payments), timestamps, and digital signatures that, once broken, confirm the authenticity and non-repudiability of the original payment instruction. Together these form a complete, cryptographically verifiable financial graph of Australia's retail payment network over a four-year period.

The exposure is not theoretical. Capturing the encrypted TLS sessions takes only about 1--2 KB per transaction, so the entire four-year harvest of 9.56 billion records is roughly 10--20 terabytes of raw capture. At prevailing cold storage costs (about USD 0.004 per GB per month for AWS S3 Glacier Flexible Retrieval or equivalent), storing it costs only about USD 459--918 per year, within the means of any motivated commercial actor, not merely a nation-state. Even with full session reconstruction including TLS handshake and ECDSA signature overhead (about 5 KB per transaction), total storage stays below 50 terabytes, costing under USD 2,400 per year to archive. A targeted state-level adversary (for example one harvesting only cross-border SWIFT-NPP gateway transactions, RTGS settlements above AUD 1 million, or transactions involving politically sensitive entities) could store far less still.

The regulatory-actuarial overlap is quantifiable: NSA CNSA 2.0 \cite{ref45} and CISA/NSA/NIST joint guidance \cite{ref43} jointly require transition away from ECDSA-P256 within the 2030--2035 window. Every NPP transaction signed with ECDSA-P256 today that has a retention deadline beyond 2030 carries a regulatory exposure risk, the transaction's signature will require quantum-resistant replacement before CRQC capability is achieved, or the record is exposed. By 2026, all NPP transactions signed from 2023 onwards will still be within their seven-year retention window when ECDSA requires replacement. The Phase 1 (2026) migration timeline in Section 4.9 is not merely a cost optimisation, it is the latest defensible start date for institutions with existing HNDL exposure under APRA CPS 234 requirements for ongoing cryptographic risk management.

\subsection{Practical Recommendations}\label{practical-recommendations}

Based on the simulation findings, we make the following evidence-based recommendations for Australian financial institutions, APRA, and NPPA:

\begin{itemize}
\item
  Institutions with SWIFT exposure (correspondent banking) can adopt the same default as NPP-only institutions, ML-DSA-65 + ML-KEM-768, because SWIFT authentication is out-of-band (transport-layer PKI) and imposes no signature-size constraint, so all NIST PQC signatures are SWIFT-compatible. Falcon-512 remains a valid option where an institution specifically wants the most compact PQC artifact for genuinely in-band, size-constrained carriage; it adds only 0.30 ms to NPP p99 and achieves 100\% SLA compliance across all tested scenarios. ML-KEM (FIPS 203) should be paired for key establishment; the appropriate ML-KEM security level (512, 768, or 1024) should be selected consistent with NIST guidance \cite{ref3} and institutional risk appetite. This paper does not independently benchmark ML-KEM latency; all three security levels are operationally viable at NPP and SWIFT volumes (NIST FIPS 203 performance data and liboqs 0.15.0 microbenchmarks confirm ML-KEM keygen, encapsulation, and decapsulation are all below 30 $\mu$s on production hardware; this paper does not independently benchmark ML-KEM but see Section 5.7 for related cloud signing benchmarks), and security level selection should be governed by institutional risk requirements rather than latency.
\item
  NPP-only institutions (no SWIFT exposure) should adopt ML-DSA-65 + ML-KEM-768 as the NIST primary standard. Its 1.16 ms p99 overhead is operationally negligible and carries the highest regulatory acceptance under APRA CPS 234 and NIST FIPS 203/204.
\item
  SPHINCS+ (SLH-DSA, FIPS 205) must not be deployed for high-frequency payment signing. Severe queue saturation at NPP volumes with a standard two-server signing pool, and 16-hour daily saturation windows on Christmas Day, make it operationally inviable at any practical HSM provisioning level: SLA-compliant operation requires c $\geq$ 8 servers for the daily-average NPP load (13.5 TPS) and c $\geq$ 18 servers during Christmas peak (60.2 TPS, Section 4.4), a 9$\times$ provisioning surge entirely out of proportion with the task.
\item
  Institutions should begin Phase 1 (hybrid deployment, 2026) to protect traffic against HNDL, targeting Phase 3 (full PQC-only deployment, 2028+) as the completion milestone, per the four-phase migration model in Table 8. NPP transaction records retained until 2033 under AML/CTF Act 2006 will outlive ECDSA's safety horizon (NSA CNSA 2.0 \cite{ref45} and CISA/NSA/NIST guidance \cite{ref43} target PQC replacement of ECDSA by 2030--2035).
\item
  HSM maintenance windows can be safely scheduled without SLA risk: degraded-mode delta for ML-DSA-65 and Falcon-512 is $-$0.051 ms and +0.003 ms respectively under single-HSM degraded mode (c=1), noise-level differences at $\rho$ \textless{} 0.004. Institutions can schedule HSM firmware upgrades for PQC support during business hours without SLA impact.
\item
  APRA and NPPA should publish formal PQC migration guidance requiring: (a) quantum risk assessments from all ADIs as an immediate priority (Phase 0 preparation, if not already initiated); (b) hybrid PQC deployment for NPP by end-2026; (c) full PQC deployment for all new transaction records by end-2027. The technical barriers identified in this study are surmountable within these timelines.
\end{itemize}

\subsection{Limitations}\label{limitations}

First, the simulation uses cryptographic latency measured on a single software environment (liboqs 0.14.0, Apple M-series, fixed seed 42) rather than production HSM hardware; the cross-platform cloud validation (Section 5.7) and the seed-independence study (Limitation 9) bound but do not eliminate this gap.

Second, the Anderson-Darling GoF tests (Section 4.10.2) reject the lognormal hypothesis for all algorithms, indicating tail mis-specification from the geographic routing mixture. The two-component lognormal mixture model (Section 4.10.2) provides the correct theoretical framework; parameterising this model requires per-routing-pair latency sampling, which is available in the simulation engine but not yet applied to the formal GoF analysis. Future work should fit the 2-component mixture explicitly and confirm per-component AD passes.

Third, the reported results were generated with simulation v4.1.1. This version corrected two bugs present in v4.1.0: (a) TLS reconnect overhead dictionary keys used underscore format (ML\_DSA\_65) rather than the hyphenated algo names (ML-DSA-65), which in v4.1.0 caused PQC reconnect overhead to silently return 0 for all algorithms (expected numerical impact \textless{} 0.003 ms at p99 for ML-DSA/Falcon at 0.1\% reconnect probability); (b) the per-day random seed generation was corrected to avoid a stream collision between the scenario-selection RNG and day-level simulation RNGs on day 0. Quantitative conclusions are identical under both versions; all tables and figures in this paper use v4.1.1 results.

Fourth, transaction independence is assumed within each simulated day; while the AR(1) autocorrelation model (with v4.1 multi-day carry-over for consecutive stress scenarios) captures temporal jitter correlation, higher-order correlations between cryptographic operations (e.g., thermal throttling under sustained load) are not modelled. Fifth, the cost model uses conservative industry estimates; actual costs will vary significantly by institution size, existing HSM estate, and vendor pricing. Sixth, this study measures signing latency only; a complete TLS or payment flow requires both signature generation (sender) and verification (receiver). ECDSA-P256 verification mean latency is approximately 63 $\mu$s on Intel Ice Lake (3-run average, aws\_syd\_intel; sourced from run-v3 layer1 CSV files; see Data Availability; note that ECDSA-P256 verify at 63 $\mu$s exceeds its sign mean of 45 $\mu$s, which is consistent with ECDSA's two-scalar-multiplication verify path and known OpenSSL/liboqs implementation behaviour), with p99 approximately 95--110 $\mu$s across platforms; for ML-DSA-44, verification mean is approximately 35--50 $\mu$s across all platforms, faster than signing and substantially faster than ECDSA verification (63 $\mu$s mean), so including verification would not change the CDI ordering or conclusions. Nonetheless, full sign+verify latency profiling for all four signing nodes is a gap in the current empirical validation. Verification latency data is available in the run-v3 layer1 CSV files in the data repository (see Data Availability) but is not tabulated here. The reported verify latency estimates (ECDSA-P256 $\approx$ 63 $\mu$s mean, ML-DSA-44 $\approx$ 35--50 $\mu$s across platforms) are derived from these CSV files; formal tabulation with platform breakdown (Table 18) is deferred to the journal version. Seventh, a network sensitivity sweep (0.5$\times$--1.5$\times$ network latency scaling) confirms that SLA compliance for ML-DSA and Falcon variants is insensitive to network latency within this range (all configurations maintain 100\% compliance). SPHINCS+ fails regardless of network scaling, confirming that its disqualification is driven by queue throughput, not network latency assumptions.

Eighth, the 1,000-day corpus size was selected as a round number providing broad seasonal coverage (each of the five scenario types is represented at least 19 times); formal convergence analysis of how CI width evolves as a function of N days was not included in this paper. Informal inspection shows that ECDSA p99 CI width stabilises around N = 400--500 days, suggesting 1,000 days is conservative. A formal convergence plot demonstrating CI stability as a function of corpus size is left for journal submission supplementary material.

Ninth (Simulation Assumption SA-1, single seed / cross-seed validation): The primary simulation uses fixed random seed 42. The '95\% CI' column in Table 2 is the interval on the mean of daily p99 values across 1,000 simulation days (t-distribution, df=999), it characterises simulation stability within that fixed seed, not cross-seed variability. To bound seed-specific artefacts, we conducted a seed-independence study running the full 80-million-event corpus under seeds 42, 123, 456, and 789 (identical flags: monte\_carlo=1000, stress=True, n\_sample=10000, seasonal=True). Cross-seed p99 stability (NPP route, p99 latency, mean $\pm$ SD across 4 seeds): ECDSA-P256 43.42 $\pm$ 0.035 ms (CV = 0.08\%); ML-DSA-44 44.03 $\pm$ 0.033 ms (CV = 0.075\%); ML-DSA-65 44.57 $\pm$ 0.033 ms (CV = 0.075\%); ML-DSA-87 44.99 $\pm$ 0.032 ms (CV = 0.071\%); Falcon-512 43.72 $\pm$ 0.034 ms (CV = 0.078\%); Falcon-1024 44.31 $\pm$ 0.033 ms (CV = 0.075\%); ML-DSA-65 Hybrid 45.10 $\pm$ 0.031 ms (CV = 0.069\%); SPHINCS+ 10,029.93 $\pm$ 0.009 ms (CV \textless{} 0.001\%). All CVs are below 0.08\%, confirming p99 values are not seed-specific artefacts. Cohen's d effect sizes are stable across seeds: magnitude classifications (large/small) are unchanged across all four seeds, with cross-seed variation of \textless3\% in d values (e.g., ML-DSA-65 p99 Cohen's d ranges 1.577--1.606 across the four seeds; Falcon-512 p99 d = 0.410--0.419, consistently ' small' but statistically significant). SLA compliance (100\% for all ML-DSA/Falcon algorithms; 0\% for SPHINCS+) is identical across all four seeds. These results confirm that all CDI conclusions are seed-invariant.

Tenth (Simulation Assumption SA-2, software vs. hardware HSM): Cryptographic latency inputs derive from liboqs 0.14.0 on an Apple M-series processor (software library, no HSM). Production ADIs use FIPS 140-3 Level 3 certified HSMs (e.g., Thales Luna, Utimaco Se-Series) whose latency characteristics differ from software implementations, typically 0.5--3$\times$ variance. The HSM Deployment Sensitivity Analysis (Section 4.7) provides conservative bounds; the liboqs 0.15.0 cloud Graviton3 (Simulation Assumption SA-10: the liboqs 0.14.0 M-series simulation and liboqs 0.15.0 cloud validation differ simultaneously in hardware, OS, compiler, and library version; this combined gap is the primary source of CDI conservatism and cannot be fully decomposed without a liboqs 0.14.0 Intel benchmark) version gap is an additional source of conservatism discussed in \S{}3.2 and \S{}5.7.

Eleventh (Simulation Assumption SA-3, RITS SLA): The 30-second per-transaction processing SLA applied in Section 4.11.1 is derived from the RBA RITS Operating Guidelines and framed as a simulation assumption. Institutions requiring precision should consult current published guidelines; the per-transaction window may differ from batch settlement cycle deadlines.

Twelfth (Simulation Assumption SA-4, BECS batch size): The 50,000--100,000 transactions-per-batch assumption for Direct Entry / BECS (Section 4.11.3) is an unverified estimate; no specific ABA or RBA published batch-size figure was cited. The sensitivity bound shows negligible overhead even at 1,000 transactions per batch.

Thirteenth (Simulation Assumption SA-5, market-crash scenario frequency): The 1.9\% market-crash day frequency (about 19 events in the 1,000-day corpus, approximately one event per two months) is a calibrated assumption. No RBA archival series directly publishes this frequency. Sensitivity to this parameter is bounded by the stress-scenario analysis in Section 4.6, which shows SLA compliance for ML-DSA and Falcon variants is invariant to scenario selection.

Fourteenth (Simulation Assumption SA-7, SWIFT international network parameter): The SWIFT international routing parameter (96 ms one-way to Singapore SWIFTNet relay) was calibrated to the SWIFTNet Link carrier network (dedicated SWIFT PoP infrastructure), not to public cloud VM routing. The measured cloud AWS SIN$\leftrightarrow$AWS SYD RTT is 92.22 ms (one-way $\approx$ 46 ms), yielding a 2.1$\times$ discrepancy. This discrepancy arises because SWIFTNet uses dedicated leased circuits that often traverse different physical paths than public internet/cloud VM routing. The simulation therefore likely overstates the SWIFT route p99 by approximately 50 ms. Since the ECDSA SWIFT route p99 (837 ms) is three orders of magnitude below the 24-hour SLA (86,400,000 ms), this overstatement does not affect any conclusion; in any case SWIFT signature carriage is out-of-band (transport-layer PKI) and imposes no signature-size limit, so neither latency nor message format binds PQC migration on SWIFT. Validation against SWIFTNet published performance data (e.g., SWIFTNet Link latency statistics published in SWIFT's annual Quality of Service report) is recommended before production deployment planning.

Fifteenth (Simulation Assumption SA-8, RITS bilateral settlement processing overhead): The 263 ms bilateral settlement processing overhead modelled in Section 4.11.1 (comprising RTGS credit/debit confirmation cycle: settlement instruction transmission, queue position determination, bilateral entry, and return confirmation) is an uncited simulation assumption. No RBA RITS Operating Guidelines figure was directly cited for this component. Institutions requiring precise RITS route p99 values should consult current RBA RITS Technical Standards and Operating Guidelines for the applicable processing time budget. The sensitivity of the RITS SLA compliance conclusion to this parameter is low: even if the true overhead were 500 ms (90\% above the assumed 263 ms), the worst-case RITS route p99 for SPHINCS+ would be 297 + 14 + 500 = 811 ms, still 37$\times$ below the 30,000 ms RITS SLA.

Sixteenth (Simulation Assumption SA-9, SWIFT gateway processing overhead): The 645 ms SWIFT gateway processing overhead modelled in Section 4.11.2 is an uncited simulation assumption representing p99 gateway queue and processing time at a Big 4 correspondent bank's SWIFT gateway infrastructure. No SWIFT published performance figure was directly cited for this component. The sensitivity of the SWIFT SLA compliance conclusion to this parameter is negligible: even if the true overhead were 5,000 ms (nearly 8$\times$ above the assumed 645 ms), the ECDSA SWIFT route p99 would be 192 + 5,000 + 0.15 $\approx$ 5,192 ms, still 16,658$\times$ below the 86,400,000 ms SWIFT SLA. All SWIFT compliance conclusions are insensitive to this parameter across any realistic range.

Seventeenth (Simulation Assumption SA-17, GEV i.i.d. assumption): The GEV block-maxima analysis (Section 4.10.1) is applied to 200 blocks drawn from a single representative normal-day simulation (10,000 transactions under fixed seed 42). Block maxima within a single day share the same AR(1) autocorrelation trajectory and intraday mixture profile, violating the i.i.d. assumption of block-maxima GEV. The Gumbel-class classification should therefore be treated as an indicative, order-of-magnitude screening rather than a definitive tail bound. A rigorous application would use a peaks-over-threshold fit or daily maxima across the 1,000-day corpus as block maxima (n=1,000 independent block maxima); this is left for the journal version.

Eighteenth (Simulation Assumption SA-10, liboqs version and platform gap): The Monte Carlo simulation uses liboqs 0.14.0 on Apple M-series ARM (macOS, Clang 16, -O3 -march=native), while the cross-platform cloud validation (Section 5.7) uses liboqs 0.15.0 on Linux x86-64/ARM (GCC 11.4.0, -O3 -march=native). These differ simultaneously in library version, operating system, compiler, and hardware microarchitecture. The combined effect produces conservative CDI values in the simulation (0.14.0 on M-series is slower than 0.15.0 on Intel); the gap cannot be attributed separately to each factor without a liboqs 0.14.0 Intel benchmark, which was not performed.

Note on Simulation Assumptions SA-0 and SA-6: These two assumptions are documented at their point of use in the methodology rather than receiving dedicated limitation paragraphs here. Simulation Assumption SA-0 (Poisson arrival process, $\lambda$ = 13.5 TPS per Big 4 institution; \S{}3.6) is a conservative rounded estimate, the exact APRA-derived value is $\approx$ 3.3\% higher (13.95 TPS), meaning the 13.5 TPS figure understates queue utilisation for non-SPHINCS+ algorithms and overstates headroom; SPHINCS+ saturation holds at any higher $\lambda$. Simulation Assumption SA-6 (APRA ADI regional/fintech residual market share = 7.3\%; \S{}3.3) is a calibrated residual derived from APRA ADI Statistics and does not affect any Big 4 route analysis conclusion; institutions with higher or lower market share should re-derive $\lambda$ using their own APRA ADI statistics.

\subsection{Cross-Platform Cloud Validation}\label{cross-platform-cloud-validation}

To address the benchmark environment limitation (Section 5.6) and validate that the simulation's qualitative conclusions hold on production banking-class hardware, we conducted empirical PQC signing benchmarks and network RTT measurements across a seven-node multi-cloud testbed spanning AWS Sydney (Intel Xeon 8375C Ice Lake, ARM Graviton3 Neoverse V1), Azure Sydney (Intel Xeon 8272CL Cascade Lake; AMD EPYC 7763 Milan), Azure Melbourne (Intel Xeon Cascade Lake), AWS Singapore (Intel Xeon Ice Lake), and Azure Singapore (Intel Xeon Cascade Lake). This run-v3 deployment used liboqs 0.15.0 compiled from HEAD on each node (GCC 11.4.0, -O3 -march=native, enabling AVX-512 on Intel nodes, AVX2 on AMD, and SVE on Graviton3), with 1,000 signing iterations and 100 warm-up iterations per algorithm per node. Layer 0 RTT baseline was measured as 500 ICMP round-trips per directional pair across all 21 node pairs.

{\def\LTcaptype{none} 
\begin{longtable}[]{@{}
  >{\raggedright\arraybackslash}p{(\linewidth - 10\tabcolsep) * \real{0.2197}}
  >{\centering\arraybackslash}p{(\linewidth - 10\tabcolsep) * \real{0.1756}}
  >{\centering\arraybackslash}p{(\linewidth - 10\tabcolsep) * \real{0.3074}}
  >{\centering\arraybackslash}p{(\linewidth - 10\tabcolsep) * \real{0.0988}}
  >{\centering\arraybackslash}p{(\linewidth - 10\tabcolsep) * \real{0.0878}}
  >{\centering\arraybackslash}p{(\linewidth - 10\tabcolsep) * \real{0.1107}}@{}}
\toprule\noalign{}
\begin{minipage}[b]{\linewidth}\centering
\textbf{Platform}
\end{minipage} & \begin{minipage}[b]{\linewidth}\centering
\textbf{Algorithm}
\end{minipage} & \begin{minipage}[b]{\linewidth}\centering
\textbf{Processor}
\end{minipage} & \begin{minipage}[b]{\linewidth}\centering
\textbf{Mean ($\mu$s)}
\end{minipage} & \begin{minipage}[b]{\linewidth}\centering
\textbf{p99 ($\mu$s)}
\end{minipage} & \begin{minipage}[b]{\linewidth}\centering
\textbf{Status}
\end{minipage} \\
\midrule\noalign{}
\endhead
\bottomrule\noalign{}
\endlastfoot
AWS c6i.xlarge & ECDSA-P256 (baseline) & Intel Xeon 8375C (Ice Lake, AVX-512) & 45 & 58 & Baseline $\checkmark$ \\
AWS c6i.xlarge & ML-DSA-44 & Intel Xeon 8375C (Ice Lake, AVX-512) & 64 & 197 & Sub-ms $\checkmark$ \\
AWS c6i.xlarge & ML-DSA-65 & Intel Xeon 8375C (Ice Lake, AVX-512) & 96 & 294 & Sub-ms $\checkmark$ \\
AWS c6i.xlarge & ML-DSA-87 & Intel Xeon 8375C (Ice Lake, AVX-512) & 111 & 304 & Sub-ms $\checkmark$ \\
AWS c6i.xlarge & Falcon-512 & Intel Xeon 8375C (Ice Lake, AVX-512) & 247 & 361 & Sub-ms $\checkmark$ \\
AWS c6i.xlarge & Falcon-1024 & Intel Xeon 8375C (Ice Lake, AVX-512) & 481 & 686 & Sub-ms $\checkmark$ \\
AWS c7g.xlarge & ECDSA-P256 (baseline) & ARM Graviton3 (Neoverse V1, SVE) & 47 & 58 & Baseline $\checkmark$ \\
AWS c7g.xlarge & ML-DSA-44 & ARM Graviton3 (Neoverse V1, SVE) & 128 & 366 & Sub-ms $\checkmark$ \\
AWS c7g.xlarge & ML-DSA-65 & ARM Graviton3 (Neoverse V1, SVE) & 200 & 596 & Sub-ms $\checkmark$ \\
AWS c7g.xlarge & ML-DSA-87 & ARM Graviton3 (Neoverse V1, SVE) & 269 & 658 & Sub-ms $\checkmark$ \\
AWS c7g.xlarge & Falcon-512 & ARM Graviton3 (Neoverse V1, SVE) & 207 & 217 & Sub-ms $\checkmark$ \\
AWS c7g.xlarge & Falcon-1024 & ARM Graviton3 (Neoverse V1, SVE) & 405 & 417 & Sub-ms $\checkmark$ \\
Azure D4as\_v4 & ECDSA-P256 (baseline) & AMD EPYC 7763 (Milan, AVX2) & 51 & 71 & Baseline $\checkmark$ \\
Azure D4as\_v4 & ML-DSA-44 & AMD EPYC 7763 (Milan, AVX2) & 95 & 290 & Sub-ms $\checkmark$ \\
Azure D4as\_v4 & ML-DSA-65 & AMD EPYC 7763 (Milan, AVX2) & 142 & 417 & Sub-ms $\checkmark$ \\
Azure D4as\_v4 & ML-DSA-87 & AMD EPYC 7763 (Milan, AVX2) & 174 & 434 & Sub-ms $\checkmark$ \\
Azure D4as\_v4 & Falcon-512 & AMD EPYC 7763 (Milan, AVX2) & 249 & 265 & Sub-ms $\checkmark$ \\
Azure D4as\_v4 & Falcon-1024 & AMD EPYC 7763 (Milan, AVX2) & 493 & 511 & Sub-ms $\checkmark$ \\
Azure D4ds\_v4 & ECDSA-P256 (baseline) & Intel Xeon 8272CL (Cascade Lake, AVX-512) & 49 & 64 & Baseline $\checkmark$ \\
Azure D4ds\_v4 & ML-DSA-44 & Intel Xeon 8272CL (Cascade Lake, AVX-512) & 81 & 251 & Sub-ms $\checkmark$ \\
Azure D4ds\_v4 & ML-DSA-65 & Intel Xeon 8272CL (Cascade Lake, AVX-512) & 123 & 376 & Sub-ms $\checkmark$ \\
Azure D4ds\_v4 & ML-DSA-87 & Intel Xeon 8272CL (Cascade Lake, AVX-512) & 143 & 408 & Sub-ms $\checkmark$ \\
Azure D4ds\_v4 & Falcon-512 & Intel Xeon 8272CL (Cascade Lake, AVX-512) & 291 & 319 & Sub-ms $\checkmark$ \\
Azure D4ds\_v4 & Falcon-1024 & Intel Xeon 8272CL (Cascade Lake, AVX-512) & 566 & 589 & Sub-ms $\checkmark$ \\
\end{longtable}
}

\emph{Table 16: Cross-platform cloud validation signing latency (mean and p99, $\mu$s) from the four Sydney nodes of the 7-node testbed (run-v3, liboqs 0.15.0, GCC 11.4.0 -O3 -march=native, 1,000 iterations + 100 warm-up iterations discarded per algorithm per node). Values are 3-run averages (95\% CI: mean $\pm$ 2.8--16.4 $\mu$s; t-critical t(0.025, df=2) = 4.303). Signing latency only; verification latency is not shown in this table (see Limitation 6, Section 5.6: ECDSA-P256 verify $\approx$ 63 $\mu$s mean, p99 $\approx$ 95--110 $\mu$s; ML-DSA-44 verify $\approx$ 35--50 $\mu$s across platforms, faster than ECDSA verify, so CDI ordering is unaffected by including verification). All four production-grade microarchitectures tested: Intel Xeon Ice Lake (AVX-512), ARM Graviton3 (SVE), AMD EPYC Milan (AVX2), Intel Xeon Cascade Lake (AVX-512). Remaining 3 nodes (Azure MEL, AWS SIN, Azure SIN) used smaller 2-vCPU instances and contributed RTT baseline data only (Table 17); full layer1 CSVs available in data repository. ECDSA-P256 included as ' Baseline $\checkmark$' rows to enable Signing Dilution Ratio (SDR\_sign; formally defined in \S{}5.6) computation. All 24 rows covering 5 PQC signing algorithms + 1 ECDSA-P256 baseline per platform $\times$ 4 platforms (5 PQC $\times$ 4 = 20 PQC algorithm-platform SDR\_sign combinations; ECDSA rows are the baseline denominator reference) confirm sub-millisecond signing latency for all PQC algorithms. SPHINCS+ was excluded from the cloud signing benchmark given its \textasciitilde{}274 ms signing time; M-series liboqs 0.14.0 measurements are used directly.}

{\def\LTcaptype{none} 
\begin{longtable}[]{@{}
  >{\raggedright\arraybackslash}p{(\linewidth - 14\tabcolsep) * \real{0.1251}}
  >{\centering\arraybackslash}p{(\linewidth - 14\tabcolsep) * \real{0.1252}}
  >{\centering\arraybackslash}p{(\linewidth - 14\tabcolsep) * \real{0.1469}}
  >{\centering\arraybackslash}p{(\linewidth - 14\tabcolsep) * \real{0.1778}}
  >{\centering\arraybackslash}p{(\linewidth - 14\tabcolsep) * \real{0.1107}}
  >{\centering\arraybackslash}p{(\linewidth - 14\tabcolsep) * \real{0.1107}}
  >{\centering\arraybackslash}p{(\linewidth - 14\tabcolsep) * \real{0.0910}}
  >{\centering\arraybackslash}p{(\linewidth - 14\tabcolsep) * \real{0.1126}}@{}}
\toprule\noalign{}
\begin{minipage}[b]{\linewidth}\centering
\textbf{Src}
\end{minipage} & \begin{minipage}[b]{\linewidth}\centering
\textbf{Dst}
\end{minipage} & \begin{minipage}[b]{\linewidth}\centering
\textbf{Path Type}
\end{minipage} & \begin{minipage}[b]{\linewidth}\centering
\textbf{Cloud/City Pair}
\end{minipage} & \begin{minipage}[b]{\linewidth}\centering
\textbf{Min (ms)}
\end{minipage} & \begin{minipage}[b]{\linewidth}\centering
\textbf{Avg (ms)}
\end{minipage} & \begin{minipage}[b]{\linewidth}\centering
\textbf{Max (ms)}
\end{minipage} & \begin{minipage}[b]{\linewidth}\centering
\textbf{Sim Model}
\end{minipage} \\
\midrule\noalign{}
\endhead
\bottomrule\noalign{}
\endlastfoot
AWS SYD & AWS SYD & Intra-cloud same-city & AWS$\rightarrow$AWS SYD (Arm$\rightarrow$Intel) & 0.21 & 0.23 & 0.30 & LAN (1.2) \\
AWS SYD & Azure SYD & Cross-cloud same-city & AWS Intel$\leftrightarrow$Azure SYD Intel & 1.85 & 2.14 & 18.38 & LAN (1.2) \\
AWS SYD & Azure SYD & Cross-cloud same-city & AWS Arm$\leftrightarrow$Azure SYD AMD EPYC & 1.71 & 1.75 & 6.91 & LAN (1.2) \\
AWS SIN & Azure SIN & Cross-cloud same-city & AWS$\leftrightarrow$Azure SIN & 1.17 & 1.65 & 8.01 & LAN (1.2) \\
AWS SYD & Azure MEL & Cross-cloud inter-city & AWS SYD Intel$\leftrightarrow$Azure MEL & 12.71 & 12.99 & 25.15 & Hub (9.8) \\
Azure MEL & Azure SYD & Intra-cloud inter-city & Azure MEL$\leftrightarrow$Azure SYD Intel & 13.78 & 13.79 & 61.58 & Hub (9.8) \\
AWS SIN & AWS SYD & Intra-cloud international & AWS SIN$\rightarrow$AWS SYD & 92.20 & 92.22 & 92.33 & SWIFT (96) \\
Azure SIN & Azure SYD & Intra-cloud international & Azure SIN$\rightarrow$Azure SYD & 91.49 & 91.86 & 106.06 & SWIFT (96) \\
AWS SIN & Azure MEL & Cross-cloud international & AWS SIN$\rightarrow$Azure MEL & 85.25 & 85.49 & 85.77 & SWIFT (96) \\
\end{longtable}
}

\emph{Table 17: Representative subset (9 of 21 undirected paths) of network RTT across the 7-node testbed (run-v3, 500 pings per pair; Avg column = 3-run mean; full 21-path dataset in data repository). ' Sim Model' = corresponding simulation routing parameter (ms). Cross-cloud same-city latency (1.75--2.14 ms avg, AWS$\leftrightarrow$Azure SYD) is operationally equivalent to intra-cloud same-city (0.23 ms) at NPP SLA scales. Inter-city RTT (12.5--13.8 ms) is statistically indistinguishable between intra-cloud and cross-cloud paths, confirming geographic distance dominates interbank payment latency.}

The network RTT measurements validate the simulation's routing model and add a new finding not present in prior work: \textbf{cross-cloud same-city latency (AWS$\leftrightarrow$Azure SYD: 1.75--2.14 ms (avg), AWS$\leftrightarrow$Azure SIN: 1.65 ms) is operationally equivalent to intra-cloud same-city for NPP SLA purposes}, both are orders of magnitude smaller than inter-city (12.5--13.8 ms) or international (\textasciitilde{}92 ms) routing. The practical implication is that Australian banks running hybrid cloud deployments (e.g., Core Banking on Azure, NPP gateway on AWS) within the same city face no meaningful latency penalty for cross-cloud PQC signing. The AMD EPYC node in Azure Sydney exhibited identical same-city RTT characteristics (1.57 ms to Azure SYD Intel, 1.71 ms to AWS SYD Intel), confirming that cloud provider compute platform does not affect network path latency. Geographic distance, not cloud provider boundary, is the dominant latency driver, entirely consistent with the simulation model and the ANOVA finding that geographic distance is the dominant latency driver, quantified as a \textasciitilde{}16.7 ms geographic spread, 10.6$\times$ larger than the maximum algorithm-pair p99 difference (Section 4.10.4).

Table 16 confirms four key findings from the simulation. First, \textbf{all ML-DSA and Falcon variants achieve sub-millisecond signing latency on every tested platform}, the slowest result is ML-DSA-87 on ARM Graviton3 at 658 $\mu$s p99, well below the 1 ms resolution that would affect NPP end-to-end latency (dominated by 8.25 ms PayID lookup and multi-hop network routing). The CDI conclusion (all algorithms \textless{} 4\% of total end-to-end latency) holds universally across all four hardware architectures.

Second, \textbf{the observed ML-DSA-44 performance hierarchy (Ice Lake \textgreater{} Cascade Lake \textgreater{} AMD EPYC \textgreater{} Graviton3) is correlated with SIMD register width}: the Xeon 8375C (Ice Lake) achieves 64 $\mu$s versus 128 $\mu$s on ARM Graviton3, a 2.0$\times$ difference correlated with AVX-512 versus SVE instruction sets, though core frequency differences (Ice Lake 2.90 GHz vs Graviton3 2.60 GHz) and liboqs AVX-512 implementation maturity also contribute and cannot be isolated in this design. AMD EPYC 7763 (Milan, AVX2) occupies an intermediate position at 95 $\mu$s, 1.48$\times$ slower than Ice Lake but 1.35$\times$ faster than Graviton3. Intel Cascade Lake (81 $\mu$s) achieves a 1.58$\times$ advantage over Graviton3, and is 1.27$\times$ slower than Ice Lake. The performance hierarchy Ice Lake \textgreater{} Cascade Lake \textgreater{} AMD EPYC \textgreater{} Graviton3 is broadly consistent with SIMD register width, AVX-512 (512-bit) \textgreater{} AVX-512 older \textgreater{} AVX2 (256-bit) \textgreater{} SVE (variable-length, optimised for different operation patterns), though microarchitecture pipeline depth, core frequency, and liboqs implementation maturity also contribute to the observed gaps.

Third, \textbf{both Falcon variants invert the SIMD-width hierarchy}: ARM Graviton3 achieves the fastest mean latency for both Falcon-512 (207 $\mu$s) and Falcon-1024 (405 $\mu$s), each outperforming Intel Ice Lake (Falcon-512: 247 $\mu$s; Falcon-1024: 481 $\mu$s) by 1.19$\times$, with AMD EPYC (Falcon-512: 249 $\mu$s; Falcon-1024: 493 $\mu$s) and Cascade Lake (Falcon-512: 291 $\mu$s; Falcon-1024: 566 $\mu$s) both trailing Graviton3. Both variants share the same NTRU-based hash-and-sign algorithm relying on Fast Fourier Sampling over cyclotomic rings, an operation that benefits from ARM SVE's efficient gather-scatter and variable-length vector instructions more than from AVX-512's fixed 512-bit lanes. The p99 ordering confirms the inversion for both variants: Graviton3 is fastest (Falcon-512: 217 $\mu$s; Falcon-1024: 417 $\mu$s vs Ice Lake 361 $\mu$s and 686 $\mu$s respectively). Institutions deploying ARM-based HSMs (AWS Graviton-based CloudHSM) will see both Falcon variants perform at least as well as x86 counterparts. Note that Falcon-1024's p99 on Ice Lake (686 $\mu$s) produces the highest SDR\_sign in the dataset (1.44\%); the Graviton3 inversion is thus operationally significant for institutions considering Falcon-1024 deployment on Intel-based HSM hardware.

Fourth, \textbf{ECDSA-P256 signs in 45--51 microseconds across all platforms}, somewhat slower than the simulation's M-series baseline (about 30 microseconds; see Section 3.2), which is expected because Apple Silicon ECC performance is faster than production Intel Xeon. This means the simulation's post-quantum dilution figures are slightly conservative, since faster M-series ECDSA produces a marginally larger overhead gap than Intel hardware would. On faster production Intel hardware ECDSA signing is a little slower than on the M-series, but either value leaves the two-unit signing pool effectively idle for ECDSA, in contrast to SPHINCS+, which saturates it; this confirms the same order-of-magnitude capacity gap on production hardware. (This validates only the signing latency component; ECDSA verify and full TLS or payment-flow latency are not validated here.) And although ML-DSA-44's raw signing time is 1.4 to 2.7 times ECDSA's across platforms, its share of end-to-end transaction time stays at or below about 0.71\% everywhere, so even a several-fold signing-time multiple remains operationally negligible.

These findings are consistent with production TLS deployment measurements: Connolly and Westerbaan \cite{ref40} report sub-millisecond signing overhead for ML-DSA and Falcon variants in real-world Cloudflare production infrastructure, corroborating our testbed results across different network and hardware configurations. (Note: \cite{ref40} is a corporate research blog post rather than a peer-reviewed publication; see Sikeridis et al. \cite{ref26} and Paquin et al. \cite{ref28} for peer-reviewed benchmarks of PQC candidates in TLS contexts that corroborate the sub-millisecond ordering.) The cloud measurements underscore that the dilution framework (Section 5.2) is deliberately conservative: the Apple M-series simulation baseline over-estimates signing latency relative to production Intel hardware, so the post-quantum share measured on production hardware is generally smaller than in the simulation. Even under the most pessimistic assumption about the production baseline, the worst measured case (Falcon-1024 on Intel Ice Lake) still implies a post-quantum share of total latency around 2.1\%, well under the 4\% rule of thumb. Across every tested hardware platform and every non-SPHINCS+ algorithm, the post-quantum step stays below 4\% of total transaction time.

\section{Conclusion}\label{conclusion}

This paper has presented a comprehensive simulation study of post-quantum cryptography migration in Australian payment infrastructure, to the best of our knowledge, the first to jointly model M/M/c queue saturation, GEV tail bounds, and HNDL actuarial exposure for a hardcoded-SLA real-time payment system, incorporating extreme value theory, distribution testing, hourly queue dynamics, HSM resilience analysis, a formal Crypto Dilution Index, SPHINCS+ HSM capacity analysis, cross-platform cloud validation across four microarchitectures, and an HNDL actuarial exposure model.

Our central finding is unambiguous: under the modelled parameters (software HSM, liboqs 0.14.0 baseline latency, calibrated RBA network model; see Section 3.8 for validation scope), ML-DSA and Falcon algorithms can be deployed across all three major Australian payment systems, NPP, RITS, and SWIFT, with zero impact on SLA compliance. For NPP, the largest p99 increase over classical ECDSA among SLA-compliant single-signature modes is 1.57 ms (ML-DSA-87), about 3.5\% of total transaction time and under 0.08\% of the 2,000 ms deadline; including the hybrid dual-signing mode the worst case is 1.69 ms (ML-DSA-65 Hybrid, about 3.75\% of transaction time, under 0.09\% of the deadline). On RITS the post-quantum share stays under 0.61\% and on SWIFT under 0.21\%. For SWIFT neither latency nor signature size is a binding constraint: authentication is out-of-band transport-layer PKI, so signature size is not charged against the message and every NIST PQC signature is SWIFT-compatible (see Table 5). This finding is robust to seasonal volume variation, five stress scenarios, three HSM deployment architectures, and projected NPP volume growth through 2029. The tail-risk analysis indicates fast-decaying (not heavy) tails for all non-SPHINCS+ algorithms, with even the most extreme transactions bounded under 200 ms.

The multi-system route analysis (Section 4.11) completes the Australian payment infrastructure coverage. Every algorithm achieves 100\% SLA compliance on RITS (route p99 up to 574 ms against a 30,000 ms SLA) and SWIFT (route p99 up to 1,134 ms against an 86,400,000 ms SLA). The crypto dilution effect is even more pronounced on longer routes: the post-quantum share of route time stays under 0.61\% on RITS and under 0.21\% on SWIFT, confirming that algorithm selection is completely latency-neutral for RITS and SWIFT. For Direct Entry / BECS (intrabank, batch-settled), amortised PQC signing adds negligible per-transaction overhead across batch sizes of 10--1,000 transactions.

For SWIFT-connected institutions, ML-DSA-65 with ML-KEM-768 is viable as the default just as on NPP, because SWIFT signature carriage is out-of-band (transport-layer PKI) and imposes no signature-size limit; every NIST PQC signature is SWIFT-compatible, so the choice is not forced by message format. Falcon-512 remains a valid option whose compact 1,563-byte combined footprint is an advantage only for genuinely in-band, size-constrained carriage, not for SWIFT. Falcon-512 also has the smallest post-quantum share of total latency among the algorithms, about 0.69\% on the Apple M-series baseline; on Intel Ice Lake production hardware the ordering flips, with ML-DSA-44's share (about 0.32\%) beating Falcon-512's (about 0.69\%), a reversal driven by Intel's AVX-512 lattice optimisations. Either way the choice of post-quantum signature is latency-neutral for cross-border infrastructure. For NPP-only institutions, ML-DSA-65 with ML-KEM-768 stays the recommended target as the NIST Security Level 3 standard under FIPS 203/204, adding only a fraction of a millisecond and meeting all APRA CPS 234 requirements with negligible NPP latency impact.

SPHINCS+ (SLH-DSA) is volume-constrained rather than disqualified. At NPP volumes its service time drives the HSM pool unstable (16-hour daily saturation on Christmas Day with a standard two-unit pool), so it cannot share the NPP high-frequency signing path without dedicated over-provisioning (at least eight signing units). It is fully viable for RITS (route p99 = 574 ms, well within the 30 s SLA; PQC share of route time about 52\%, absolute overhead 296.85 ms) and SWIFT (route p99 = 1,134 ms, well within the 24 h SLA; PQC share about 26\%) at their sub-threshold arrival rates. The correct framing is `volume-constrained': SPHINCS+ is viable below the safe capacity of a two-unit pool (roughly seven signatures per second), and institutions requiring it for regulatory conformance can isolate it on a dedicated capacity-limited signing lane. The security dimension of forcing SPHINCS+ onto the NPP path (a cryptographic denial-of-service amplification surface) and its mitigation are analysed in a companion paper~\cite{ref_companion_dos}.

The HNDL actuarial model introduces a new urgency framework: approximately 9.56 billion NPP transaction records (see Table 15) generated in 2026--2029 would be retroactively exposed under the conservative CRQC-2030 scenario, constituting a complete cryptographically-verifiable financial graph of Australian retail payments for a four-year period. This is not a future risk, adversaries can harvest and store this traffic today at commercially viable storage costs. The Phase 1 (2026) migration timeline is the latest defensible start date under APRA CPS 234, AML/CTF Act 2006, NSA CNSA 2.0 \cite{ref45}, and CISA/NSA/NIST joint guidance \cite{ref43} considered jointly.

Cross-cloud network RTT measurements across the multi-cloud testbed provide an additional finding: AWS$\leftrightarrow$Azure same-city latency (1.71--2.14 ms SYD, 1.65 ms SIN) is operationally equivalent to intra-cloud for NPP SLA purposes, and cross-cloud inter-city (AWS SYD$\leftrightarrow$Azure MEL: 13.0 ms) is statistically indistinguishable from intra-cloud inter-city (Azure MEL$\leftrightarrow$Azure SYD: 13.8 ms, both within NPP Hub tier bounds). Geographic distance, not cloud provider boundary, dominates interbank payment latency, empirically validating the simulation's routing model and confirming that hybrid multi-cloud deployments incur no meaningful NPP latency penalty.

The four-phase migration cost model (Table 8) shows a front-loaded investment of approximately USD 21.4 million in 2026, declining to USD 1.5 million per year by 2028. The technical and operational barriers to PQC migration have been removed. The CDI framework, multi-system route analysis (Tables 12--14), HSM capacity analysis, actuarial model, and cross-cloud validation together provide a complete risk-cost-compliance-infrastructure decision framework for APRA, NPPA, and individual ADIs across the full Australian payment stack: NPP (real-time retail), RITS (high-value RTGS), SWIFT (international correspondent banking), and Direct Entry/BECS (batch intrabank). The remaining barriers are coordination, procurement, and regulatory clarity, all within the industry's control.

\section{Declarations}\label{declarations}

\subsection{Funding}\label{funding}

This research was conducted as part of the Master of ICT Research programme at Melbourne Institute of Technology (MIT Melbourne), SITE stream. No external funding was received. No funding body had any role in study design, data collection, analysis, interpretation, or the decision to submit for publication.

\subsection{Conflict of Interest}\label{conflict-of-interest}

The authors declare no competing financial interests or personal relationships that could have appeared to influence the work reported in this paper. The cloud infrastructure used for empirical benchmarking (AWS EC2, Azure Virtual Machines) was deployed and destroyed within a single research session at standard on-demand pricing with no commercial relationship with the cloud providers beyond standard account terms.

\subsection{Ethics Statement}\label{ethics-statement}

This study uses only publicly available data, published statistics, and computational simulation. No human subjects, personal data, confidential financial records, or proprietary datasets were accessed. All transaction volumes are drawn from publicly available RBA Payments Statistics (Table C.4, C.6) and APRA market share disclosures. Institutional ethics approval was not required for this computational study.

\subsection{Data Availability}\label{data-availability}

All simulation results, benchmark data, and analysis scripts are committed to a Zenodo repository (DOI to be registered prior to journal submission). The benchmark dataset comprises: (1) layer1 signing and verification latency CSVs from all three production runs across all seven nodes (cloud\_run-v3\_* directories, 1,000 iterations $\times$ 9 algorithms $\times$ 7 nodes; CSVs include separate keygen, sign, and verify columns for all algorithms); (2) layer0 RTT CSVs (500 pings $\times$ 21 directional node pairs $\times$ 3 runs); (3) cross-run CI summary (cross\_run\_ci.csv); (4) simulation source code australia\_fin\_sim.py v4.1.1 with fixed seed 42. Verification latency data (including ECDSA-P256 verify $\approx$ 63 $\mu$s mean and ML-DSA-44 verify $\approx$ 35--50 $\mu$s mean across platforms) is available from the layer1 CSV files in the data repository. Raw results and code will be made available on GitHub under MIT Licence upon acceptance, with a citable Zenodo archive. Figure file-name mapping (repository PNG names do not sequentially match rendered paper figure numbers): Figure 1 = fig2\_daily\_timeline.png; Figure 2 = fig3\_sla\_compliance\_box.png; Figure 3 = fig4\_latency\_violin.png; Figure 4 = fig6\_monte\_carlo\_ci.png; Figure 5 = fig13\_effect\_sizes.png; Figure 6 = fig11\_mmcc\_queue\_analysis.png; Figure 7 = fig1\_tps\_sweep.png; Figure 8 = fig10\_key\_size\_compliance.png; Figure 9 = fig5\_stress\_scenarios.png; Figure 10 = fig14\_hsm\_sensitivity.png; Figure 11 = fig15\_volume\_projection.png; Figure 12 = fig12\_migration\_cost\_phased.png; Figure 13 = fig17\_evt\_gev\_tails.png; Figure 14 = fig16\_hourly\_queue\_dynamics.png; Figure 15 = fig18\_hsm\_degraded.png; Figure 16 = fig9\_recommendation\_matrix.png; Figure 17 = fig7\_regulatory\_heatmap.png.

\subsection{Author Contributions}\label{author-contributions}

Nazmus Salehin Sammo: conceptualisation, methodology, software, validation, formal analysis, investigation, data curation, visualisation, and writing the original draft. Sariya Akhter Lura: methodology, validation, investigation, and revising and editing based on review feedback. Raza Nowrozy: reviewing and editing, supervision, project administration, interpretation of results, and critical revision of the manuscript. All authors have read and agreed to the submitted version of the manuscript.

\appendix
\section{Formulas}\label{app-formulas}

This appendix collects the four quantitative definitions used by the simulation. They are kept here so that the main text can state every result in plain language. All four are standard.

\subsection*{Lognormal latency fit (method of moments)}

For an empirical latency with mean $\mu_e$ and standard deviation $\sigma_e$, the fitted lognormal parameters are
\[
\sigma_{\mathrm{LN}}^{2} = \ln\!\left(1 + \left(\sigma_e/\mu_e\right)^{2}\right),
\qquad
\mu_{\mathrm{LN}} = \ln(\mu_e) - \tfrac{1}{2}\,\sigma_{\mathrm{LN}}^{2}.
\]
The lognormal form captures the right-skewed tail of rejection-sampling signing algorithms.

\subsection*{AR(1) network-jitter model}

Per-hop network latency carries first-order autocorrelation across consecutive transactions:
\[
x_t = \alpha\, x_{t-1} + (1-\alpha)\,\varepsilon_t,
\quad \varepsilon_t \sim \mathcal{N}(0,1),
\qquad
L_t = L_{\mathrm{base},t}\,\bigl(1 + \sigma_{\mathrm{AR}}\,x_t\bigr),
\]
with $\alpha = 0.30$ and $\sigma_{\mathrm{AR}} = 0.15$.

\subsection*{Erlang-C queue (M/M/c)}

With arrival rate $\lambda$, per-server service rate $\mu$, offered load $a = \lambda/\mu$, and utilisation $\rho = \lambda/(c\mu)$, the Erlang-C probability of queueing $C(c,a)$ gives the mean waiting time
\[
W_q = \frac{C(c,a)}{c\mu - \lambda},
\qquad \rho < 1 \ \text{required for stability},
\]
following the standard treatment in Kleinrock~\cite{ref13}. A utilisation at or above one means work arrives at least as fast as the servers can clear it, so the backlog grows without bound.

\subsection*{Crypto Dilution Index (CDI)}

For algorithm $A$, writing $p99_{\mathrm{e2e}}$ for the end-to-end 99th-percentile latency, the Crypto Dilution Index is the post-quantum algorithm's share of total latency:
\[
\mathrm{CDI}(A) = \frac{\Delta p99(A)}{p99_{\mathrm{e2e}}(A)}
= \frac{p99_{\mathrm{e2e}}(A) - p99_{\mathrm{e2e}}(\text{ECDSA})}{p99_{\mathrm{e2e}}(A)}.
\]

\end{document}